\documentclass[a4paper,fleqn,usenatbib]{mnras}

\usepackage[T1]{fontenc}
\usepackage{ae,aecompl}
\usepackage{graphicx,longtable,multirow,enumitem,amsmath,amssymb,tabularx,lscape}
\usepackage{xspace}

\usepackage{natbib,bm}
\usepackage[usenames]{color}

\usepackage[utf8]{inputenc}

\newcommand{\zlens}{0.6575}
\newcommand{\zsrc}{1.662}
\newcommand{\zgtwo}{0.7450}
\newcommand{\zgthree}{0.6548}
\newcommand{\zgseven}{0.6573}
\newcommand{\zgrp}{0.6588}
\newcommand{\zfggrp}{0.4956}
\newcommand{\dt}{4784_{-248}^{+399}~\mathrm{Mpc}}
\newcommand{\dtprec}{6.6}
\newcommand{\ulcdm}{71.6_{-4.9}^{+3.8}~\mathrm{km~s^{-1}~Mpc^{-1}}}
\newcommand{\equalweight}{72.2_{-4.8}^{+4.3}~\mathrm{km~s^{-1}~Mpc^{-1}}}
\newcommand{\combprec}{3.5}
\newcommand{\rani}{r_{\rm ani}}
\newcommand{\reff}{r_{\rm eff}}
\newcommand{\appropto}{\mathrel{\vcenter{
  \offinterlineskip\halign{\hfil$##$\cr
    \propto\cr\noalign{\kern2pt}\sim\cr\noalign{\kern-2pt}}}}}

\newcommand{\bd}{\begin{displaymath}}
\newcommand{\ed}{\end{displaymath}}
\newcommand{\be}{\begin{equation}}
\newcommand{\ee}{\end{equation}}
\newcommand{\beaa}{\begin{eqnarray*}}
\newcommand{\eeaa}{\end{eqnarray*}}
\newcommand{\bea}{\begin{eqnarray}}
\newcommand{\eea}{\end{eqnarray}}

\def\hequad{HE\,0435$-$1223}

\def\wfilens{WFI2033$-$4723}
\def\blens{B1608$+$656}
\def\rxjlens{RXJ1131$-$1231}
\def\sdsslens{SDSS 1206$+$4332}
\def\pglens{PG 1115$+$080}

\def\Om{\Omega_{\rm m}}
\def\OL{\Omega_{\Lambda}}

\def\tdist{D_{\Delta t}}
\def\tdistmod{D_{\Delta t}^{\rm model}}
\def\Dd{D_{\rm d}}
\def\Dds{D_{\rm ds}}
\def\Ds{D_{\rm s}}

\def\kext{\kappa_{\rm ext}}
\def\gext{\gamma_{\rm ext}}

\def\zd{z_{\rm d}}
\def\zs{z_{\rm s}}

\def\hst{\textit{HST}}

\def\GLEE{{\sc Glee}\xspace}

\def\kms {\rm km\,s^{-1}}

\newcommand{\sref}[1]{Section~\ref{#1}}
\newcommand{\fref}[1]{Figure~\ref{#1}}

\newcommand{\eref}[1]{Equation~(\ref{#1})}

\title[\wfilens~Lens Model, $\tdist$ and $H_0$]{H0LiCOW XII. Lens mass model of \wfilens~and blind measurement of its time-delay distance and $H_0$}

\author[C. E. Rusu et al.]{\parbox{\textwidth}{
Cristian E. Rusu,$^{1,2,3}$\thanks{Subaru Fellow; \texttt{E-mail: cerusu@naoj.org}}
Kenneth C. Wong,$^{4,1}$
Vivien Bonvin,$^{5}$
Dominique Sluse,$^{6}$
Sherry H. Suyu,$^{7,8,9}$
Christopher D. Fassnacht,$^{3}$
James H. H. Chan,$^{5}$
Stefan Hilbert,$^{10,11}$
Matthew W. Auger,$^{12}$
Alessandro Sonnenfeld,$^{13,4}$
Simon Birrer,$^{14}$
Frederic Courbin,$^{5}$
Tommaso Treu,$^{14}$
Geoff~C.-F.~Chen,$^{3}$
Aleksi Halkola,$^{15}$
L{\'e}on V. E. Koopmans,$^{16}$
Philip J. Marshall$^{17}$
and Anowar J. Shajib$^{14}$
}
\\
\\
\parbox{\textwidth}{
$^{1}$National Astronomical Observatory of Japan, 2-21-1 Osawa, Mitaka, Tokyo 181-8588, Japan\\
$^{2}$Subaru Telescope, National Astronomical Observatory of Japan, 650 N Aohoku Pl, Hilo, HI 96720\\
$^{3}$Department of Physics, University of California, Davis, 1 Shields Avenue, Davis, CA 95616, USA\\
$^{4}$Kavli Institute for the Physics and Mathematics of the Universe (Kavli IPMU, WPI), University of Tokyo, Chiba 277-8583, Japan\\
$^{5}$Laboratoire d'Astrophysique, Ecole Polytechnique F{\'e}d{\'e}rale de Lausanne (EPFL), Observatoire de Sauverny, CH-1290 Versoix, Switzerland\\
$^{6}$STAR Institute, Quartier Agora - All\'ee du six Ao\^ut, 19c B-4000 Li\`ege, Belgium\\
$^{7}$Max-Planck-Institut f{\"u}r Astrophysik, Karl-Schwarzschild-Str.~1, 85748 Garching, Germany\\
$^{8}$Physik-Department, Technische Universit\"at M\"unchen, James-Franck-Stra\ss{}e~1, 85748 Garching, Germany\\
$^{9}$Institute of Astronomy and Astrophysics, Academia Sinica, 11F of ASMAB, No.1, Section 4, Roosevelt Road, Taipei 10617, Taiwan\\
$^{10}$Exzellenzcluster Universe, Boltzmannstr. 2, 85748 Garching, Germany\\
$^{11}$Ludwig-Maximilians-Universit{\"a}t, Universit{\"a}ts-Sternwarte, Scheinerstr. 1, 81679 M{\"u}nchen, Germany\\
$^{12}$Institute of Astronomy, University of Cambridge, Madingley Rd, Cambridge, CB3 0HA, UK\\
$^{13}$Leiden Observatory, Leiden University, Niels Bohrweg 2, 2333 CA Leiden, The Netherlands\\
$^{14}$Department of Physics and Astronomy, University of California, Los Angeles, CA 90095-1547, USA\\
$^{15}$Py{\"o}rrekuja 5A, 04300 Tuusula, Finland\\
$^{16}$Kapteyn Astronomical Institute, University of Groningen, PO Box 800, NL-9700 AV Groningen, The Netherlands\\
$^{17}$Kavli Institute for Particle Astrophysics and Cosmology, Stanford University, 452 Lomita Mall, Stanford, CA 94035, USA\\
}}

\date{Accepted XXX. Received YYY; in original form ZZZ}
\pubyear{2019}

\begin{document}
\label{firstpage}
\pagerange{\pageref{firstpage}--\pageref{lastpage}}
\maketitle

\begin{abstract}
We present the lens mass model of the quadruply-imaged gravitationally lensed quasar \wfilens, and perform a blind cosmographical analysis based on this system. Our analysis combines (1) time-delay measurements from 14 years of data obtained by the COSmological MOnitoring of GRAvItational Lenses (COSMOGRAIL) collaboration, (2) high-resolution {\it Hubble Space Telescope} imaging, (3) a measurement of the velocity dispersion of the lens galaxy based on ESO-MUSE data, and (4) multi-band, wide-field imaging and spectroscopy characterizing the lens environment. We account for all known sources of systematics, including the influence of nearby perturbers and complex line-of-sight structure, as well as the parametrization of the light and mass profiles of the lensing galaxy. After unblinding, we determine the effective time-delay distance to be $\dt$, an average precision of $\dtprec\%$. This translates to a Hubble constant $H_{0} = \ulcdm$, assuming a flat $\Lambda$CDM cosmology with a uniform prior on $\Omega_\mathrm{m}$ in the range [0.05, 0.5]. This work is part of the $H_0$ Lenses in COSMOGRAIL's Wellspring (H0LiCOW) collaboration, and the full time-delay cosmography results from a total of six strongly lensed systems are presented in a companion paper (H0LiCOW XIII).
\end{abstract}



\section{Introduction} \label{sec:intro}

The flat $\Lambda$CDM cosmological model, characterized by spatial flatness, dark energy in the form of a cosmological constant, and cold dark matter, is considered to be the standard cosmological model today. Although this model is known as the concordance model, sources of tension have nonetheless begun to appear as the results of different cosmological experiments have grown in precision. Most notably, the tension between the measurements of the Hubble constant from the analysis of the cosmic microwave background (CMB) by the {\it Planck} mission (under the strict assumption of flat $\Lambda$CDM) and of Type Ia supernovae standard candles calibrated using the local distance ladder by the Supernovae, H0, for the Equation of State of Dark Energy collaboration \citep[SH0ES;][]{riess2016} has recently increased from $3.4\sigma$ \citep{planck2015,riess2016} to $4.4\sigma$ \citep{planck2018,riess2019}. The latest results are $H_0=67.4\pm0.5~\mathrm{km~s^{-1}~Mpc^{-1}}$ from {\it Planck},
and $H_0=74.03\pm1.42~\mathrm{km~s^{-1}~Mpc^{-1}}$ from SH0ES.

At present, sources of systematic error in either of these measurements that are significant enough to explain the discrepancy have not been demonstrated to exist. This opens up the intriguing possibility of having to extend the standard cosmological model by allowing for curvature, more general dark energy, or increasing the number of neutrinos \citep[see Figure 13~in][]{riess2019}, or to consider exotic alternatives, such as a vacuum phase transition \citep{valentino2018}, early dark energy models \citep{poulin2018}, self-interacting neutrinos \citep{kreisch2019} or decaying dark matter \citep{vattis2019}. The various parameters of such extensions are highly degenerate with the value of $H_{0}$, and therefore a high-precision determination, with a technique independent of, and therefore not subject to the same systematics of either {\it Planck} or SH0ES, is in demand \citep[e.g.,][]{hu2005,suyu2012c,weinberg2013}. Some proposed independent methods, such as water masers \citep[e.g.,][]{gao2016,braatz2018}, extragalactic background light attenuation \citep[e.g.,][]{dominguez2019}, and gravitational waves \citep[e.g.,][]{feeney2019}, etc. have yet to resolve the $H_{0}$ discrepancy, as their precision is not yet comparable to {\it Planck} or SH0ES.

The time-delay cosmography technique uses gravitational lens time delays to measure $H_{0}$. This technique rests on the fact that light rays from a multiply-imaged source will take different paths as they propagate through spacetime, with different geometrical lengths and gravitational potential depths. This will introduce an offset in arrival times, which can be measured through monitoring, if the source brightness varies in time. The measured time delays are used to infer the ``time delay distance", primarily sensitive to $H_{0}$, which therefore provides a one-step way of measuring $H_{0}$ \citep[e.g.,][]{vanderriest1989,keeton1997,oguri2007,suyu2010b}. Although proposed more than half a century ago by \citet{refsdal1964} in the context of lensed supernovae, the original idea has only recently been implemented \citep{grillo2018}. Far more common is the use of gravitationally lensed quasars, given the sample of 250 such systems known to date \citep[e.g.,][]{lemon2019}. 

In practice, an accurate measurement of $H_{0}$ through this method requires extensive observational data for each system, as well as the development of advanced modeling techniques \citep{suyu2010a,suyu2012b,tewes2013a,birrer2015,bonvin2016,birrer2018}, and has only become feasible in the current decade. Our collaboration, $H_{0}$ Lenses in COSMOGRAIL's Wellspring \citep[H0LiCOW;][hereafter H0LiCOW I]{suyu2017} is designed to perform such measurements. We have precise, long-term time-delay measurements from the COSmological MOnitoring of GRAvItational Lenses \citep[COSMOGRAIL;][]{courbin2005,eigenbrod2005,bonvin2018} project.  We use deep high-resolution imaging from the {\it Hubble Space Telescope}
(\hst) or adaptive optics that provide constraints on the lens model not only from the point-source positions, but also from the extended arcs of the lensed quasar host galaxy. Finally, we have velocity dispersion measurements of the lens galaxies and characterizations of their environments and line of sight \citep[LOS; e.g.,][]{collett2013,greene2013,mccully2014,mccully2017,tihhonova2018}, in order to reduce the mass-sheet degeneracy \citep[e.g.,][]{falco1985,schneider2013}. 

With four lenses, we measured $H_0=72.5_{-2.3}^{+2.1}~\mathrm{km~s^{-1}~Mpc^{-1}}$ with a precision of 3.0\% \citep[][hereafter H0LiCOW IX]{birrer2019} including systematic uncertainties, achieving our previous goal of the program of reaching $< \combprec\%$ precision from the five separate lenses in the base H0LiCOW sample (see H0LiCOW I) and finding good agreement with SH0ES. We have thus shown that we are on track to measure $H_0$ with a precision of 1\% from a future sample of $\sim40$ lenses with comparable precision per system \citep[e.g.,][]{treu2016,shajib2018}, a result which will have significant implications for understanding the current tension with the CMB value. Time-delay cosmography is therefore a very effective technique, in the sense that only a relatively small number of systems is required to achieve a tight precision. The efficiency is similar to that expected for gravitational wave detections with optical counterparts \citep{chenhy2018}.  As we work towards the 1\% precision goal from a sample of lenses, it is important to keep systematics in the inference of $H_0$ from individual systems within the 1\% threshold, in order to insure accuracy, and also to test for biases by using multiple codes \citep{birrer2019} and data challenges \citep{liao2015,ding2018}.

In this paper, we present the results of a detailed lens modeling
analysis of the gravitational lens \wfilens~(J2000: $20^{\mathrm{h}}33^{\mathrm{m}}41\mbox{\ensuremath{.\!\!^{\mathrm{s}}}}9$,
$-47^{\circ}23\arcmin43\farcs4$), a quadruply-lensed quasar
discovered by \citet{Morgan2004}. The source redshift is $\zs = \zsrc$ \citep{Sluse2012}, and the main deflector is a massive
elliptical galaxy at a redshift of $\zd = \zlens \pm 0.0002$ \citep[][hereafter H0LiCOW X]{sluse2019}, updating the $\zd = 0.661 \pm 0.001$ measurement from \citet{Eigenbrod2006b}). \citet{bonvin2019b} (hereafter COSMOGRAIL XVIII) measure the time delays between the quasar images based on 14 years of monitoring, and H0LiCOW X study the environment and LOS to the lens, based on multi-band imaging and targeted spectroscopy. Our work supersedes the models presented in \citet{vuissoz2008} (hereafter COSMOGRAIL VII), which are based on monitoring of shorter duration and constrained only by the positions of the quasar images.   

This is the fifth H0LiCOW system analyzed
in this manner, following \blens~\citep{suyu2010b}, \rxjlens~\citep{suyu2013, suyu2014}, \hequad~\citep[][hereafter H0LiCOW IV]{wong2017}, and \sdsslens~(H0LiCOW IX), with a sixth lens, \pglens, analyzed simultaneously \citep{chen2019}. A H0LiCOW milestone paper \citep{wong2019} presents the results of a conjoined cosmographical analysis of these lenses.

This paper is organized as follows. We give a brief overview of
using time-delay lenses for cosmography in \sref{sec:theory}.  In
\sref{sec:data}, we describe the observational data used in our
analysis. We describe our lens modeling procedure in
\sref{sec:lensmod}. In \sref{sec:lensenv} we quantify the effect of the lens environment in terms of an external convergence. The time-delay distance results and their implications for cosmology are presented in
\sref{sec:results}. We summarize our main conclusions in
\sref{sec:conclusions}. 


\section{Summary of Time-Delay Cosmography} \label{sec:theory}

\subsection{Time-delay distance} \label{subsec:tddist}
When a source is gravitationally lensed by a foreground mass, the arrival time of photons traveling from
the source to the observer depends on both the path length and the gravitational potential traversed by the light rays.  For a single lens plane, the excess time delay of an image at a position $\bm{\theta} = (\theta_{1}, \theta_{2})$ on the sky with a corresponding source position $\bm{\beta} = (\beta_{1}, \beta_{2})$
relative to the case of no lensing is
\begin{equation} \label{eq:td_ex}
t(\bm{\theta}, \bm{\beta}) = \frac{\tdist}{c} \left[ \frac{(\bm{\theta} - \bm{\beta})^{2}}{2} - \psi(\bm{\theta}) \right],
\end{equation}
where $\tdist$ is the time-delay distance and
$\psi(\bm{\theta})$ is the lens potential.  The time-delay distance $\tdist$ \citep{refsdal1964,schneider1992,suyu2010b} is defined\footnote{For historical reasons, the time-delay distance is written in terms of angular diameter distances.  A more natural definition is $\tdist \equiv {\hat{D}_{\mathrm{d}} \hat{D}_{\mathrm{s}}}/{\hat{D}_{\mathrm{ds}}}$ where $\hat{D}$  are the proper distances that the photons have travelled.} as
\begin{equation} \label{eq:ddt}
\tdist \equiv (1+\zd) \frac{\Dd \Ds}{\Dds},
\end{equation}
where $\zd$ is the lens redshift, $\Dd$, $\Ds$, and $\Dds$ are the angular diameter
distances between the lens and the observer, the source and the observer, and the lens and the source, respectively.  $\tdist$ has units of distance and is inversely proportional to $H_{0}$, with weak dependence on other cosmological parameters.

The time delay between two images, $i$ and $j$, of a lensed source is the difference of their excess time delays,
\begin{equation} \label{eq:td}
\Delta t_{ij} = \frac{\tdist}{c} \left[ \frac{(\bm{\theta}_{i} - \bm{\beta})^{2}}{2} - \psi(\bm{\theta}_{i}) - \frac{(\bm{\theta}_{j} - \bm{\beta})^{2}}{2} + \psi(\bm{\theta}_{j}) \right],
\end{equation}
where $\bm{\theta}_{i}$ and $\bm{\theta}_{j}$ are the positions of images $i$ and $j$, respectively, in the image plane.
If the source is variable on short timescales (on the order of weeks to months), it is possible to monitor the lensed image fluxes at positions $\bm{\theta}_{i}$ and
$\bm{\theta}_{j}$ and measure the time delay, $\Delta t_{ij}$,
between them \citep[e.g.,][]{vanderriest1989,schechter1997,fassnacht1999,fassnacht2002,kochanek2006,courbin2011}.  The lens potentials at the image positions, $\psi(\bm{\theta}_{i})$ and $\psi(\bm{\theta}_{j})$, as well as at the
source position, $\bm{\beta}$, can be determined from by modeling the system.  In this way, lenses with measured time delays and accurate lens models can constrain $\tdist$, and in turn, $H_{0}$.

If there are multiple deflectors at different redshifts, the
observed time delays depend on combinations of the angular
diameter distances among the observer, the multiple deflectors, and
the source.  The observed image positions are determined by the multi-plane lens equation
\citep[e.g.,][]{blandford1986,kovner1987,schneider1992,petters2001,collett2014,mccully2014}, but there is no longer a unique time-delay distance associated with the system.
However, if the lensing is dominated by the mass in a single redshift
plane, the observed time delays are mostly sensitive to the
time-delay distance (Equation~\ref{eq:ddt}), with the
deflector redshift set to the redshift of the main lens plane.  This approximation is valid for
\wfilens~(see Appendix~\ref{app:cosmo}), and thus our results can be interpreted in terms of the ``effective" time-delay distance,
$\tdist(\zd,\zs)$.  Hereafter, $\tdist$ refers to the
effective time-delay distance unless otherwise indicated.

A complicating factor in determining the time delay is the ``microlensing time delay", an effect first described by \citet{tie2018}.  Stars and compact objects in the lens galaxy can act as microlenses, which causes a differential magnification of the accretion disk of the lensed quasar.  Since the microlensing effect is different at the positions of the various lensed images and varies over time as the microlenses move, this may create an additional bias and scatter in the measurement of the time delay between different images.  The microlensing time delay depends on a number of assumptions about the accretion disk size, its orientation and inclination, and the propagation of radiation through the disk.  The effect tends to be small, of order $\sim$days or shorter, and can be modeled and accounted for under proper assumptions \citep{bonvin2018}.  This effect can also be mitigated by using the relative offsets between the measured time delays and those expected from lens modeling \citep{chen2018}.

Another difficulty is due to
the fact that external perturbations from mass along the LOS can affect to the lens
potential that light rays pass through.  These perturbations
not only can affect the lens model of the system, but also lead to
additional focusing and defocusing of the light rays, which also
affect the measured time delays \citep[e.g.,][]{seljak1994}.  If unaccounted for, these 
perturbers can lead to biased inferences of $\tdist$.  If the
effects of LOS perturbers are small enough that higher-order terms are unimportant \citep{keeton2003,mccully2014}, they can be
approximated by an external convergence term in the lens plane.  The true $\tdist$ is related to the $\tdistmod$ inferred from a mass model by
\begin{equation} \label{eq:ddtkappa}
\tdist = \frac{\tdistmod}{1-\kext} \to H_0 = (1 - \kext) H_0^\mathrm{model}.
\end{equation}
Here, $\kext$ cannot, in general, be constrained from the lens model due
to the mass-sheet degeneracy
\citep[e.g.,][]{falco1985,gorenstein1988,saha2000}, in which the
addition of a uniform mass sheet and a rescaling of the source plane
coordinates can affect the inferred $\tdist$ but
leaves other observables unchanged.

This degeneracy can be substantially mitigated by estimating
the mass distribution along the LOS
\citep[e.g.,][]{fassnacht2006,momcheva2006,momcheva2015,williams2006,wong2011} and assuming that the physical mass of the deflector profile goes to zero at large radius. However, perturbers that are very massive or projected very close to the
lens may need to be included explicitly in the mass model since their higher-order effects need to be accounted for
\citep{mccully2017}.  In contrast, the lens profile is also
degenerate with the time-delay distance in that the radial profile
slope is tightly correlated with the time-delay distance
\citep[e.g.,][]{kochanek2002,wucknitz2002,suyu2012a}.  This degeneracy can affect
models with the same form of mass density profile (e.g., a
power-law density profile), as well as models with different forms of density
profiles (described analytically or not).  Furthermore, this 
degeneracy can mimic the effects of the mass-sheet degeneracy because
different profiles can approximate or exactly match mass-sheet
transformations of one form or another
\citep[e.g.,][]{schneider2013,schneider2014,unruh2017}.  These
degeneracies can be reduced by combining the lensing data with stellar
kinematics information \citep[e.g.,][]{treu2002,koopmans2003,auger2010,suyu2014,yildirim2019}, and by
making reasonable assumptions about the mass profile. Including
a velocity dispersion measurement in the modeling helps constrain any
internal uniform mass component from a local galaxy group that the
dynamics is sensitive to \citep{koopmans2004}.

\subsection{Joint Inference} \label{subsec:joint}
Our inference of $\tdist$ generally follows that of previous H0LiCOW analyses \citep[][H0LiCOW IV, IX]{suyu2013}. Our observational data are denoted by $\bm{d_{\mathrm{HST}}}$ for the
\hst~imaging data, $\bm{\Delta t}$ for the time delays, $\bm{\sigma}$ for the
velocity dispersion of the lens galaxy, and $\bm{d_{\mathrm{LOS}}}$ for
the LOS mass distribution determined from our
photometric and spectroscopic data. We want to determine the posterior
probability distribution function (PDF) of the model parameters
$\bm{\xi}$ given the data, $P(\bm{\xi} | \bm{d_{\rm HST}, \Delta t, \sigma,
d_{\mathrm{LOS}},A})$.  The vector~$\bm{\xi}$ includes the lens model
parameters~$\bm{\nu}$,
the cosmological parameters $\bm{\pi}$,
and nuisance parameters representing the external convergence ($\kext$; Section~\ref{sec:lensenv}) and anisotropy radius for the lens stellar velocity ellipsoid ($r_{\rm ani}$; Section~\ref{subsec:kinematics}).
${\bm A}$ denotes a discrete set of assumptions about the form of the
model, which includes the data modeling region, the source reconstruction grid, the treatment of
the various deflector mass distributions, etc.  In general, ${\bm A}$ is not fully captured by continuous parameters.
From Bayes' theorem, we have
\begin{eqnarray} \label{eq:pdf}
&& P(\bm{\xi} | \bm{d_{\rm HST}, \Delta t, \sigma, d_{\mathrm{LOS}}, A}) \nonumber \\
        &\propto& P(\bm{d_{\rm HST}, \Delta t, \sigma, d_{\mathrm{LOS}}} | \bm{\xi,A}) P(\bm{\xi} | \bm{A}),
\end{eqnarray}
where ${P(\bm{d_{\rm HST}, \Delta t, \sigma, d_{\mathrm{LOS}}}} | \bm{\xi,
A})$ is the joint likelihood function and $P(\bm{\xi} | \bm{A})$ is the
prior PDF for the parameters given our assumptions.  Since the data sets
are independent, the likelihood can be separated,
\begin{eqnarray} \label{eq:pdf_sep}
P(\bm{d_{\rm HST}, \Delta t, \sigma, d_{\mathrm{LOS}}} | \bm{\xi, A}) &=& P(\bm{d_{\rm HST}} | \bm{\xi, A}) \nonumber \\
  &&  \times P(\bm{\Delta t} | \bm{\xi, A}) \nonumber \\
  &&  \times P(\bm{\sigma} | \bm{\xi, A})  \nonumber \\
  &&  \times P(\bm{d_{\mathrm{LOS}}} | \bm{\xi, A}).
\end{eqnarray}
We can calculate the
individual likelihoods separately and combine them as in
\eref{eq:pdf_sep} to get the final posterior PDF for a given
set of assumptions.

For each of our main lens models in \sref{subsubsec:plmodel} and \sref{subsubsec:composmodel}, we have a range of systematics tests (Section~\ref{subsubsec:sys_tests}) where we vary the content of $\bm{A}$ and repeat the inference
of ${\bm \xi}$.  These tests are important for checking
the magnitude of various known but unmodeled systematic effects, but
leave us with the question of how to combine the results.  We follow H0LiCOW IX in using the Bayesian Information Criterion (BIC) to weight the various models in our final inference  (Section~\ref{subsec:bic}).  This effectively combines our various assumptions $\bm{A}$ using the BIC so that we obtain $P(\bm{\xi} | \bm{d_{\rm HST}, \Delta t, \sigma, d_{\mathrm{LOS}}})$.  We can further marginalise over the non-cosmological parameters ($\bm{\nu}$, $\kext$, $r_{\rm ani}$) and obtain the posterior probability distribution of the cosmological parameters $\bm{\pi}$:
\begin{eqnarray} \label{Pcosmo}
  & & P(\bm{\pi} | \bm{d_{\rm HST}, \Delta t, \sigma, d_{\mathrm{LOS}}}) \nonumber \\
  & = &  \int {\rm d}{\bm \nu}\, {\rm d}\kext\, {\rm d}r_{\rm ani}P(\bm{\xi} | \bm{d_{\rm HST}, \Delta t, \sigma, d_{\mathrm{LOS}}}).
\end{eqnarray}

In the lens model, we actually vary $H_{0}$, keeping other parameters fixed at $w = -1$, $\Omega_{\mathrm{m}} = 0.3$, and $\Omega_{\Lambda} = 0.7$.  This assumes a fixed curvature of the expansion history of the Universe, but not the absolute scale (represented by $H_{0}$ or $\tdist$).  This is done because there is not a unique $\tdist$ when accounting for multiple lens planes, but we convert this to an ``effective" $\tdist$ that is insensitive to assumptions of the cosmological model (see Appendix~\ref{app:cosmo}). Specifically, given the lens/quasar redshifts and $\bm{\pi}$ (i.e., $H_{0}$ and the other fixed cosmological parameters), we can compute the effective time-delay distance $\tdist(\bm{\pi}, \zd, \zs)$ to obtain the posterior probability distribution of $\tdist$, $P(\tdist |  \bm{d_{\rm HST}, \Delta t, \sigma, d_{\mathrm{LOS}}})$.



\section{Data} \label{sec:data}

The data we use to infer $\tdist$ consists of 1) the \hst~imaging used for lens modeling, which we present in \sref{subsec:hst}; 2) the spectroscopy of the lensing galaxy, used to measure its stellar
velocity dispersion, and 3) targeted spectroscopy of the LOS environment, both of which we present in \sref{subsec:spec}; 4) wide-field multi-band imaging, which we present in \sref{subsec:phot} and we use to infer $\kappa_\mathrm{ext}$ in \sref{sec:lensenv}; and 5) the time delays measured by COSMOGRAIL, presented in \sref{subsec:td}.

\subsection{{\it HST} Imaging} \label{subsec:hst}
The \hst~images we use to model \wfilens~consist of Wide Field
Camera 3 (WFC3) F160W band observations (Program \#12889; PI:
Suyu), as well as archival Advanced Camera for Surveys
(ACS) observations in the F814W filter (Program \#9744; PI:
Kochanek). The latter program also contains imaging in the F555W filter, which we do not use, because the signal-to-noise ratio from the lensed images is low and does not add much information to the lens model.

The details of the observations from Program \#12889 are presented in H0LiCOW I.
Using a combination of short (74 s) and long (599--699 s) exposures, we
obtain the brightness distribution of the lens system covering a large dynamic range (of the bright lensed AGN, its much fainter host galaxy, and the foreground lens galaxy). The WFC3 images are drizzled using {\sc
  DrizzlePac}\footnote{{\sc DrizzlePac} is a product of the Space
  Telescope Science Institute, which is operated by AURA for
  NASA.} to a final pixel scale of
$0\farcs08$, whereas the ACS images are reduced using {\sc
  MultiDrizzle}\footnote{MultiDrizzle is a product of the Space
  Telescope Science Institute, which is operated by AURA for NASA.} onto a final pixel scale of 0\farcs05. More details of the reduction are presented in H0LiCOW IV.

\begin{figure*}
\includegraphics[width=\textwidth]{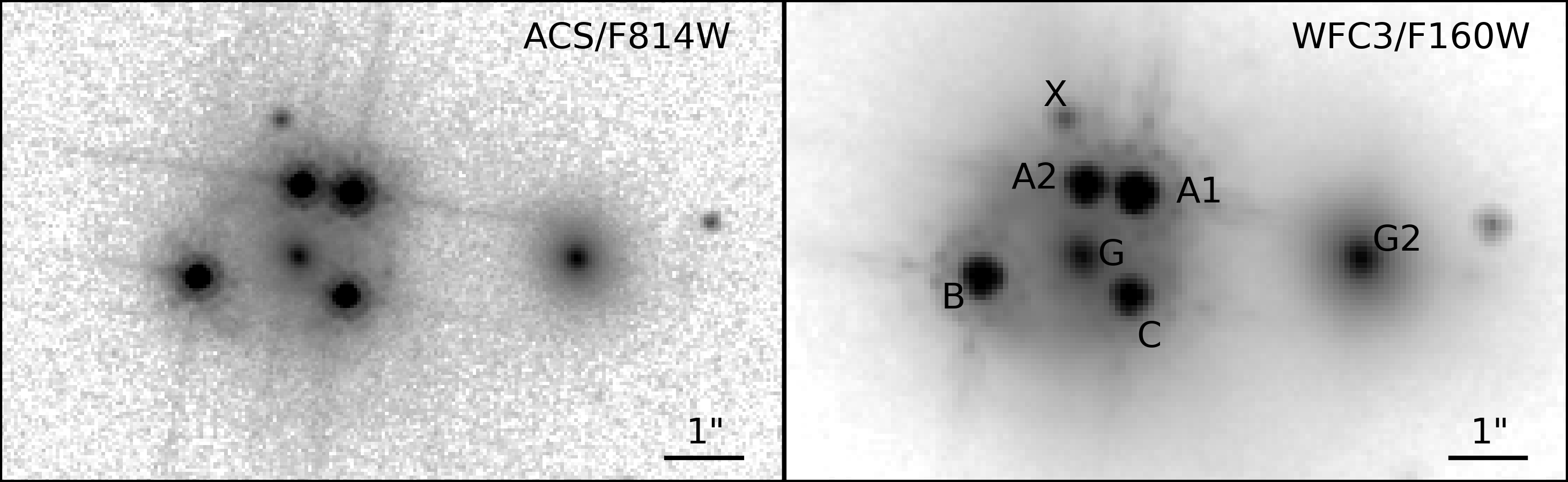}
\caption{\hst~images of \wfilens.  Shown are cutouts of the lens
  system used for lens modeling in the ACS/F814W (left) and WFC3/F160W (right) bands.  The images are $10\farcs4 \times 6\farcs4$.  The scale is indicated in the bottom right of each
  panel.  The main lens galaxy (G), lensed quasar images (A1, A2, B, and C), satellite galaxy (X), and nearby perturber (G2) are marked.  The small object to the west of G2 is a foreground star.}
\label{fig:images}
\end{figure*}

We create cutouts of the reduced \hst~images and define an arcmask around the lens in each of the two filters, which encloses the region where we reconstruct the lensed arc from the extended quasar host galaxy. We expand the cutout to the west of the lens to include the nearby galaxy G2, which is a bright perturber at $z = \zgtwo$ whose light profile needs to be modeled, as it may contaminate the signal within the arcmask.  The cutout region is $10\farcs4 \times 6\farcs4$, which corresponds to a $208 \times 128$ pixel cutout for the F814W image and a $130\times 80$ pixel cutout for the F160W image. These
cutouts are shown in Figure~\ref{fig:images}.

The reconstruction of the point spread function (PSF) for each \hst~exposure, as well as of the weight images and bad pixel masking for each cutout, are analogous to the procedure described in H0LiCOW IV. As detailed in that paper, we note that in order to avoid biasing the modeling due to large residuals from PSF mismatch near the AGN image centers, we rescale the weights in those regions by a power law model such that a pixel originally given a noise value of $p_{i}$ is rescaled to a noise value of $A \times p_{i}^{b}$. The constants $A$ and $b$ are chosen for each band such that the normalized residuals in the AGN image regions are approximately consistent with the normalized residuals in the rest of the arc region.  For completeness, we show the residuals for models using the weight images without this power-law weighting in Appendix~\ref{app:noplwht}.  The strong residuals in these images motivates our decision to adopt this rescaling.

We note that although the background noise for the WFC3 IR camera depends on the number of non-destructive reads, we check that the number of reads in the lensed arc region is the same as for the blank sky patch used for estimating the background noise, so this procedure is valid.  Since most of the lens model constraints come from the parts of the lensing arcs away from the centers of
the AGN images, we check that these arcs do not have pixels that were flagged as bad in too many exposures,
which would otherwise affect our lens mass model.



\subsection{Spectroscopic data} \label{subsec:spec}

 Our spectroscopic observations, presented in H0LiCOW X, reveal that the lens is part of a galaxy
group at $z_{\mathrm{grp}} = \zgrp$ with a velocity dispersion of $\sigma = 500 \pm 80~\mathrm{km~s}^{-1}$ measured from 22 member galaxies, which is independently confirmed
by \citet{wilson2016} based on a spectroscopic study by \citet{momcheva2006,momcheva2015}.

We summarize hereafter the characteristics of the spectroscopic data used. A more exhaustive description of the data acquisition and analysis is provided in H0LiCOW X. \wfilens~was observed with the ESO-MUSE integral field spectrograph \citep{Bacon2010} during several observing runs between 2014-06-19 and 2016-07-20. The velocity dispersion measurement of the lensing galaxy was based on a total of 3$\times$2400 s exposures with the lensing galaxy located close to the center of the 1\arcmin$\times$1\arcmin~field of view (FOV). The data cubes are characterized by a $0\farcs2 \times 0\farcs2$ spatial sampling, a wavelength coverage in the optical range from 4800\,\AA~to 9350\,\AA, a spectral sampling of 1.25\,\AA~per pixel, and a resolving power $R\sim 1800-3600$ \citep[i.e. 2.5\,\AA~spectral resolution; ][]{Richard2017}. The analysis has been carried out on the combined datacube characterized by a median seeing of 1\arcsec. To deblend the lensing galaxy and the quasar images, we modeled each monochromatic slice with a model of the system composed of four Moffat \citep{moffat1969} components for the quasar lensed images, and one de Vaucouleurs \citep{vaucouleurs1948} model for the lensing galaxy. After removing the quasar images from the datacube, we extracted the lensing galaxy spectrum within a square aperture of 9 pixels = 1\farcs8 side-length.

The velocity dispersion was obtained following the same procedure as \citet{suyu2010b,suyu2013}, resulting in an inference of $\sigma_{\mathrm{LOS}} = 250$~km~s$^{-1}$ with a statistical uncertainty of $\approx 10~\mathrm{km~s^{-1}}$. The order of the polynomial continuum and spectral regions masked for the fit introduce additional systematic uncertainties. The various choices we made have been treated as nuisance parameters over which we have marginalised to derive our final velocity dispersion PDF (see H0LiCOW X). The overall uncertainty, accounting for the random and systematic errors, reaches $\sigma_{\sigma_{\mathrm{LOS}}} = 19$~km~s$^{-1}$. We integrate this measurement in our cosmographic inference in \sref{subsec:kinematics}.

In addition to ESO MUSE spectroscopy of the galaxies located in the vicinity of the lens, we have also obtained multi-object spectroscopy of the galaxies in the FOV with the ESO FORS  \citep{Appenzeller1998} and the Gemini GMOS \citep{Hook2004} instruments. In total, we used 10 masks, with about 35 long-slits (6\arcsec~length) per mask positioned on targets located within 2\arcmin~from the lens. For each mask, we obtained 40 minutes long exposures, and used a setup allowing to cover most of the optical wavelength range (typically 4500-9000\AA) with a resolving power of $\approx440$ (FORS) / 1100 (GMOS).

\subsection{Photometric data} \label{subsec:phot}

\begin{figure*}
\includegraphics[width=\textwidth]{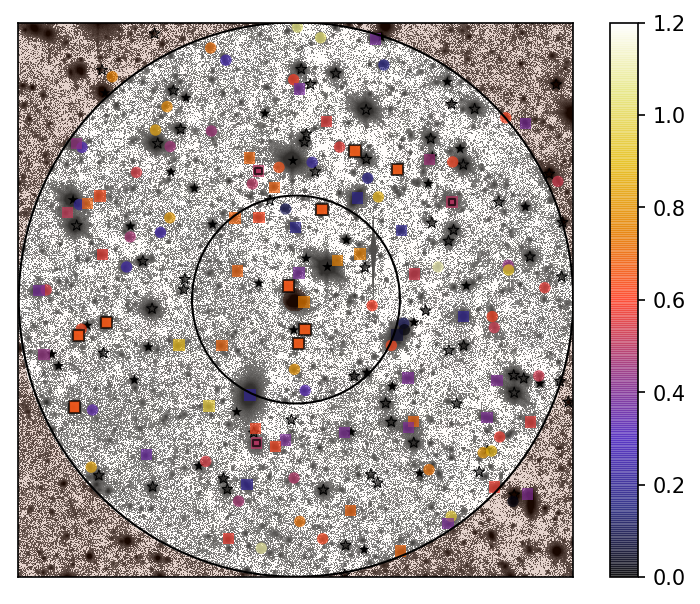}
\caption{$240\arcsec \times 240\arcsec$
  region around \wfilens , overlaying the catalogue data from H0LiCOW X on top of the deepest image available, WFI $R-$band (see \sref{subsec:td} and COSMOGRAIL XVIII for details). The $\leq5\arcsec$- and $\geq120\arcsec$-radius apertures are masked. The $45\arcsec$- and $120\arcsec$-radius apertures are marked by black circles. Detected sources with $i\leq22.5$, corresponding to the limit used in our weighted number counts analysis, are marked: stars are marked with black star symbols, filled if confirmed spectroscopically and empty otherwise; galaxies are marked with squares if spectroscopic redshifts are available, and with circles otherwise. The color scale corresponds to the spectroscopic redshift, if available, and to the photometric redshift, otherwise. Galaxies spectroscopically confirmed to be members of the galaxy group which includes the lensing galaxy are marked with squares with black contours, and those part of the group at $z=0.49$ are marked with smaller square contours.
  For a larger FOV and more details on the available LOS spectroscopy, see Figure~2 in H0LiCOW X.
  \label{fig:fov}}
\end{figure*}

Our photometric data consists of wide-field optical wavelength data from the Dark Energy Survey\footnote{\url{https://www.darkenergysurvey.org}} (DES), ultraviolet data from the DES Camera \citep{flaugher15} on the Blanco Telescope, VLT/HAWK-I \citep{pirard04,kissler08} near-infrared data, and archival IRAC \citep{fazio04} infrared data from the Spitzer Space Telescope. These data and their products, consisting of the galaxy-star classification, photometric redshifts and stellar masses of all galaxies with $i<23$ mag within a $120\arcsec$ radius around \wfilens, are described in H0LiCOW X. In \sref{sec:lensenv}, where we measure the relative density of the environment of \wfilens, we use a conservative cut of $i<22.5$ mag in order to ensure that the galaxy catalogue, with a $5\sigma$ limiting magnitude of $\sim23.13$, is complete.\footnote{While shallow magnitude limits may bias the $\kappa_\mathrm{ext}$ distribution we determine in \sref{sec:lensenv}, Figure~6 in \citet{collett2013} shows that the expected bias is at a level of $\sim0.25\%$, which is acceptable given our goal of inferring $H_0$ with biases below the $1\%$ level.} We show the $4\arcmin\times4\arcmin$ FOV, with the galaxy catalogue overlapped, in Figure~\ref{fig:fov}.

\subsection{Time delays} \label{subsec:td}

\subsubsection{Time-delay measurements} \label{subsubsec:td_meas}
COSMOGRAIL XVIII presents the most comprehensive analysis of the time delays of \wfilens~so far, with the analysis of four different data sets spanning across 14 years of monitoring, for a total of $\sim 447$ hours of observations. 
The data were acquired in the scope of the COSMOGRAIL collaboration, using three different telescopes in the Southern hemisphere; the C2 and ECAM instruments mounted on the 1.2m Leonhard Euler Swiss telescope and the WFI instrument mounted on the ESO/MPIA 2.2m telescope, both located at La Silla Observatory in Chile, and the 1.3m Small and Moderate Aperture Research System (SMARTS) at the Cerro Tololo Inter-American Observatory (CTIO) in Chile. 

The data is split in four data sets, one per instrument (C2 and ECAM on the Euler telescope, WFI on the 2.2m telescope and SMARTS), each being reduced independently.
The photometry of the four images of \wfilens\ is recovered using the MCS deconvolution scheme \citep{magain1998, cantale2016a}. The light curves obtained are presented in Figure 2 of COSMOGRAIL Paper XVIII. For three of the four data sets (C2, ECAM and SMARTS), the deconvolution scheme is not able to properly resolve the flux coming from the A1 and A2 images. Thus, the A1 and A2 fluxes are summed into a virtual light curve A, under the assumption that the time delay between A1 and A2 is zero. The WFI data set being composed of exposures of better quality, the deconvolution scheme manages to properly resolve the A1 and A2 images. A virtual light curve A=A1+A2 is also constructed for WFI in order to compare it to the other data sets.


The time-delay measurements between each pair of light curves are made with the {\tt PyCS} software \citep{tewes2013a, bonvin2016} and follow the formalism introduced by \citet{bonvin2018}. We use two different curve shifting techniques.
Both techniques share a common framework to assess their own uncertainties, based on a statistical analysis of the residuals of the real data that prevents, by construction, involuntary fine-tuning of the curve-shifting technique parameters to recover a biased value of the time delays.

Each data set is analyzed independently. The time-delay estimates obtained are in good agreement with each other and a Bayes Factor analysis states that they can be combined without loss of consistency. In this work, we use the combined time-delay estimates with respect to image B. For our fiducial set of models, we use the B-A1 and B-A2 time delays estimated from the WFI data set (see Figure 4 of COSMOGRAIL XVIII), and the B-C time delay estimated by combining all the data sets together (labeled ``PyCS-mult'' on Figure 3 of COSMOGRAIL XVIII). They read $\Delta t_{\mathrm{B-A1}} = -36.2^{+1.6}_{-2.3}$, $\Delta t_{\mathrm{B-A2}} = -37.3^{+2.6}_{-3.0}$ and $\Delta t_{\mathrm{B-C}} = -59.4 \pm 1.3$. Although using different time delays from different combinations of data sets might appear subjective, we recall that i) only the WFI data set is of good enough quality to resolve the A1 and A2 images, thus bringing an additional independent constraint to the modeling and solving the potential issue of where to anchor a time-delay estimate related to a virtual image A, and ii) all the time-delay estimates and combination of time-delay estimates are statistically consistent with each other.

\subsubsection{Microlensing time-delay} \label{subsubsec:ml_td}
Our time-delay measurements do not include the contribution from the microlensing time delay \citep{tie2018, bonvin2019}, a time-dependent reweighting of the geometrical delay (originating from the extended spatial structure of the source) by the microlensing pattern affecting each image independently. As a result, an excess microlensing time delay adds to the excess cosmological time delay of each lensed image, and the measured time delays between pairs of images can deviate from the cosmological time delays by a noninegligible amount. The amplitude of the effect depends mainly on the mass of the central black hole of \wfilens\ \citep{Sluse2012, motta2017}, and its estimation relies on the assumption that the accretion disk can be modeled as a thin-disk \citep{shakura1973} - which, so far, is disfavored by the data \citep[see e.g.][]{morgan2018} - and that the emission of the accretion disk follows an idealized lamp-post model \citep{cackett2007, starkey2016}.

In Figure 6 of COSMOGRAIL XVIII, we compute the amplitude of the microlensing time delay for various disk sizes. Although the measured time delays do not show any discrepancies that would be evidence for a microlensing time delay, it cannot be ruled out either.  We thus chose to include it by default in our models, noting that the effect is much smaller than our other uncertainties. We follow the framework presented in \citet{chen2018} and assume the accretion disk size of \citet{morgan2018} with $r = R_{0}$.  We also test the effect of ignoring the microlensing time delay for one of our models, finding that it changes the $\tdist$ accuracy by $< 1\%$ (Section~\ref{subsubsec:sys_tests}).



\section{Lens Modeling} \label{sec:lensmod}
In this section, we describe our procedure to simultaneously model the
images in the two \hst~bands, and the time delays, in order to infer the lens
model parameters and $\tdist$.

\subsection{Overview} \label{subsec:overview}
We perform our lens modeling using \GLEE, a software package developed by
S.~H.~Suyu and A.~Halkola \citep{suyu2010a,suyu2012b}.  The lensing
mass distribution is described by a parameterized profile.  The
extended host galaxy of the source
is modeled separately on a $50\times50$ pixel grid with curvature regularization \citep{suyu2006}.
The lensed quasar images are modeled as point sources on the image plane convolved with the PSF.  The quasar image amplitudes are allowed to freely vary and are independent from the extended host galaxy light distribution to allow for variability due to microlensing, time delays, and
substructure.  The lens galaxy light distribution is modeled using either
S\'{e}rsic profiles or Chameleon profiles.  The S\'{e}rsic profile is defined as
\begin{equation} \label{eq:sersic}
I(\theta_{1}, \theta_{2}) = A~\mathrm{exp} \left[ -k \left( \left( \frac{\sqrt{\theta_{1}^{2} + \theta_{2}^{2} / q_{\mathrm{L}}^{2}}}{\reff} \right)^{1/n} - 1 \right) \right],
\end{equation}
where $A$ is the amplitude, $k$ is a constant such that $\reff$~is the effective (half-light) radius, $q_{\mathrm{L}}$ is the axis ratio, and $n$ is the S\'{e}rsic index.  The Chameleon profile (also known as the pseudo-Jaffe profile) is defined as the difference of two non-singular $r^{-2}$ elliptical profiles \citep{kassiola1993,dutton2011}, which are a good approximation to S\'{e}rsic profiles.  

We represent the galaxy light distribution as the sum of two S\'{e}rsic (or two Chameleon) profiles plus a point source (to account for possible AGN emission from the lens galaxy) with a common centroid.  Since the light of G2 can also influence the model, we represent its light distribution as a single S\'{e}rsic profile plus a point source with a common centroid, although we mask its central regions since we only care about light from G2 that could affect the lens galaxy or arc light.  There is a small nearby perturber (``X" in Figure~\ref{fig:images}), which we also represent as a single S\'{e}rsic profile plus a point source with a common centroid.
Model parameters of the lens and
source are constrained through Markov Chain Monte Carlo (MCMC)
sampling.

\begin{figure}
\includegraphics[width=0.48\textwidth]{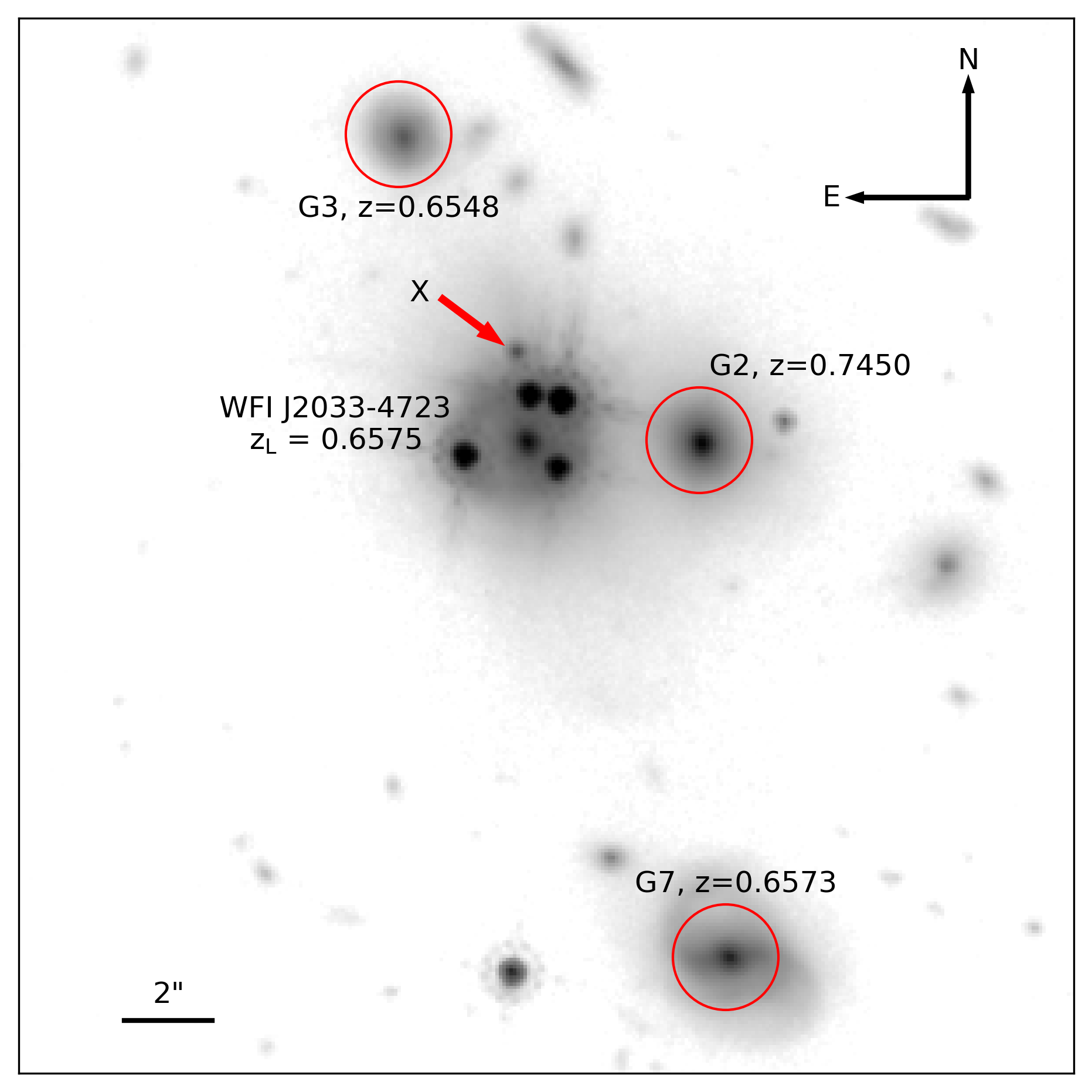}
\caption{ \hst/WFC3 F160W image of a $24\arcsec \times 24\arcsec$
  field around \wfilens.  The angular scale is indicated in the bottom
  left corner.  The three most significant nearby perturbers are marked with red circles,
  and the redshifts of the perturbers are indicated.  G2, G3, and G7 are included
  explicitly in our model, as they are the most massive and nearest in
  projection to \wfilens.  The small object X is indicated by a red arrow, and is assumed to be at the lens redshift in our models.
  \label{fig:field}}
\end{figure}

In accounting for perturbers at different redshifts from the main lens galaxy, we use the full multi-plane lens equation \citep[e.g.,][]{blandford1986,kovner1987,schneider1992,petters2001,collett2014,mccully2014} in our modeling.  We vary $H_{0}$ directly in our models and use this
distribution to calculate the effective model time-delay distance $\tdistmod$.  In calculating $\tdistmod$, we assume $\Om = 0.3$, $\OL = 0.7$, and $w = -1$.  Relaxing these assumptions by allowing these cosmological parameters to vary freely shifts the resulting $\tdistmod$ distributions by $< 1$\% in previous analyses (see H0LiCOW IV), and we also verify that this is true for \wfilens~(Appendix~\ref{app:cosmo}). Thus, this approximation has no measurable effect on the inferred time delay distance, which can then be applied to constrain any arbitrary cosmology.

\subsection{Mass Models} \label{subsec:massmodel}
Our primary mass models for the lens galaxy are a singular power-law
elliptical mass distribution \citep[SPEMD;][]{barkana1998}, and a model consisting of a baryonic component that traces the light distribution plus a separate dark matter component (hereafter the ``composite" model).  We also
include an external shear in the strong lens plane.  Non-linear couplings due to multi-plane effects are small and thus ignored.

We explicitly include the nearby perturber X in the lens model, linking its mass centroid to that of its light. Although we do not have a spectroscopic confirmation of the redshift of X, it is likely a satellite galaxy that is physically associated with the lens galaxy, given its small size and proximity. We also see evidence in the F160W image for possible tidal features emanating from X in the direction away from the lens galaxy, suggesting that it may be an infalling satellite. We therefore assume that X is at the same redshift as the lens and parameterize it as a singular isothermal sphere (SIS). In our models, X generally has a much smaller mass than the main lens, and therefore it has a minor influence on the potential, even if it is located at a different redshift.

We also explicitly include three nearby massive perturbing
galaxies \citep[denoted G2, G3, and G7, following the naming convention of][]{vuissoz2008} in Figure~\ref{fig:field} that are
projected close the lens.   G2 is close enough
that its influence may not be adequately described by external shear
\citep[H0LiCOW II; see also][]{mccully2017}, and H0LiCOW X showed that G3 ($z = \zgthree$) and G7 ($z = \zgseven$) may have a non-negligible higher-order influence on the model as well. Our updated estimation of the influence of these galaxies, computed in terms of the flexion shift considered in H0LiCOW X  but taking into account the galaxy morphologies and velocity dispersions measured in that paper, shows that G2, with $\log M_\star\sim11.15$, is in fact the only galaxy with significant impact on the modeling. Nonetheless, based on their proximity to the lensing galaxy, we choose to explicitly model G3 ($\log M_\star\sim10.17$) and G7 ($\log M_\star\sim11.16$) as well. G2 is modeled as a singular
isothermal ellipsoid, which is a reasonable assumption since higher-order moments of the potential will have a small effect at the position of the main lens galaxy.  G3 and G7 are modeled as SIS.  The
  relative Einstein radii of G2, G3, and G7 are calculated from their measured velocity dispersions (H0LiCOW X), assuming isothermal profiles.
  The ratio of their Einstein radii is fixed, but with a global scaling allowed to vary freely, as in H0LiCOW IV, IX.  This is done to
  prevent the model from optimizing the perturbers' Einstein radii in
  a way that would be inconsistent with their measured redshifts and
  velocity dispersions.  The centroid of G2 is linked to the centroid of its light distribution in the F160W band in the modeling, while the centroids of G3 and G7 are fixed to their measured positions in the F160W image.  We set the redshifts of G3 and G7 equal to the lens redshift of $z = \zlens$ in our model, as their redshifts are consistent with this value within the range allowed by peculiar velocities.
All masses are treated using the full multi-plane lens equation, as detailed by Suyu et al., in preparation.


Our constraints on the primary lens model include the positions of the
lensed quasar images, the measured time delays, and the surface
brightness of the pixels in the ACS/F814W and WFC3/F160W images that are fit simultaneously.  The quasar positions are fixed to the positions of the point sources on the image plane (after they have stabilized) and are given a Gaussian uncertainty of width $0\farcs004$ to account for offsets due to substructure in the lens or LOS, which is small enough to satisfy astrometric requirements for cosmography \citep{birrer2019b}. The quasar flux ratios are not used as constraints, as they can be affected by microlensing.
We first model the lens seperately in each band to iteratively
update the respective PSFs using the lensed AGN images themselves,
similar to \citet{chen2016}, but with the PSF corrections and source
intensity reconstructed simultaneously in our case (H0LiCOW IV, IX) rather than separately.  We keep these ``corrected" PSFs fixed and use them in our final models that simultaneously use the
surface brightness distribution in both bands as constraints.  We then use the positions of the quasar images to align the images in the two {\it HST} bands.  We do not enforce any similarity of pixel values at the same spatial position across different bands (i.e., the flux at any position in one band is independent of the other band).
We also directly include the effect of microlensing time delays, as described in Section~\ref{subsubsec:ml_td}, although our tests show that this has a very small effect on our results (Section~\ref{subsubsec:sys_tests}).
In our MCMC sampling, we vary the light parameters of the lens galaxy, G2, X, and quasar
images, the mass parameters of the lens galaxy, X, G2, G3, and G7, the external shear, and $H_{0}$.  The quasar image
positions are linked across both bands, but the other light
parameters are allowed to vary independently.

\subsubsection{Power-law Model} \label{subsubsec:plmodel}
Our fiducial SPEMD model uses the double S\'{e}rsic parameterization for the lens galaxy light and has the additional free parameters:
\begin{enumerate}[label=(\roman*)]
\item position ($\theta_{1}$,$\theta_{2}$) of the centroid (allowed to vary independently from the centroid of the light distribution)
\item Einstein radius $\theta_{\mathrm{E}}$
\item minor-to-major axis ratio $q$ and associated position angle $\theta_{q}$
\item 3-dimensional slope of the power-law mass distribution $\gamma^{\prime}$
\item position of X, linked to its light centroid
\item Einstein radius of X
\item position of G2, linked to its light centroid
\item global scaling parameter that controls the Einstein radii of G2, G3, and G7
\item minor-to-major axis ratio $q$ and associated position angle $\theta_{q}$ of G2
\item external shear $\gext$ and associated position angle $\theta_{\gamma}$\footnote{$\theta_{\gamma}$ is defined to be the direction of the shear itself, i.e. orthogonal to the direction of the mass producing the shear.}
\item the Hubble constant, $H_{0}.$
\end{enumerate}
We conservatively assume uniform priors on the model parameters over a wide physical range.  Although the lens is not drawn from a random population, but rather with some selection function that could, in principle, bias the inferred time-delay distance, this selection function is not well known and these biases are negligible for this type of analysis \citep[e.g.,][]{collett2016}.  The parameters that are exceptions to our choice of uniform priors are that the global scaling parameter for the Einstein radii of the perturbers is given a Gaussian prior such that the expected mean and uncertainty of G2's Einstein radius is constrained by its measured velocity dispersion, and that the position angle $\theta_{q}$ of G2's is given a Gaussian prior based on the fit of its light profile.  We anchor the scaling parameter to G2 as it is the perturber with the most precisely-measured velocity dispersion, and its proximity to the lens makes it the most significant of the three massive perturbing galaxies.

\begin{figure*}
\includegraphics[width=\textwidth]{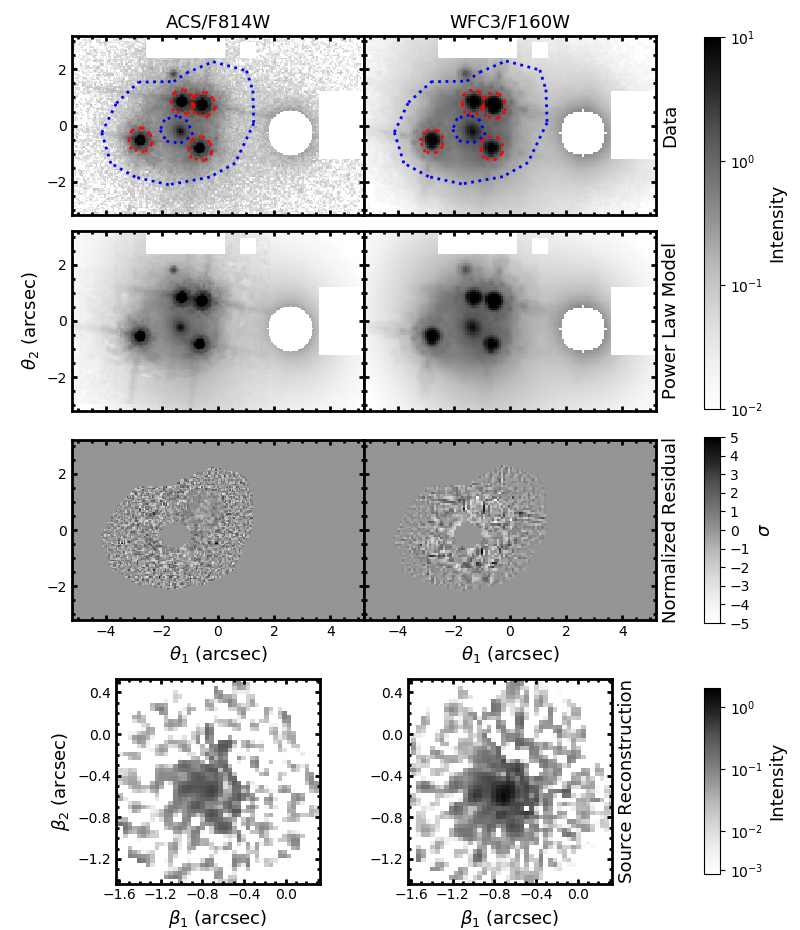}
\caption{SPEMD lens model results for ACS/F814W (left)
  and WFC3/F160W (right).  Shown are the observed image (top row), the
  reconstructed image predicted by the model (second row), the
  normalized residual within the arcmask region (defined as the difference between the data and model, normalized by the estimated uncertainty of each pixel; third row), and the reconstructed source (bottom
  row).  This uses the weight image with the power-law rescaling near the AGN images.  We show the normalized residuals without this rescaling in Appendix~\ref{app:noplwht}.  In the top row, the blue dotted lines indicate the arcmask (donut-shaped) region used for
  fitting the extended source, the red dotted lines indicate the AGN mask region where the power-law weighting is applied, and the region outside the blue dotted arcmask is used to further constrain the foreground lens light and (partly) the AGN light (but not the AGN host galaxy light since its corresponding lensed arcs are below the noise level in this outer region).  The white regions indicate areas of the image that are masked out during the modeling.  The color bars show the scale in the respective panels.  The results shown here are for the fiducial SPEMD model, but the results for the other systematics tests (Section~\ref{subsubsec:sys_tests}) are qualitatively similar.
\label{fig:model_results_pl}}
\end{figure*}

Figure~\ref{fig:model_results_pl} shows the data and the lens model results in both bands for our fiducial SPEMD model, as well as the source reconstructions.  Our model reproduces the surface brightness structure of the lensed AGN and host galaxy in both bands simultaneously.

\subsubsection{Composite Model} \label{subsubsec:composmodel}
We follow \citet{suyu2014} and H0LiCOW IV to construct the composite model, consisting of a baryonic component linked to the light profile of the lens galaxy, plus a dark matter component.  The composite model assumes the double Chameleon light profile for the lens galaxy in the
WFC3/F160W band scaled by an overall mass-to-light (M/L) ratio.  We use the Chameleon light profiles for the composite model because it is straightforward to link the parameters describing the light distribution to those of the mass distribution, as they are fundamentally just a combination of isothermal profiles.  We use the F160W band because it probes the rest-frame near-infrared and thus
should be the best tracer of stellar mass.  Although we include a point source in the light profile, we assume that this is due to low-level AGN emission from the lens galaxy, and do not associate it with a massive component in the model. This point source is roughly $\sim2\%$ of the total light in the F160W band, so its inclusion would have a minor impact on our results. We keep the double S\'{e}rsic parameterization for the lens galaxy light in the F814W band to maintain consistency with the SPEMD models.  The dark matter component
is modeled as an elliptical NFW \citep{navarro1996} potential with the
centroid linked to the light centroid in the F160W band, as non-contracted NFW profiles are a good representation of the dark matter halos of massive elliptical galaxies \citep{dutton2014}.

Our fiducial composite model has the same free parameters (v) to (xi) as the SPEMD model in \sref{subsubsec:plmodel}, as well as the additional parameters:
\begin{enumerate}[label=(\alph*)]
\item M/L ratio for the baryonic component
\item NFW halo normalization $\kappa_\mathrm{0,h}$ \citep[defined as $\kappa_\mathrm{0,h} \equiv 4\kappa_\mathrm{s}$;][]{golse2002}
\item NFW halo scale radius $r_\mathrm{s}$
\item NFW halo minor-to-major axis ratio $q$ and associated position angle $\theta_{q}$
\end{enumerate}
We set a Gaussian prior of $r_\mathrm{s} = 11\farcs9 \pm 1\farcs6$ based on the results of
\citet{gavazzi2007} for lenses in the Sloan Lens ACS Survey \citep[SLACS;][]{bolton2006} sample, which encompasses the redshift and stellar mass of \wfilens.
All other parameters are given uniform priors, again with the exception of the Gaussian prior on global scaling parameter based on G2's Einstein radius, as well as the Gaussian prior on G2's position angle.  The relative amplitudes of the two Chameleon profiles that represent the stellar light distribution of the
lens galaxy can vary, but the relative
amplitudes of these two components in the mass profiles are fixed.  To account for this, we
iteratively run a series of MCMC chains and update the relative amplitudes of the two mass components to
match that of the light components after each chain.  We iterate until the
inferred $H_{0}$ stabilizes, then combine the chains after this
point into a single distribution to represent the fiducial composite model.  The other composite models use fixed relative amplitudes of the mass components based on the latest iteration of the fiducial composite model.

\begin{figure*}
\includegraphics[width=\textwidth]{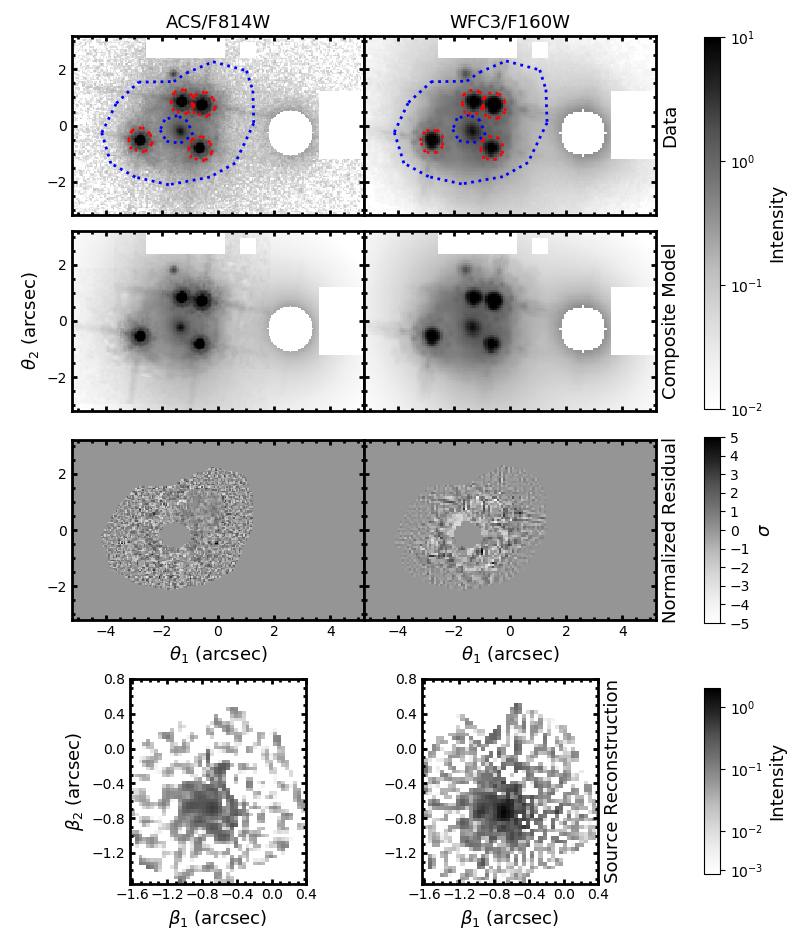}
\caption{Same as Figure~\ref{fig:model_results_pl}, but for the fiducial composite model.
\label{fig:model_results_comp}}
\end{figure*}

Figure~\ref{fig:model_results_comp} shows the data and the lens model results in both bands for the fiducial composite model described in this section, as well as the source reconstruction.

\subsubsection{Systematics Tests} \label{subsubsec:sys_tests}
In this section we describe a range of tests of the effects
of various systematics in our modeling, stemming from different assumptions in the way we constructed the model that might affect the posterior.  In
addition to the basic fiducial models described above, we perform inferences for both the SPEMD and composite models given
the following sets of assumptions:
\begin{itemize}
\item A model with the arcmask region increased by one pixel on both
  the inner and outer edges.  To compensate for the larger arcmask region, we increase
  the source plane resolution to $60\times60$ pixels in all bands.
\item A model where the region near the AGN images scaled by the
  power-law weighting is increased by one pixel around the outer edge.  Increasing these regions by more pixels would start to greatly reduce the area of the arcmask where we fit the extended source.
\item A model where the regions near the AGN images are given zero
  weight rather than being scaled by a power-law weighting.
\item A model that includes the group at $z = \zgrp$ (of which the lens galaxy is a member) as a spherical NFW halo.  The halo centroid and mass are given Gaussian priors based on the calculations of H0LiCOW X. The scale radius is given a Gaussian prior of $r_{\mathrm{s,g}} = 32\farcs0 \pm 8\farcs0$ from a calculation of its virial mass and radius (H0LiCOW X) and a halo concentration based on the results of \citet{diemer2018}.  The redshift of the group is set to the lens redshift ($z = \zlens$), as the difference can be explained by peculiar velocity.
\item A model that includes both the group at $z = \zgrp$ (again set to the lens redshift) and a foreground group at $z = \zfggrp$ which may have a significant effect on the lens potential based on H0LiCOW X, who estimate its flexion shift \citep[following the definition in][]{mccully2014}.  The foreground group's centroid, mass, and scale radius are given Gaussian priors in the same way as for the group at the lens redshift.  The scale radius prior from H0LiCOW X and \citet{diemer2018} is $r_{\mathrm{s,gf}} = 34\farcs8 \pm 9\farcs3$.
\end{itemize}

In addition to the above models for both the SPEMD and composite models, we run one additional SPEMD model:
\begin{itemize}
\item A model where the light profile of the lens galaxy in both bands is
  represented by the sum of two Chameleon profiles rather than the
  sum of two S\'{e}rsic profiles.
\end{itemize}

As described in Section~\ref{subsec:bic}, we combine the MCMC chains
from all of these tests, weighted by the BIC (e.g., H0LiCOW IX). We calculate the relative BIC for the SPEMD models and composite models separately, then give the combined distributions equal weight in the final inference so that we are not biased by the parameterization of the mass profile.

\begin{figure}
\includegraphics[width=0.5\textwidth]{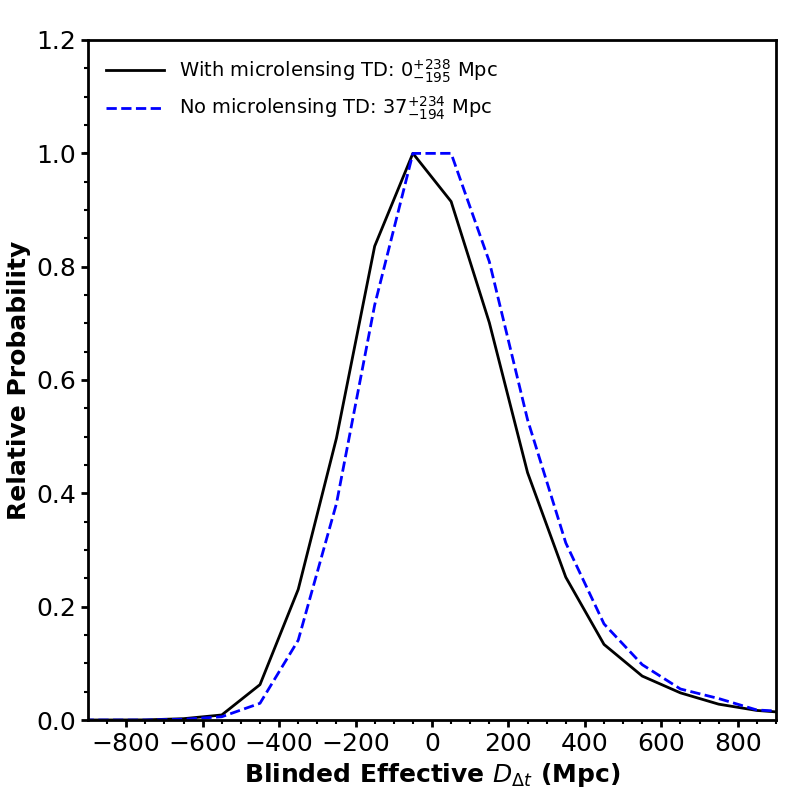}
\caption{
PDF of $\tdist$ for the fiducial SPEMD model with (black) and without (blue) the microlensing time delay effect.  The median of the blinded effective time-delay distance PDF is insensitive to the microlensing time delay effect to within 1\%.
\label{fig:dt_mltd}}
\end{figure}

We also run a test to verify that the microlensing time delay does not significantly impact our results.  We test our fiducial SPEMD model without including the microlensing time delay effect and compare the blinded effective time-delay distance to the model with this effect included, in Figure~\ref{fig:dt_mltd}.  We find that the microlensing time delay affects the inferred $\tdist$ at $< 1\%$, so its inclusion in our models, given our assumptions about the disk size, does not have an appreciable effect.

\subsubsection{Comparison of Power Law and Composite Models} \label{subsubsec:pl_comp}
\begin{figure*}
\includegraphics[width=\textwidth]{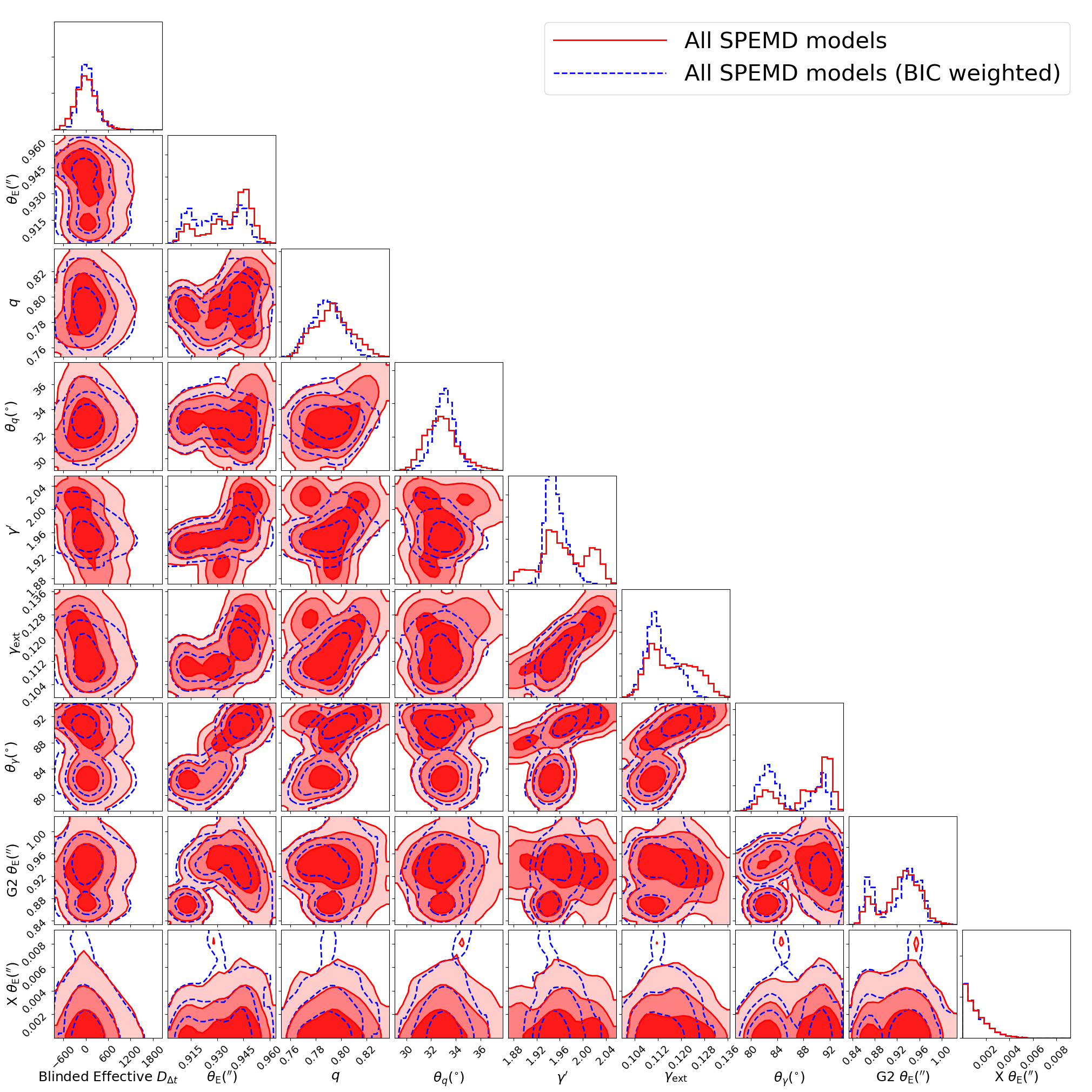}
\caption{
Marginalized parameter distributions from our SPEMD lens model
results.  We show the combined results
from our systematics tests (shaded red contours) with each model weighted equally, as well as the BIC-weighted model results (dashed blue contours).  The contours represent
the 68.3\%, 95.4\%, and 99.7\% quantiles.
\label{fig:corner_spemd}}
\end{figure*}

The marginalized parameter distributions of the SPEMD model are shown in Figure~\ref{fig:corner_spemd}.  We show the combined distributions of all SPEMD models where each model is given equal weight, as well as the BIC-weighted distribution.  The parameter statistics for each model are given in Appendix~\ref{app:params}.  There are some minor variations in the model parameters from model to model, but the $\tdist$ distributions are generally consistent.  We note that the model with Chameleon light profiles for the lens galaxy is somewhat offset toward a lower $\tdist$ (see Section~\ref{sec:results}).  This model is disfavored by our BIC weighting, so this has a minimal effect on our final results.  This does not necessarily mean that the Chameleon profiles in general are a bad fit to the lens galaxy light, as the composite models (which use the Chameleon light profile by default in the F160W band) are not similarly offset.

The multi-modal distributions in some of the parameters arises primarily from differences in the posterior PDFs of different models corresponding to the various systematics tests, not from bimodality within individual lens models.  We note that despite this multi-modal behavior, the effective $\tdist$ distribution remains stable and unimodal, suggesting that the cosmological inference is robust to the various systematics tests.

The model that includes both group halos has the highest BIC weighting for both the SPEMD and composite models.  To check that the addition of the $z=\zfggrp$~group contributes meaningful information to the modeling, we run a test where the centroid of this group is given a prior located at a similar distance but rotated by $90^{\circ}$ and $135^{\circ}$ on the sky relative to the lens.  We compare the BIC weight values of these test models to that of the model with just the group at the lens redshift and the original model with both groups.  These test cases show a lower BIC weight than the original model with both groups, suggesting that the addition of the foreground group with the actual centroid prior is contributing information, although the small BIC difference is within the typical BIC variance, so it is difficult to draw a firm conclusion.  The $\tdist$ distributions remain robust within the uncertainties for each of these test cases.

The offset between the mass centroid and the light centroid in the F160W band for the SPEMD model is typically $\sim0\farcs02 - 0\farcs03$ (roughly $150-200$ pc for a flat $\Lambda$CDM cosmology with $h = 0.7$ and $\Om = 0.3$) such that the mass centroid is slightly southeast of the light centroid.  This might be partially explained by the influence of object X, although we note that in our SPEMD models, the mass of X is consistent with zero.  The centroids of the light profiles in F814W and F160W are consistent with each other at the $\sim0\farcs002$ level for both models.  The SPEMD models are able to fit the quasar positions to an rms of $\sim 0\farcs01$, while the composite models have a larger rms of $\sim 0\farcs025$.  Despite these differences, the SPEMD and composite models' $\tdist$ distributions are not drastically different, and by weighting them equally in the final inference, we are accounting in part for the astrometric uncertainty.

\begin{figure*}
\includegraphics[width=\textwidth]{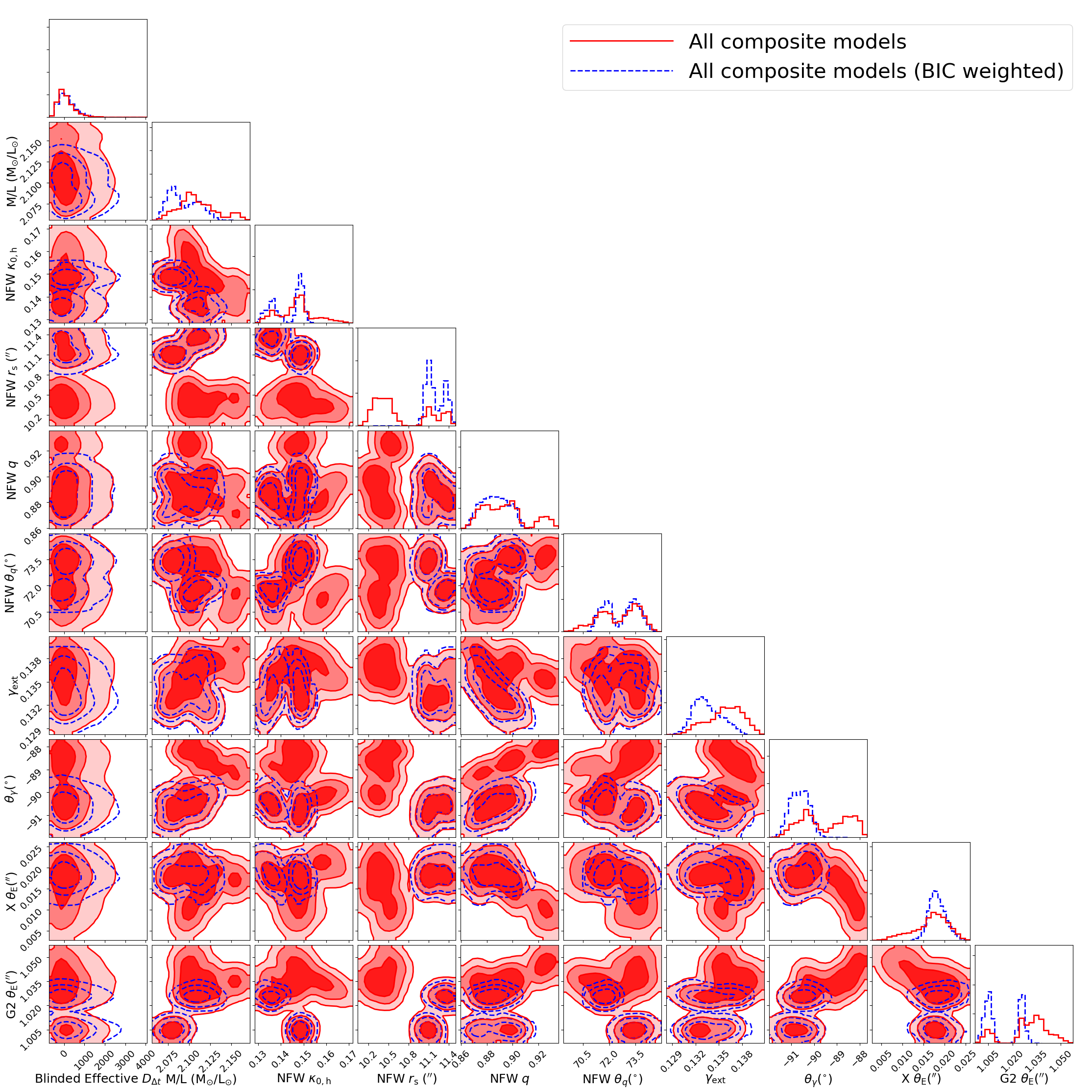}
\caption{
Marginalized parameter distributions from our composite lens model
results.  We show the BIC-weighted model (dashed blue contours) and the combined results
from our systematics tests (shaded red contours).  The contours represent
the 68.3\%, 95.4\%, and 99.7\% quantiles.
\label{fig:corner_compos}}
\end{figure*}

We show the marginalized parameter distributions of the composite model in Figure~\ref{fig:corner_compos}.  Again, we show the uniformly-combined distributions as well as the BIC-weighted composite model separately, and the parameter statistics for each model are given in Appendix~\ref{app:params}.  As with the SPEMD model, there are small variations in the model parameters, but the $\tdist$ inference is consistent.

We compare the physical parameters of our BIC-weighted SPEMD model
to the composite model.  The results are shown in
Table~\ref{tab:lensmod_params}, with the parameter statistics for all composite models given in Appendix~\ref{app:params}.

\renewcommand*\arraystretch{1.5}
\begin{table}
\caption{Lens Model Parameters \label{tab:lensmod_params}}
\begin{minipage}{\linewidth}
\begin{tabular}{l|c}
\hline
Parameter &
BIC-weighted Marginalized Constraints
\\
\hline
\multicolumn{2}{l}{Singular Power Law Ellipsoid Model}
\\
\hline
$\theta_{\mathrm{E}}~(\arcsec)$\footnote{Spherical-equivalent Einstein radius} &
$0.929_{-0.016}^{+0.015}$
\\
$q$ &
$0.79_{-0.01}^{+0.01}$
\\
$\theta_{q}$ ($^{\circ}$) &
$33.1_{-0.9}^{+0.8}$
\\
$\gamma^{\prime}$ &
$1.95_{-0.01}^{+0.02}$
\\
$\gext$ &
$0.112_{-0.004}^{+0.006}$
\\
$\theta_{\gamma}$ ($^{\circ}$) &
$83.6_{-2.1}^{+7.1}$
\\
X $\theta_{\mathrm{E}}~(\arcsec)$ &
$0.001_{-0.001}^{+0.001}$
\\
G2 $\theta_{\mathrm{E}}~(\arcsec)$ &
$0.932_{-0.062}^{+0.027}$
\\
G2 $q$ &
$0.66_{-0.02}^{+0.04}$
\\
G2 $\theta_{q}$ ($^{\circ}$) &
$38.5_{-4.1}^{+4.4}$
\\
\hline
\multicolumn{2}{l}{Composite Model}
\\
\hline
Stellar M/L ($\mathrm{M_{\odot}/L_{\odot}}$)\footnote{M/L within $\theta_{\mathrm{E}}$ for rest-frame $V$ band.  The point source component of the lens light is assumed to be from low-level AGN emission as opposed to stellar light and is not included in the calculation.  The given uncertainties are a combination of statistical effects and a systematic uncertainty equal to the difference between the calculated M/L with and without the point source contribution.  The stellar mass is calculated assuming $H_{0} = 70~\mathrm{km~s^{-1}~Mpc^{-1}}$, $\Omega_{\mathrm{m}} = 0.3$, $\Omega_{\Lambda} = 0.7$, but changes in the cosmology affect the M/L by a negligible amount.} &
$2.1_{-0.2}^{+0.2}$
\\
Chameleon1 $q$ &
$0.762_{-0.003}^{+0.003}$
\\
Chameleon1 $\theta_{q}$ ($^{\circ}$) &
$23.3_{-0.2}^{+0.2}$
\\
Chameleon2 $q$ &
$0.771_{-0.003}^{+0.002}$
\\
Chameleon2 $\theta_{q}$ ($^{\circ}$) &
$26.2_{-0.5}^{+0.3}$
\\
NFW $\kappa_{0,\mathrm{h}}$ &
$0.147_{-0.012}^{+0.002}$
\\
NFW $r_{\mathrm{s}}~(\arcsec)$ &
$11.15_{-0.09}^{+0.21}$
\\
NFW $q$ &
$0.89_{-0.01}^{+0.01}$
\\
NFW $\theta_{q}$ ($^{\circ}$) &
$72.9_{-1.3}^{+0.8}$
\\
$\gext$ &
$0.133_{-0.001}^{+0.002}$
\\
$\theta_{\gamma}$ ($^{\circ}$) &
$89.4_{-0.4}^{+0.4}$
\\
X $\theta_{\mathrm{E}}~(\arcsec)$ &
$0.018_{-0.002}^{+0.002}$
\\
G2 $\theta_{\mathrm{E}}~(\arcsec)$ &
$1.008_{-0.004}^{+0.019}$
\\
G2 $q$ &
$0.93_{-0.01}^{+0.01}$
\\
G2 $\theta_{q}$ ($^{\circ}$) &
$39.4_{-10.3}^{+1.6}$
\\
\hline
\end{tabular}
\\
{\footnotesize Reported values are medians, with errors corresponding to the 16th and 84th percentiles.}
\\
{\footnotesize Angles are measured east of north.}
\end{minipage}
\end{table}
\renewcommand*\arraystretch{1.0}

\medskip

\subsection{Kinematics} \label{subsec:kinematics}
We compute
the LOS stellar velocity dispersion of the strong lens galaxy
through the spherical Jeans equation \citep[see also][]{treu2002,koopmans2003}, similar to previous H0LiCOW analyses \citep[e.g.,][H0LiCOW IV]{suyu2010b}.
\citet{yildirim2019} recently showed that the assumption of spherical Jeans equation is applicable to time-delay cosmography with a single aperture-averaged lens velocity dispersion without significant bias, as in our case of \wfilens.  
For a given lens model, we obtain the 3D mass profile of the lens galaxy by taking the spherical deprojection of the circularized surface mass density profile. The resulting 3D profile assumes analytical forms for both the SPEMD and the composite model.
The 3D distribution of tracers is obtained by applying the same procedure to the surface brightness distribution of the lens galaxy, modeled as a \citet{hernquist1990} profile.  We also tested a \citet{jaffe1983} profile, which has been shown to produce similar results \citep{suyu2010b}, and find that the results change by less than 1\%.
We parametrize the orbital anisotropy profile as an Osipkov-Merritt model \citep{osipkov1979,merritt1985}
\begin{equation}
\frac{\sigma_\theta^2}{\sigma_r^2} = 1 - \frac{r^2}{\rani^2 + r^2},
\end{equation}
where $\sigma_\theta$ and $\sigma_r$ are the tangential and radial velocity dispersions, respectively.  Given values of the lens mass parameters in \sref{subsec:massmodel}, the external convergence $\kext$ in \sref{sec:lensenv}, and the anisotropy radius $\rani$, we then calculate the LOS velocity dispersion profile by numerically integrating the solutions of the spherical Jeans equation as given by \citet{mamonlokas2005}.
Finally, we calculate the integral over the spectroscopic slit of the seeing-convolved brightness-weighted LOS velocity dispersion $\sigma^{\rm P}$ \citep[Equation (20) of][]{suyu2010b} and compare to the measurements to calculate the likelihood of the kinematics data,
\begin{eqnarray} \label{eq:kinlike}
  & & P(\sigma_{\mathrm{LOS}}|\boldsymbol{\nu},\bm{\pi},\kext,\rani) \nonumber \\
  & = &  \frac{1}{\sqrt{2\pi}\sigma_{\sigma_{\mathrm{LOS}}}} \exp\left[ -\frac{(\sigma^{\rm P}(\boldsymbol{\nu},\bm{\pi},\kext,\rani) -
    \sigma_{\mathrm{LOS}})^2}{2\sigma_{\sigma_{\mathrm{LOS}}}^2}\right],
\end{eqnarray}
where $\sigma_{\mathrm{LOS}} = 250\,\kms$ and $\sigma_{\sigma_{\mathrm{LOS}}}=19\,\kms$ (H0LiCOW X).  We adopt a uniform prior
on $\rani$ in a range from 0.5 to 5 times the effective radius, $\reff$,
which we calculate to be $\reff=1\farcs41$ from our lens light fitting in the F160W filter.  We fit to the double S\'{e}rsic light profile, as the Chameleon profile does not provide an accurate representation of the galaxy light distribution at large radii \citep{dutton2011}.  The point source contributes a very small amount to the galaxy light, but not enough to impact this calculation.  We note that the choice of filter affects $\reff$, but the impact is small and results in a negligible effect ($\lesssim 0.1\%$) on the final inference.

We use importance sampling \citep[e.g.,][]{lewis2002} to simultaneously combine the velocity dispersion and external
convergence distributions in \sref{sec:lensenv} with the $\tdistmod$ inferred from our lens
model.  Specifically, for each set of lens and cosmological parameters
$\{\boldsymbol{\nu},\boldsymbol{\pi}\}$ from our lens model MCMC chain, we draw a $\kext$ sample from the distribution in \sref{sec:lensenv}
and a sample of $\rani$ from the uniform distribution [0.5,5]$\reff$.
With these, we can then compute the kinematics likelihood in \eref{eq:kinlike} for the joint sample $\{\boldsymbol{\nu}, \boldsymbol{\pi}, \kext,
\rani\}$ and use this to weight the joint sample.  From the effective
{\it model} time-delay distance computed from our multi-plane lensing
($\tdistmod$) and the external convergence ($\kext$), we can then
compute the effective time-delay distance ($\tdist$) via \eref{eq:ddtkappa}, keeping its absolute value blinded until we finalize our analysis.  The resulting distribution of $\tdist$ encapsulates
the cosmological information from \wfilens.

\subsection{BIC Weighting}
\label{subsec:bic}
We weight our models using the BIC, defined as
\begin{equation} \label{eq:bic}
\mathrm{BIC} = \mathrm{ln}(n) k - 2 \mathrm{ln}(\hat{L}).
\end{equation}
$n$ is the number of data points, which is the number of pixels in the image region across both bands that are outside the fiducial AGN mask (so that we are comparing equal areas), plus eight (for the four AGN image positions), plus three (for the time delays), plus one (for the velocity dispersion).  $k$ is the number of free parameters, which is 
the number of parameters in the lens model that are given uniform priors, plus two (for the source position), plus one (for the anisotropy radius to predict the velocity dispersion).  $\hat{L}$ is the maximum likelihood of the model, which is the product of the AGN position likelihood, the time-delay likelihood, the pixelated image plane likelihood, and the kinematic likelihood.  The image plane likelihood is the Bayesian evidence of the pixelated source intensity reconstruction using the arcmask imaging data \citep[which marginalizes over the source surface brightness pixel parameters and is thus the likelihood of the lens/cosmological parameters excluding the source pixel parameters; see][]{suyu2010a} times the likelihood of the lens model parameters within the image plane region that excludes the arcmask.  
We evaluate the BIC using the fiducial weight image and arcmask, as the majority of the models were optimized with these.  This may penalize the model with a larger AGN mask and the 60x60 pixel source grid model with a larger arcmask, but choosing any other region would penalize the fiducial model and all the other models that used the same regions, so this choice is fair to the largest number of models.  We note that our computation of the BIC described above uses all available data sets (lensing image, time delays and lens velocity dispersion) for model comparison.\footnote{Ranking the lensing mass models based on BIC values computed from only the lensing data would lead to insignificant changes to the final BIC-weighted $\tdist$ distributions.}. The kinematics have a relatively small impact in comparison to the other terms, and does not strongly favor either the SPEMD or composite model.

We estimate the variance in the BIC, $\sigma_{\mathrm{\mathrm{BIC}}}^{2}$, by running the fiducial model with source resolutions of [47, 48, 49, 50, 51, 52, 53, 54, 56, 58, 60] pixels on a side (the $50\times 50$ pixel case is just the original fiducial model), keeping the arcmask the same.  Changing the source resolution in this way shifts the inferred $\tdist$ values stochastically, but there is no overall trend with resolution, and the degree of the shifts are smaller than the scatter among the different models we run.  We calculate the BIC for each of these models and take the variance of this set of models as the variance on the BIC, $\sigma_{\mathrm{BIC}}^{2}$.  We find $\sigma_{\mathrm{BIC}}^{2} \sim 41$ for the SPEMD models and $\sigma_{\mathrm{BIC}}^{2} \sim 55$ for the composite models.  In Appendix~\ref{app:srcres}, we show the BIC and BIC weight values for these source resolution tests.

To avoid biases due to our choice of lens model parameterization, we split the samples into the SPEMD and composite models and calculate the relative BIC and weighting for each set separately, similar to H0LiCOW IX.  Specifically, we weight a model with a given BIC of $x$ by a function $f_{\mathrm{BIC}}(x)$, defined as the convolution
\begin{equation} \label{eq:bicwht}
f_{\mathrm{BIC}}(x) = h(x,\sigma_{\mathrm{BIC}}) * \mathrm{exp} \left( -\frac{x - \mathrm{BIC_{min}}}{2} \right),
\end{equation}
where $\mathrm{BIC_{min}}$ is the smallest BIC value within a set of models (SPEMD or composite), and $h$ is a Gaussian centered on $x$ with a variance of $\sigma_{\mathrm{BIC}}^{2}$.  We follow the analytic calculation of \citet{yildirim2019} in evaluating the convolution integral in \eref{eq:bicwht}.  Once we have a weighted $\tdist$ distribution for the SPEMD models and another for the composite models, we combine these two with equal weight in the final inference.


\subsection{Blind Analysis} \label{subsec:blind}
We perform our analysis blindly using a similar procedure as for previous H0LiCOW analyses \citep[][H0LiCOW IV, IX]{suyu2013}.
In practice, this is done by subtracting the median of certain parameter PDFs from the distribution when displaying and analyzing results.  In particular, we blind the values of $\tdist$ or $H_{0}$.  This blinding procedure still allows us to measure their precision and relative offsets, as well as their correlation with other lens model parameters, but without knowing their values. Blinding also eliminates confirmation bias and the tendency for experimenters to stop their analysis when they obtain a value consistent with an ``expected" value, and forces us to be confident in our checks of systematic errors before finalizing our result.  After completing our analysis, writing this paper draft with blinded $\tdist$ distributions, and coming to an agreement among the coauthors to unblind the results on 7 May 2019, we unblinded and did not make any further changes to the models.  There is also no iteration between the lens modeling and time-delay measurements.  Throughout this paper, we show blinded $\tdist$ distributions until Section~\ref{sec:results}, where we reveal the absolute $\tdist$ values from our inference.


\section{Estimating the external convergence} \label{sec:lensenv}

We estimate the external convergence $\kappa_{\mathrm{ext}}$ using the weighted number counts technique, introduced in \citet{greene2013} and \citet[][hereafter H0LiCOW III]{rusu2017} and reframed as approximate Bayesian computation in H0LiCOW IX. For details of the numerical implementation, and a justification of the applicability in cosmography, we refer the reader to H0LiCOW III. Briefly, we use the catalogue of galaxies and associated physical properties (redshifts and stellar masses) around \wfilens~from \sref{subsec:phot} and H0LiCOW X, as well as a control catalogue of the same properties from non-contiguous regions of the sky, large enough to overcome sample variance, from the Canada-France-Hawaii Telescope Lensing Survey \citep[CFHTLenS;][]{heymans2012}. We compute relative galaxy number counts within the matching limiting magnitude and within the same apertures of $45\arcsec$- and $120\arcsec$-radii\footnote{In order to ensure a fair comparison of the number counts between the lens fields and the control field, in H0LiCOW X we performed detections in the same $i$-filter, where the images have similar seeing. To account for the coarser pixel scale in the DES data compared to CFHTLenS, we used more aggressive deblending parameters. This has no noticeable effect, with the exceptions of regions around bright stars. We ignore the negligible differences between the $ugriz$ filter curves in DES and CFHTLenS.}, using physically motivated weights $\zeta_q$ introduced by \citet{greene2013} and H0LiCOW III. Here $q$ stands for the redshift $z$, stellar mass $M_\star$, the inverse of the distance $r$ between each galaxy and the lens or the center of the aperture, etc. A full list of the weights and of the corresponding measured relative number counts is shown in Table~\ref{tab:overdens}.

\begin{table*}
 \centering
 \begin{minipage}{155mm}
  \caption{Weighted galaxy count ratios $\overline{\zeta_q}$ for \wfilens}
  \setlength\extrarowheight{3pt}
  \begin{tabular}{@{}lcccccc@{}}
  \hline
& $45\arcsec$ & $45\arcsec$ & $45\arcsec$ & $120\arcsec$ & $120\arcsec$ & $120\arcsec$ \\
Weight $q$ & fiducial & fiducial + & fiducial + & fiducial & fiducial + & fiducial +  \\
& & $z=0.49$ group & $z=0.49,0.66$ groups & & $z=0.49$ group & $z=0.49,0.66$ groups \\
 \hline
$1$               			      & $1.44^{+0.06}_{-0.08}$ & $1.20^{+0.03}_{-0.10}$ & $1.17^{+0.04}_{-0.07}$ & $1.55^{+0.07}_{-0.11}$ & $1.34^{+0.05}_{-0.08}$ & $1.27^{+0.04}_{-0.08}$  \\
$z$                               	      & $1.70^{+0.09}_{-0.13}$ & $1.36^{+0.07}_{-0.07}$ & $1.34^{+0.07}_{-0.08}$ & $1.65^{+0.12}_{-0.12}$ & $1.37^{+0.09}_{-0.10}$ & $1.31^{+0.08}_{-0.09}$  \\
$M_\star$                   	      & $1.45^{+0.11}_{-0.28}$ & $0.78^{+0.10}_{-0.04}$ & $0.80^{+0.10}_{-0.03}$ & $2.39^{+0.11}_{-0.29}$ & $1.68^{+0.30}_{-0.09}$ & $1.48^{+0.27}_{-0.09}$  \\
$M^2_\star$                  	      & $1.42^{+0.26}_{-0.49}$ & $0.51^{+0.12}_{-0.06}$ & $0.54^{+0.13}_{-0.05}$ & $3.69^{+0.43}_{-0.78}$ & $2.12^{+0.76}_{-0.17}$ & $1.75^{+0.64}_{-0.14}$  \\
$M^3_\star$                            & $1.40^{+0.45}_{-0.65}$ & $0.33^{+0.13}_{-0.06}$ & $0.37^{+0.14}_{-0.06}$ & $5.75^{+0.95}_{-1.75}$ & $2.73^{+1.53}_{-0.38}$ & $2.12^{+1.18}_{-0.31}$  \\
$1/r$                        	              & $1.33^{+0.01}_{-0.08}$ & $1.07^{+0.04}_{-0.06}$ & $1.04^{+0.04}_{-0.05}$ & $1.55^{+0.07}_{-0.11}$ & $1.35^{+0.05}_{-0.09}$ & $1.25^{+0.05}_{-0.08}$  \\
$z/r$                        	              & $1.49^{+0.03}_{-0.14}$ & $1.17^{+0.06}_{-0.06}$ & $1.14^{+0.06}_{-0.05}$ & $1.59^{+0.09}_{-0.10}$ & $1.33^{+0.06}_{-0.08}$ & $1.24^{+0.05}_{-0.07}$  \\
$M_\star/r$                              & $1.69^{+0.34}_{-0.33}$ & $0.69^{+0.11}_{-0.03}$ & $0.75^{+0.10}_{-0.05}$ & $2.08^{+0.21}_{-0.13}$ & $1.53^{+0.29}_{-0.09}$ & $1.38^{+0.27}_{-0.08}$  \\
$M^2_\star/r$                          & $1.97^{+0.73}_{-0.64}$ & $0.49^{+0.10}_{-0.07}$ & $0.56^{+0.11}_{-0.08}$ & $3.24^{+0.52}_{-0.55}$ & $1.89^{+0.85}_{-0.20}$ & $1.52^{+0.78}_{-0.11}$  \\
$M^3_\star/r$                   	      & $2.07^{+0.93}_{-1.00}$ & $0.32^{+0.12}_{-0.07}$ & $0.38^{+0.14}_{-0.07}$ & $5.30^{+1.00}_{-1.49}$ & $2.53^{+1.78}_{-0.44}$ & $1.86^{+1.42}_{-0.30}$  \\
$M^2_{\star,\mathrm{rms}}$   & $1.19^{+0.11}_{-0.23}$ & $0.71^{+0.08}_{-0.04}$ & $0.74^{+0.08}_{-0.04}$ & $1.92^{+0.11}_{-0.21}$ & $1.46^{+0.24}_{-0.06}$ & $1.32^{+0.23}_{-0.05}$  \\
$M^3_{\star,\mathrm{rms}}$   & $1.12^{+0.11}_{-0.21}$ & $0.69^{+0.08}_{-0.04}$ & $0.72^{+0.08}_{-0.04}$ & $1.79^{+0.09}_{-0.20}$ & $1.40^{+0.22}_{-0.07}$ & $1.28^{+0.21}_{-0.06}$  \\
$M^2_\star/r_\mathrm{,rms}$ & $1.40^{+0.24}_{-0.25}$ & $0.70^{+0.07}_{-0.06}$ & $0.75^{+0.07}_{-0.06}$ & $1.80^{+0.14}_{-0.16}$ & $1.38^{+0.28}_{-0.08}$ & $1.23^{+0.29}_{-0.04}$  \\
$M^3_\star/r_\mathrm{,rms}$ & $1.27^{+0.17}_{-0.25}$ & $0.68^{+0.08}_{-0.05}$ & $0.73^{+0.07}_{-0.06}$ & $1.74^{+0.11}_{-0.18}$ & $1.36^{+0.27}_{-0.08}$ & $1.23^{+0.26}_{-0.07}$  \\
$M_\star/r^3$                          & $1.13^{+0.26}_{-0.27}$ & $0.46^{+0.05}_{-0.04}$ & $0.48^{+0.06}_{-0.03}$ & $1.86^{+0.12}_{-0.23}$ & $1.37^{+0.14}_{-0.16}$ & $1.25^{+0.10}_{-0.15}$  \\
$M_\star/r^2$                          & $1.41^{+0.41}_{-0.29}$ & $0.54^{+0.09}_{-0.02}$ & $0.58^{+0.10}_{-0.04}$ & $2.04^{+0.14}_{-0.18}$ & $1.47^{+0.11}_{-0.09}$ & $1.31^{+0.11}_{-0.10}$  \\
$\sqrt{M_\star}/r$                    & $1.44^{+0.15}_{-0.19}$ & $0.81^{+0.06}_{-0.03}$ & $0.83^{+0.06}_{-0.04}$ & $1.78^{+0.06}_{-0.11}$ & $1.40^{+0.05}_{-0.06}$ & $1.29^{+0.04}_{-0.07}$  \\
$\sqrt{M_\mathrm{h}}/r$                         & $1.57^{+0.29}_{-0.39}$ & $1.08^{+0.57}_{-0.22}$ & $1.07^{+0.69}_{-0.21}$ & $1.81^{+0.09}_{-0.28}$ & $1.39^{+0.19}_{-0.06}$ & $1.27^{+0.18}_{-0.05}$  \\
\hline
\end{tabular}
\\
{\footnotesize Medians of weighted galaxy count ratios for \wfilens , inside two different aperture radii and down to $i\leq22.5$ mag. Weighted counts are themselves defined in terms of medians, following the third columns in Table~4 of H0LiCOW III. The errors include, in quadrature, scatter from 10 samplings of redshift and stellar mass for each galaxy in the \wfilens~field, scatter from the four disjoint CFHTLenS fields, and also from photometric redshifts measured with two different codes, as well as detections in the $i$ or $i+r$ bands. See H0LiCOW X for details. The weighted counts are computed after removing from counting the galaxies G2, G3 and G7 (corresponding to the fiducial lensing model from \sref{subsec:massmodel}; see Figure~\ref{fig:field}), and alternatively, by removing in addition to these the galaxies part of the group at the lens redshift ($z=0.66$), as well as the ones part of the groups at both $z=0.66$ and $z=0.49$ (see \sref{subsubsec:sys_tests}). See text for details of the selection of group members without spectroscopic redshifts.}
\label{tab:overdens}
\end{minipage}
\end{table*}

\begin{figure}
\includegraphics[width=\columnwidth]{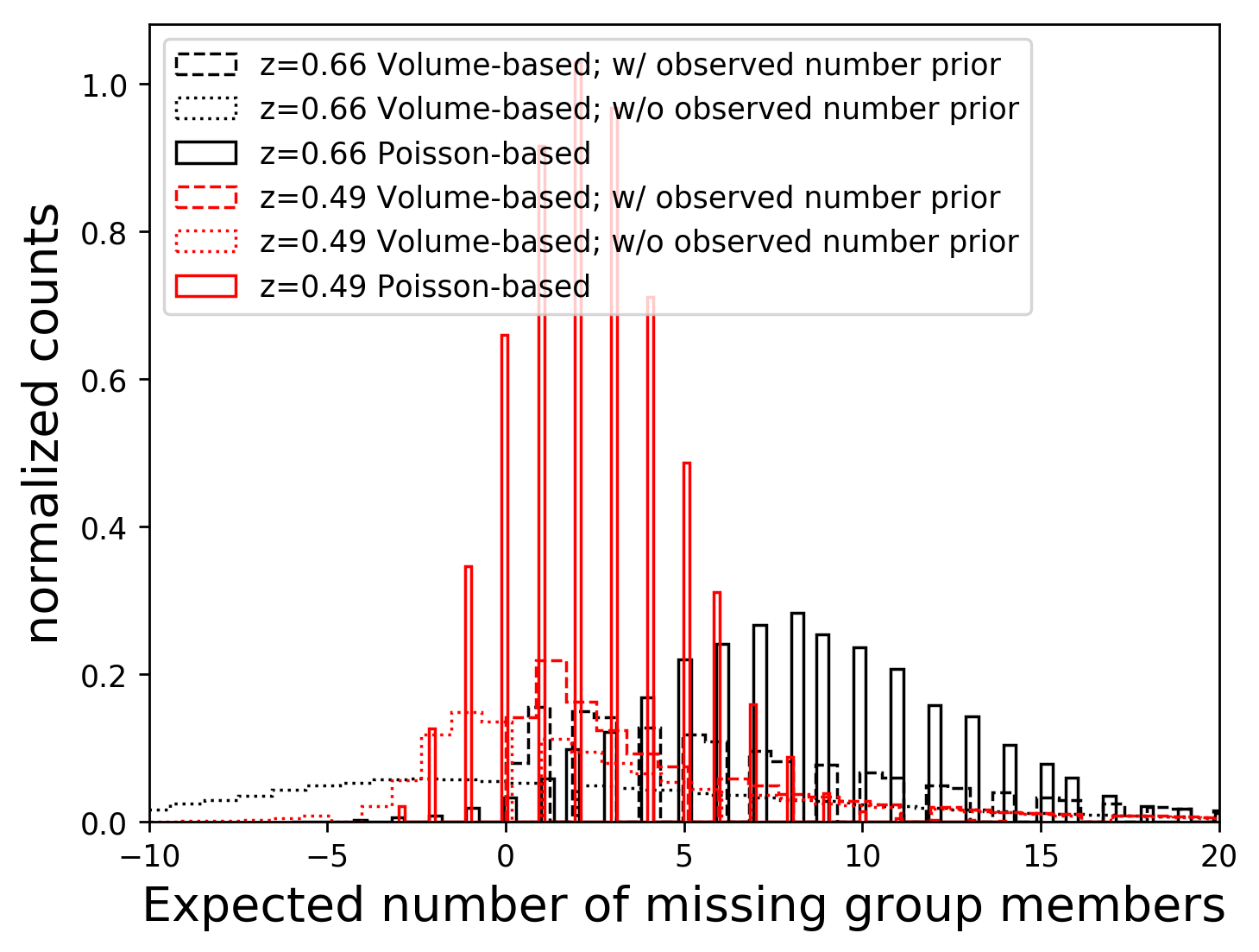}
\caption{Estimated number of missing galaxy group members inside the $\leq120\arcsec$-radius from the lens system, due to spectroscopic incompleteness, computed with two methods, for the two galaxy groups. For the volume-based method, we plot the distribution both with and without imposing the prior knowledge of the number of galaxies which are spectroscopically confirmed to be part of the groups, which is equivalent to truncating the distributions below 0.
  \label{fig:missingspec}}
\end{figure}

We perform the calculation above three times. These correspond, first, to the fiducial mass model in \sref{subsec:massmodel}, which incorporates the nearby galaxies G2, G3 and G7 from Figure~\ref{fig:field}. Since the effect of these galaxies has already been accounted for, we remove them from the input catalogue before running the computation. Second and third, we also remove one or both of the galaxy groups found in H0LiCOW X to impact the mass modeling beyond the tidal shear term, and therefore taken into account in the systematics tests presented in \sref{subsubsec:sys_tests}. However, since our spectroscopic completeness down to the limiting magnitude of $i<22.5$ is only $\sim50\%$, and also non-uniform, decreasing with radius from the lens (see Figure~4 in H0LiCOW X), it is likely that there are other galaxies part of these groups, in addition to the ones spectroscopically confirmed. If we were to keep these galaxies in the number counts, our inferred $\kappa_{\mathrm{ext}}$ would be an overestimate, when coupled to the models from \sref{subsubsec:sys_tests} which already include these galaxy groups. We use two different methods to account for these galaxies statistically. Briefly, in the first method we use the measured spectroscopic completeness and the total number of galaxies within the $120\arcsec$-radius aperture, as well as the number of confirmed group members inside the same aperture, and we apply Poisson statistics to infer the distribution of galaxy numbers we miss due to spectroscopic incompleteness. In the second method, we use the velocity dispersions of the two groups measured in H0LiCOW X, as well as the virial radii from \citet{wilson2016}, and we calculate the expected number of galaxies inside the virial radius, from the empirical relation in  \citet{andreon2010}. Then, based on the projected distance between the group centroid and the lens measured in H0LiCOW X, we estimate the expected number of galaxies at the intersection of the sphere of virial radius and the $120\arcsec$-radius cylinder centered on the lens. Subtracting from this the number of galaxy members spectroscopically confirmed, we arrive at a distribution of the number of missing galaxies. We show the resulting distributions from both methods in Figure~\ref{fig:missingspec}. For each group, the distributions from both methods overlap significantly, giving consistent results. The expected median number of galaxies missing from the group at $z=0.66$ is 6-8, and from the group at $z=0.49$ it is 3. Finally, we extract at random, 20 times, a number from these distributions, and remove these galaxies, picked at random from within our catalogue of galaxies around the lens, with photometric redshifts compatible with the group redshifts, before computing the weighted number counts. The resulting scatter is included in the values reported in Table~\ref{tab:overdens}.

Our measured relative weighted number counts, in particular for the fiducial model and inside the $120\arcsec$-radius aperture, show that the field of \wfilens~is overdense. This was also remarked in the study by \citet{fassnacht2011}, where a non-weighted number count overdensity of 1.33 was obtained. While a direct comparison with this values would be biased because of the different limiting magnitude, detection filter, and the fact that \citet{fassnacht2011} count all nearby galaxies without exception, our $\sim1.44$ unweighted relative number count inside the same aperture of $45\arcsec$-radius is consistent, within $1.5\sigma$. 

\begin{figure*}
\includegraphics[width=\textwidth]{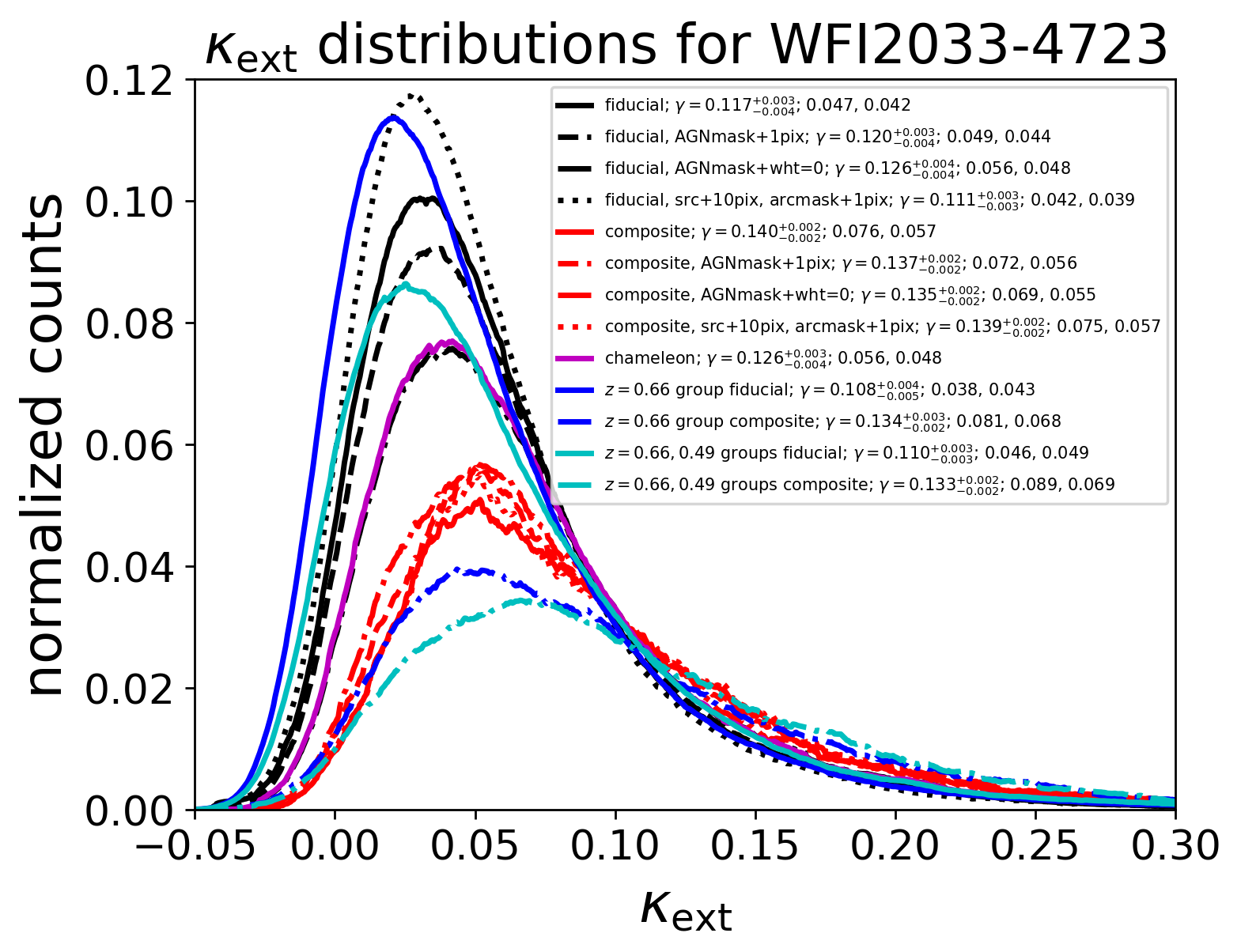}
\caption{ Distributions of $P(\kappa_{\mathrm{ext}}|\zeta^{45\arcsec}_\mathrm{1},\zeta^{45\arcsec}_\mathrm{1/r},\zeta^{120\arcsec}_\mathrm{1},\zeta^{120\arcsec}_\mathrm{1/r},\gamma)$ for the various lensing models described in Section~\ref{subsubsec:sys_tests}. The value of the corresponding external shear, which is a parameter of each lensing model and is used as constraint for the weighted number counts, is given in the legend. Following the shear values, the next two numbers in the legend are the pairs of (50\%th, (84\%th$-$16\%th)/2) percentiles, which measure the median and the spread of the distributions. The size of the histogram bin is $\Delta\kappa_\mathrm{ext}=0.00055$. As the original distributions are noisy, we plot their convolution with a large smoothing window of length $50\times\Delta\kappa_\mathrm{ext}$.
  \label{fig:kappa}}
\end{figure*}

To convert the measured relative weighted number counts into a $\kappa_{\textrm{ext}}$ distribution based on these constraints, we follow \citet[][]{suyu2010b,suyu2013,greene2013}; H0LiCOW III, IX; \citet{chen2019}, and use the results of ray-tracing by \citet{hilbert2009} through the Millennium Simulation \citep[MS;][]{springel2005}, in the form of a convergence and shear map ($\kappa$,$\gamma$) covering each simulated sky location. Our technique is justified by the results of \citet{suyu2010b,hilbert2009}, which showed that the distribution of $\kappa$ for LOS containing strong lenses is very similar to that over all LOS. With a catalogue of galaxies painted on top of the dark matter halos from the MS, following the semi-analytical models in \citet{delucia2007}, and containing realistic simulated photometry, we follow a similar procedure and compute relative weighted number counts at each spatial location throughout the MS (see H0LiCOW III for details). Finally, we compute
\begin{eqnarray} \label{eq:kappa}
P(\kappa_\mathrm{ext}|\mathbf{d}_\mathbf{LOS}) \equiv P(\kappa_\mathrm{ext}|\zeta_q,...) & \nonumber \\=\int\prod_qd\zeta_qP_\mathrm{MS}(\kappa_\mathrm{ext}|\zeta_q^\mathrm{MS}\equiv\zeta_q,...)P(\zeta_q,...|\mathbf{d}_\mathbf{LOS}) & 
\end{eqnarray}
where we combine multiple weighted number count constraints $\zeta_q$, including from both $45\arcsec$- and $120\arcsec$-radius apertures. Following H0LiCOW III, we treat the external shear $\gamma_{\textrm{ext}}$ computed from the lens models in \sref{sec:lensmod} at the location of the lens (in the case of the MS, at the center of each aperture) analogously to the weighted number count constraints.\footnote{Here and in \citet{chen2019} we modify the way we implement the $\gamma_{\textrm{ext}}$ constraint described in H0LiCOW III, in the sense that we no longer normalize by the number of LOS in each small division of the constraint range. This is because the shear values derived in \sref{sec:lensmod} use a flat prior, and the distribution of $\gamma$ in the MS maps naturally introduces a cosmological prior. This effect is negligible, except for the case of very large $\gamma_{\textrm{ext}}$ uncertainties.} In \sref{app:kappa} we explore various combinations of constraints, show that our technique is free of biases, and describe several tests we ran. We settle on the combination of $P(\kappa_{\mathrm{ext}}|\zeta^{45\arcsec}_\mathrm{1},\zeta^{45\arcsec}_\mathrm{1/r},\zeta^{120\arcsec}_\mathrm{1},\zeta^{120\arcsec}_\mathrm{1/r},\gamma)$, which employs our most robust constraints. In Figure~\ref{fig:kappa} we show the resulting distributions, corresponding to the various mass models explored in \sref{sec:lensmod}, and their associated shear values\footnote{It is unexpected that the two distributions of the composite model, which correspond to the case where one or both galaxy groups are explicitly modeled, have larger medians than the standard composite model (by $\lesssim0.2\sigma$ or at $\lesssim1\%$ level), even though they are constrained by smaller values of shear and weighted counts. We attribute this to noise, and we have checked that the excess is consistent with variations between similar distributions for these models, employing a different choice of weighted count constraints.}.

\section{Results} \label{sec:results}

\begin{figure*}
\includegraphics[width=\textwidth]{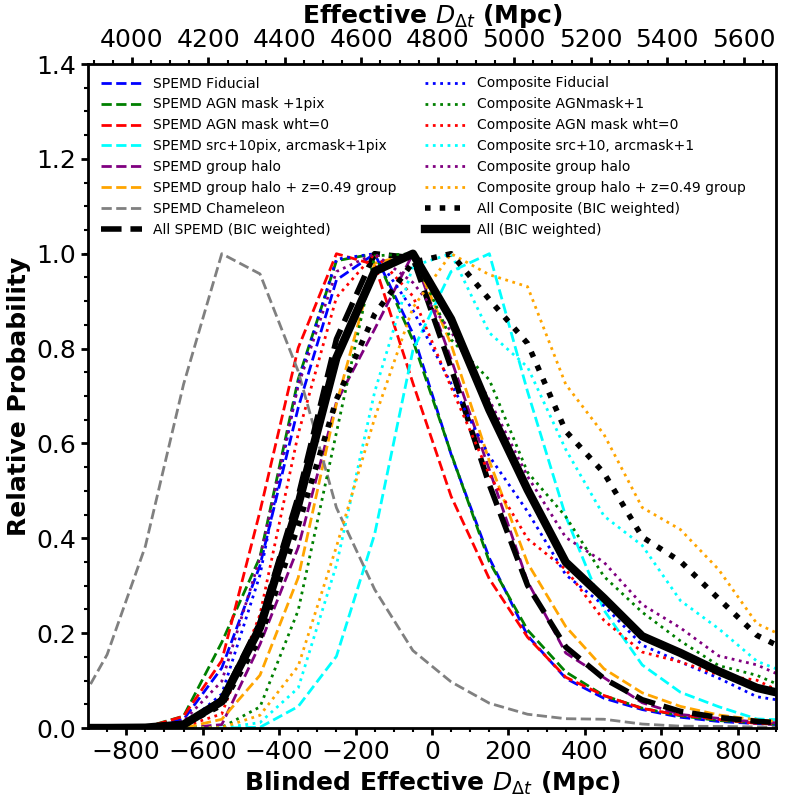}
\caption{
PDF of $\tdist$ for \wfilens .  The SPEMD and composite models are weighted by BIC, then are each given equal weight in the final inference. 
\label{fig:dt_bic}}
\end{figure*}

After conducting the analysis described in \sref{sec:lensmod} and \sref{sec:lensenv}, and combining with the time delays from \sref{subsec:td}, we plot the final BIC-weighted $\tdist$ distributions in Figure~\ref{fig:dt_bic}, with the blinded values shown on the bottom axis and the unblinded values shown on the top axis.  We report the median and 68\% quantiles of $\tdist$ for each of the models Table~\ref{tab:dt}, along with the $\Delta$BIC and associated weighting that each model receives.  Our constraint on $\tdist$ for \wfilens~is $\tdist = \dt$, a $\sim\dtprec\%$ precision measurement.

\renewcommand*\arraystretch{1.5}
\begin{table*}
\caption{Effective time-delay distance and BIC weighting for different lens models \label{tab:dt}}
\begin{minipage}{\linewidth}
\begin{tabular}{l|lll}
\hline
Model &
$\tdist$ (Mpc) &
$\Delta$BIC &
BIC weight
\\
\hline
SPEMD fiducial &
$4640_{-195}^{+238}$ &
17 &
0.674
\\
SPEMD AGN mask + 1pix &
$4631_{-199}^{+247}$ &
49 &
0.235
\\
SPEMD AGN mask weight=0 &
$4611_{-195}^{+258}$ &
1984 &
0.000
\\
SPEMD arcmask+1pix, 60x60 source &
$4894_{-190}^{+219}$ &
306 &
0.000
\\
SPEMD lens group halo &
$4732_{-203}^{+228}$ &
16 &
0.686
\\
SPEMD lens + z=0.49 group halo &
$4740_{-190}^{+251}$ &
0 &
1.000
\\
SPEMD chameleon light profiles &
$4312_{-185}^{+241}$ &
1962 &
0.000
\\
All SPEMD (BIC-weighted) &
$4703_{-203}^{+245}$ &
$-$ &
$-$
\\
\hline
Composite fiducial &
$4731_{-246}^{+416}$ &
202 &
0.000
\\
Composite AGN mask + 1pix &
$4836_{-235}^{+418}$ &
206 &
0.000
\\
Composite AGN mask weight=0 &
$4743_{-238}^{+417}$ &
637 &
0.000
\\
Composite arcmask+1pix, 60x60 source &
$4944_{-258}^{+437}$ &
559 &
0.000
\\
Composite lens group halo &
$4779_{-283}^{+475}$ &
16 &
0.770
\\
Composite lens + z=0.49 group halo &
$5009_{-305}^{+471}$ &
0 &
1.000
\\
All Composite (BIC-weighted) &
$4913_{-320}^{+489}$ &
$-$ &
$-$
\\
\hline
All &
$4784_{-248}^{+399}$ &
$-$ &
$-$
\\
\hline
\end{tabular}
\\
{\footnotesize Reported values are medians, with errors corresponding to the 16th and 84th percentiles.}
\end{minipage}
\end{table*}
\renewcommand*\arraystretch{1.0}

From this inferred $\tdist$, we can calculate cosmological parameters for flat $\Lambda$CDM or other cosmologies.  For flat $\Lambda$CDM with uniform priors on $H_{0}$ (within the range [0, 150] km s$^{-1}$ Mpc$^{-1}$) and $\Om$ (within the range [0.05, 0.5]), this translates into a constraint on the Hubble constant of $H_{0} = \ulcdm$.  Within the uncertainties, our result is consistent with the previous measurements of $H_{0}$ from H0LiCOW. After \blens~\citep{suyu2010b}, this is the H0LiCOW lens producing the second smallest uncertainty on $H_{0}$, comparable to what is expected for the lensed supernovae ``Refsdal'' \citep{grillo2018}.  We note that weighting all of our models equally and ignoring the BIC weighting (as was done with some previous H0LiCOW lenses) gives $H_{0} = \equalweight$, which is a shift of $< 1\%$ in the median value. Our result can be combined with the other five lenses in the H0LiCOW sample to give stronger constraints for a variety of cosmological models. The full cosmological analysis is presented in \citet{wong2019}. Our measurement is also consistent with recent ones from other techniques, not only from {\it Planck} and SH0ES, but also from the Carnegie Supernova Project \citep{burns2018}, the Megamaser Cosmology project \citep{braatz2018}, DES clustering and weak lensing + baryon acoustic oscillations + Big Bang nucleosynthesis experiments \citep{abbott2018}, the inverse distance ladder \citep[e.g.,][]{aubourg2015,macaulay2019}, extragalactic background light attenuation \citep[e.g.,][]{dominguez2019}, etc. 

We note that \wfilens~has been used in the past to measure $H_{0}$. Based on three years of monitoring and constrained by the relative quasar image positions measured from archival {\it HST} imaging, \citet{vuissoz2008} infer $H_{0}=67_{-10}^{+13}~\mathrm{km~s^{-1}~Mpc^{-1}}$, using non-parametric modeling, in good agreement with our result, but with significantly larger uncertainties.  


\section{Conclusions} \label{sec:conclusions}
We  have analyzed the gravitational lens \wfilens, performing a blind cosmographic analysis to determine the time-delay distance of this system.  We use deep $\hst$ imaging, precise time-delay measurements, a measurement of the lens galaxy's velocity dispersion, and deep wide-area spectroscopic and photometric data to constrain the mass distribution along the LOS.  By accurately modeling the lens and accounting for systematic uncertainties, we constrain the effective time-delay distance to be $\tdist = \dt$, a precision of $\dtprec\%$.  This translates to a Hubble constant of $H_{0} = \ulcdm$ in a flat $\Lambda$CDM cosmology with uniform priors on $H_{0}$ and $\Om$, a value consistent with measurements from other techniques, as well as previous H0LiCOW lenses. A joint analysis of all six H0LiCOW lenses and our constraints on different cosmologies is presented in \citet{wong2019}.

As with all galaxy-scale lenses where time-delay cosmography is applied, we expect that our results can be improved with future, higher resolution adaptive optics imaging \citep{chen2016}, spatially resolved kinematics \citep{shajib2018}, and a more tailored, non-statistical treatment of the external convergence \citep{mccully2017}.


\section*{Acknowledgements}
%
We thank Adriano Agnello, Roger Blandford, Xuheng Ding, Georges Meylan, Danka Paraficz, Chiara Spiniello, Malte Tewes and Olga Tihhonova for their contributions to the H0LiCOW project.
H0LiCOW and COSMOGRAIL are made possible thanks to the continuous work of all observers and technical staff obtaining the monitoring observations, in particular at the Swiss Euler telescope at La Silla Observatory. Euler is supported by the Swiss National Science Foundation.
C.E.R. and C.D.F. are funded through the NSF grant AST-1312329,
``Collaborative Research: Accurate cosmology with strong gravitational
lens time delays,'' and the {\it HST} grant GO-12889.
This work was supported by World Premier International Research Center Initiative (WPI Initiative), MEXT, Japan.
K.C.W. is supported in part by an EACOA Fellowship awarded by the East Asia Core
Observatories Association, which consists of the Academia Sinica
Institute of Astronomy and Astrophysics, the National Astronomical
Observatory of Japan, the National Astronomical Observatories of the
Chinese Academy of Sciences, and the Korea Astronomy and Space Science
Institute.
V.B. and F.C. acknowledge support from the Swiss National Science Foundation (SNSF) and through European Research Council (ERC) under the European Union's Horizon 2020 research and innovation programme (COSMICLENS: grant agreement No 787866)
D.S. acknowledges funding support from a {\it {Back to Belgium}} grant from the Belgian Federal Science Policy (BELSPO).
S.H.S. thanks the Max Planck Society for support through the Max
Planck Research Group. This research was supported in part by Perimeter Institute for Theoretical Physics. Research at Perimeter Institute is supported by the Government of Canada through the Department of Innovation, Science and Economic Development and by the Province of Ontario through the Ministry of Research, Innovation and Science.
S.H. acknowledges support by the DFG cluster of excellence \lq{}Origin and Structure of the Universe\rq{} (\href{http://www.universe-cluster.de}{\texttt{www.universe-cluster.de}}).
C.D.F. and G.C.-F.C. acknowledge support for this work from the National Science Foundation under Grant No. AST-1715611.
T.T. thanks the Packard Foundation for generous support through a Packard
Research Fellowship, the NSF for funding through NSF grant AST-1450141,
``Collaborative Research: Accurate cosmology with strong gravitational lens time delays".
G.~C.-F.~Chen acknowledges support from the Ministry of Education in Taiwan via Government Scholarship to Study Abroad (GSSA)
L.V.E.K. is supported in part through an NWO-VICI career grant (project number 639.043.308).
P.J.M. acknowledges support from the U.S.\ Department of Energy under
contract number DE-AC02-76SF00515.

This paper is based on observations made with the NASA/ESA
{\it Hubble Space Telescope}, obtained at the Space Telescope Science
Institute, which is operated by the Association of Universities for
Research in Astronomy, Inc., under NASA contract NAS 5-26555. These
observations are associated with Programs \#12889 and \#9744.
Support for program \#12889 was provided by NASA through a grant from the Space Telescope Science Institute, which is operated by the Association of Universities for Research in Astronomy, Inc., under NASA contract NAS 5-26555.

This work is partly based on the analysis in \citep{sluse2019}, which used public archival data from the Dark Energy Survey (DES). Funding for the DES Projects has been provided by the U.S. Department of Energy, the U.S. National Science Foundation, the Ministry of Science and Education of Spain, the Science and Technology Facilities Council of the United Kingdom, the Higher Education Funding Council for England, the National Center for Supercomputing Applications at the University of Illinois at Urbana-Champaign, the Kavli Institute of Cosmological Physics at the University of Chicago, the Center for Cosmology and Astro-Particle Physics at the Ohio State University, the Mitchell Institute for Fundamental Physics and Astronomy at Texas A\&M University, Financiadora de Estudos e Projetos, Funda{\c c}{\~a}o Carlos Chagas Filho de Amparo {\`a} Pesquisa do Estado do Rio de Janeiro, Conselho Nacional de Desenvolvimento Cient{\'i}fico e Tecnol{\'o}gico and the Minist{\'e}rio da Ci{\^e}ncia, Tecnologia e Inova{\c c}{\~a}o, the Deutsche Forschungsgemeinschaft, and the Collaborating Institutions in the Dark Energy Survey.
The Collaborating Institutions are Argonne National Laboratory, the University of California at Santa Cruz, the University of Cambridge, Centro de Investigaciones Energ{\'e}ticas, Medioambientales y Tecnol{\'o}gicas-Madrid, the University of Chicago, University College London, the DES-Brazil Consortium, the University of Edinburgh, the Eidgen{\"o}ssische Technische Hochschule (ETH) Z{\"u}rich,  Fermi National Accelerator Laboratory, the University of Illinois at Urbana-Champaign, the Institut de Ci{\`e}ncies de l'Espai (IEEC/CSIC), the Institut de F{\'i}sica d'Altes Energies, Lawrence Berkeley National Laboratory, the Ludwig-Maximilians Universit{\"a}t M{\"u}nchen and the associated Excellence Cluster Universe, the University of Michigan, the National Optical Astronomy Observatory, the University of Nottingham, The Ohio State University, the OzDES Membership Consortium, the University of Pennsylvania, the University of Portsmouth, SLAC National Accelerator Laboratory, Stanford University, the University of Sussex, and Texas A\&M University.
Based in part on observations at Cerro Tololo Inter-American Observatory, National Optical Astronomy Observatory, which is operated by the Association of Universities for Research in Astronomy (AURA) under a cooperative agreement with the National Science Foundation.

This work is based on observations obtained with MegaPrime/MegaCam, a joint project of CFHT and CEA/IRFU, at the Canada-France-Hawaii Telescope (CFHT) which is operated by the National Research Council (NRC) of Canada, the Institut National des Sciences de l'Univers of the Centre National de la Recherche Scientifique (CNRS) of France, and the University of Hawaii. This research used the facilities of the Canadian Astronomy Data Centre operated by the National Research Council of Canada with the support of the Canadian Space Agency. CFHTLenS data processing was made possible thanks to significant computing support from the NSERC Research Tools and Instruments grant program.

Data analysis was in part carried out on common use data analysis computer system at the Astronomy Data Center, ADC, of the National Astronomical Observatory of Japan. This work made use of \texttt{Astropy}, a community-developed core Python package for Astronomy \citep{astropy2013,astropy2018}. Plots were produced with \texttt{Matplotlib} \citep{hunter2007}.


\bibliography{wfi2033lensmodel}

\begin{thebibliography}{}
\makeatletter
\relax
\def\mn@urlcharsother{\let\do\@makeother \do\$\do\&\do\#\do\^\do\_\do\%\do\~}
\def\mn@doi{\begingroup\mn@urlcharsother \@ifnextchar [ {\mn@doi@}
  {\mn@doi@[]}}
\def\mn@doi@[#1]#2{\def\@tempa{#1}\ifx\@tempa\@empty \href
  {http://dx.doi.org/#2} {doi:#2}\else \href {http://dx.doi.org/#2} {#1}\fi
  \endgroup}
\def\mn@eprint#1#2{\mn@eprint@#1:#2::\@nil}
\def\mn@eprint@arXiv#1{\href {http://arxiv.org/abs/#1} {{\tt arXiv:#1}}}
\def\mn@eprint@dblp#1{\href {http://dblp.uni-trier.de/rec/bibtex/#1.xml}
  {dblp:#1}}
\def\mn@eprint@#1:#2:#3:#4\@nil{\def\@tempa {#1}\def\@tempb {#2}\def\@tempc
  {#3}\ifx \@tempc \@empty \let \@tempc \@tempb \let \@tempb \@tempa \fi \ifx
  \@tempb \@empty \def\@tempb {arXiv}\fi \@ifundefined
  {mn@eprint@\@tempb}{\@tempb:\@tempc}{\expandafter \expandafter \csname
  mn@eprint@\@tempb\endcsname \expandafter{\@tempc}}}

\bibitem[\protect\citeauthoryear{{Abbott} et~al.,}{{Abbott}
  et~al.}{2018}]{abbott2018}
{Abbott} T.~M.~C.,  et~al., 2018, \mn@doi [\mnras] {10.1093/mnras/sty1939},
  \href {http://adsabs.harvard.edu/abs/2018MNRAS.480.3879A} {480, 3879}

\bibitem[\protect\citeauthoryear{{Andreon} \& {Hurn}}{{Andreon} \&
  {Hurn}}{2010}]{andreon2010}
{Andreon} S.,  {Hurn} M.~A.,  2010, \mn@doi [\mnras]
  {10.1111/j.1365-2966.2010.16406.x}, \href
  {http://adsabs.harvard.edu/abs/2010MNRAS.404.1922A} {404, 1922}

\bibitem[\protect\citeauthoryear{{Appenzeller} et~al.,}{{Appenzeller}
  et~al.}{1998}]{Appenzeller1998}
{Appenzeller} I.,  et~al., 1998, The Messenger, 94, 1

\bibitem[\protect\citeauthoryear{{Astropy Collaboration} et~al.,}{{Astropy
  Collaboration} et~al.}{2013}]{astropy2013}
{Astropy Collaboration} et~al., 2013, \mn@doi [\aap]
  {10.1051/0004-6361/201322068}, \href
  {http://adsabs.harvard.edu/abs/2013A%26A...558A..33A} {558, A33}

\bibitem[\protect\citeauthoryear{{Astropy Collaboration} et~al.,}{{Astropy
  Collaboration} et~al.}{2018}]{astropy2018}
{Astropy Collaboration} et~al., 2018, \mn@doi [\aj] {10.3847/1538-3881/aabc4f},
  \href {http://adsabs.harvard.edu/abs/2018AJ....156..123A} {156, 123}

\bibitem[\protect\citeauthoryear{{Aubourg} et~al.,}{{Aubourg}
  et~al.}{2015}]{aubourg2015}
{Aubourg} {\'E}.,  et~al., 2015, \mn@doi [\prd] {10.1103/PhysRevD.92.123516},
  \href {http://adsabs.harvard.edu/abs/2015PhRvD..92l3516A} {92, 123516}

\bibitem[\protect\citeauthoryear{{Auger}, {Treu}, {Bolton}, {Gavazzi},
  {Koopmans}, {Marshall}, {Moustakas}  \& {Burles}}{{Auger}
  et~al.}{2010}]{auger2010}
{Auger} M.~W.,  {Treu} T.,  {Bolton} A.~S.,  {Gavazzi} R.,  {Koopmans}
  L.~V.~E.,  {Marshall} P.~J.,  {Moustakas} L.~A.,   {Burles} S.,  2010,
  \mn@doi [\apj] {10.1088/0004-637X/724/1/511}, \href
  {http://adsabs.harvard.edu/abs/2010ApJ...724..511A} {724, 511}

\bibitem[\protect\citeauthoryear{{Bacon} et~al.,}{{Bacon}
  et~al.}{2010}]{Bacon2010}
{Bacon} R.,  et~al., 2010, in Ground-based and Airborne Instrumentation for
  Astronomy III. p. 773508, \mn@doi{10.1117/12.856027}

\bibitem[\protect\citeauthoryear{{Barkana}}{{Barkana}}{1998}]{barkana1998}
{Barkana} R.,  1998, \mn@doi [\apj] {10.1086/305950}, \href
  {http://adsabs.harvard.edu/abs/1998ApJ...502..531B} {502, 531}

\bibitem[\protect\citeauthoryear{{Birrer} \& {Amara}}{{Birrer} \&
  {Amara}}{2018}]{birrer2018}
{Birrer} S.,  {Amara} A.,  2018, \mn@doi [Physics of the Dark Universe]
  {10.1016/j.dark.2018.11.002}, \href
  {http://adsabs.harvard.edu/abs/2018PDU....22..189B} {22, 189}

\bibitem[\protect\citeauthoryear{{Birrer} \& {Treu}}{{Birrer} \&
  {Treu}}{2019}]{birrer2019b}
{Birrer} S.,  {Treu} T.,  2019, arXiv e-prints, \href
  {http://adsabs.harvard.edu/abs/2019arXiv190410965B} {}

\bibitem[\protect\citeauthoryear{{Birrer}, {Amara}  \& {Refregier}}{{Birrer}
  et~al.}{2015}]{birrer2015}
{Birrer} S.,  {Amara} A.,   {Refregier} A.,  2015, \mn@doi [\apj]
  {10.1088/0004-637X/813/2/102}, \href
  {http://adsabs.harvard.edu/abs/2015ApJ...813..102B} {813, 102}

\bibitem[\protect\citeauthoryear{{Birrer} et~al.,}{{Birrer}
  et~al.}{2019}]{birrer2019}
{Birrer} S.,  et~al., 2019, \mn@doi [\mnras] {10.1093/mnras/stz200}, \href
  {http://adsabs.harvard.edu/abs/2019MNRAS.484.4726B} {484, 4726}

\bibitem[\protect\citeauthoryear{{Blandford} \& {Narayan}}{{Blandford} \&
  {Narayan}}{1986}]{blandford1986}
{Blandford} R.,  {Narayan} R.,  1986, \mn@doi [\apj] {10.1086/164709}, \href
  {http://adsabs.harvard.edu/abs/1986ApJ...310..568B} {310, 568}

\bibitem[\protect\citeauthoryear{{Bolton}, {Burles}, {Koopmans}, {Treu}  \&
  {Moustakas}}{{Bolton} et~al.}{2006}]{bolton2006}
{Bolton} A.~S.,  {Burles} S.,  {Koopmans} L.~V.~E.,  {Treu} T.,   {Moustakas}
  L.~A.,  2006, \mn@doi [\apj] {10.1086/498884}, \href
  {http://adsabs.harvard.edu/abs/2006ApJ...638..703B} {638, 703}

\bibitem[\protect\citeauthoryear{{Bonvin}, {Tewes}, {Courbin}, {Kuntzer},
  {Sluse}  \& {Meylan}}{{Bonvin} et~al.}{2016}]{bonvin2016}
{Bonvin} V.,  {Tewes} M.,  {Courbin} F.,  {Kuntzer} T.,  {Sluse} D.,   {Meylan}
  G.,  2016, \mn@doi [\aap] {10.1051/0004-6361/201526704}, \href
  {http://adsabs.harvard.edu/abs/2016A%26A...585A..88B} {585, A88}

\bibitem[\protect\citeauthoryear{{Bonvin} et~al.,}{{Bonvin}
  et~al.}{2018}]{bonvin2018}
{Bonvin} V.,  et~al., 2018, \mn@doi [\aap] {10.1051/0004-6361/201833287}, \href
  {http://adsabs.harvard.edu/abs/2018A%26A...616A.183B} {616, A183}

\bibitem[\protect\citeauthoryear{{Bonvin}, {Tihhonova}, {Millon}, {Chan},
  {Savary}, {Huber}  \& {Courbin}}{{Bonvin} et~al.}{2019a}]{bonvin2019}
{Bonvin} V.,  {Tihhonova} O.,  {Millon} M.,  {Chan} J.~H.-H.,  {Savary} E.,
  {Huber} S.,   {Courbin} F.,  2019a, \mn@doi [\aap]
  {10.1051/0004-6361/201833405}, \href
  {http://adsabs.harvard.edu/abs/2019A%26A...621A..55B} {621, A55}

\bibitem[\protect\citeauthoryear{{Bonvin} et~al.,}{{Bonvin}
  et~al.}{2019b}]{bonvin2019b}
{Bonvin} V.,  et~al., 2019b, \mn@doi [\aap] {10.1051/0004-6361/201935921},
  \href {https://ui.adsabs.harvard.edu/abs/2019A%26A...629A..97B} {629, A97}

\bibitem[\protect\citeauthoryear{{Braatz} et~al.,}{{Braatz}
  et~al.}{2018}]{braatz2018}
{Braatz} J.,  et~al., 2018, in {Tarchi} A.,  {Reid} M.~J.,   {Castangia} P.,
  eds,  IAU Symposium Vol. 336, Astrophysical Masers: Unlocking the Mysteries
  of the Universe. pp 86--91, \mn@doi{10.1017/S1743921317010249}

\bibitem[\protect\citeauthoryear{{Burns} et~al.,}{{Burns}
  et~al.}{2018}]{burns2018}
{Burns} C.~R.,  et~al., 2018, \mn@doi [\apj] {10.3847/1538-4357/aae51c}, \href
  {http://adsabs.harvard.edu/abs/2018ApJ...869...56B} {869, 56}

\bibitem[\protect\citeauthoryear{{Cackett}, {Horne}  \& {Winkler}}{{Cackett}
  et~al.}{2007}]{cackett2007}
{Cackett} E.~M.,  {Horne} K.,   {Winkler} H.,  2007, \mn@doi [\mnras]
  {10.1111/j.1365-2966.2007.12098.x}, \href
  {http://adsabs.harvard.edu/abs/2007MNRAS.380..669C} {380, 669}

\bibitem[\protect\citeauthoryear{{Cantale}, {Courbin}, {Tewes}, {Jablonka}  \&
  {Meylan}}{{Cantale} et~al.}{2016}]{cantale2016a}
{Cantale} N.,  {Courbin} F.,  {Tewes} M.,  {Jablonka} P.,   {Meylan} G.,  2016,
  \mn@doi [\aap] {10.1051/0004-6361/201424003}, \href
  {http://adsabs.harvard.edu/abs/2016A%26A...589A..81C} {589, A81}

\bibitem[\protect\citeauthoryear{{Chen} et~al.,}{{Chen}
  et~al.}{2016}]{chen2016}
{Chen} G.~C.-F.,  et~al., 2016, \mn@doi [\mnras] {10.1093/mnras/stw991}, \href
  {http://adsabs.harvard.edu/abs/2016MNRAS.462.3457C} {462, 3457}

\bibitem[\protect\citeauthoryear{{Chen} et~al.,}{{Chen}
  et~al.}{2018a}]{chen2018}
{Chen} G.~C.-F.,  et~al., 2018a, \mn@doi [\mnras] {10.1093/mnras/sty2350},
  \href {http://adsabs.harvard.edu/abs/2018MNRAS.481.1115C} {481, 1115}

\bibitem[\protect\citeauthoryear{{Chen}, {Fishbach}  \& {Holz}}{{Chen}
  et~al.}{2018b}]{chenhy2018}
{Chen} H.-Y.,  {Fishbach} M.,   {Holz} D.~E.,  2018b, \mn@doi [\nat]
  {10.1038/s41586-018-0606-0}, \href
  {http://adsabs.harvard.edu/abs/2018Natur.562..545C} {562, 545}

\bibitem[\protect\citeauthoryear{{Chen} et~al.,}{{Chen}
  et~al.}{2019}]{chen2019}
{Chen} G.~C.-F.,  et~al., 2019, \mn@doi [\mnras] {10.1093/mnras/stz2547}, \href
  {https://ui.adsabs.harvard.edu/abs/2019MNRAS.tmp.2193C} {}

\bibitem[\protect\citeauthoryear{{Collett} \& {Auger}}{{Collett} \&
  {Auger}}{2014}]{collett2014}
{Collett} T.~E.,  {Auger} M.~W.,  2014, \mn@doi [\mnras]
  {10.1093/mnras/stu1190}, \href
  {http://adsabs.harvard.edu/abs/2014MNRAS.443..969C} {443, 969}

\bibitem[\protect\citeauthoryear{{Collett} \& {Cunnington}}{{Collett} \&
  {Cunnington}}{2016}]{collett2016}
{Collett} T.~E.,  {Cunnington} S.~D.,  2016, \mn@doi [\mnras]
  {10.1093/mnras/stw1856}, \href
  {http://adsabs.harvard.edu/abs/2016MNRAS.462.3255C} {462, 3255}

\bibitem[\protect\citeauthoryear{{Collett} et~al.,}{{Collett}
  et~al.}{2013}]{collett2013}
{Collett} T.~E.,  et~al., 2013, \mn@doi [\mnras] {10.1093/mnras/stt504}, \href
  {http://adsabs.harvard.edu/abs/2013MNRAS.432..679C} {432, 679}

\bibitem[\protect\citeauthoryear{{Courbin}, {Eigenbrod}, {Vuissoz}, {Meylan}
  \& {Magain}}{{Courbin} et~al.}{2005}]{courbin2005}
{Courbin} F.,  {Eigenbrod} A.,  {Vuissoz} C.,  {Meylan} G.,   {Magain} P.,
  2005, in {Mellier} Y.,  {Meylan} G.,  eds,  IAU Symposium Vol. 225,
  Gravitational Lensing Impact on Cosmology. pp 297--303,
  \mn@doi{10.1017/S1743921305002097}

\bibitem[\protect\citeauthoryear{{Courbin} et~al.,}{{Courbin}
  et~al.}{2011}]{courbin2011}
{Courbin} F.,  et~al., 2011, \mn@doi [\aap] {10.1051/0004-6361/201015709},
  \href {http://adsabs.harvard.edu/abs/2011A%26A...536A..53C} {536, A53}

\bibitem[\protect\citeauthoryear{{De Lucia} \& {Blaizot}}{{De Lucia} \&
  {Blaizot}}{2007}]{delucia2007}
{De Lucia} G.,  {Blaizot} J.,  2007, \mn@doi [\mnras]
  {10.1111/j.1365-2966.2006.11287.x}, \href
  {http://adsabs.harvard.edu/abs/2007MNRAS.375....2D} {375, 2}

\bibitem[\protect\citeauthoryear{{Di Valentino}, {Linder}  \& {Melchiorri}}{{Di
  Valentino} et~al.}{2018}]{valentino2018}
{Di Valentino} E.,  {Linder} E.~V.,   {Melchiorri} A.,  2018, \mn@doi [\prd]
  {10.1103/PhysRevD.97.043528}, \href
  {http://adsabs.harvard.edu/abs/2018PhRvD..97d3528D} {97, 043528}

\bibitem[\protect\citeauthoryear{{Diemer}}{{Diemer}}{2018}]{diemer2018}
{Diemer} B.,  2018, \mn@doi [\apjs] {10.3847/1538-4365/aaee8c}, \href
  {http://adsabs.harvard.edu/abs/2018ApJS..239...35D} {239, 35}

\bibitem[\protect\citeauthoryear{{Ding} et~al.,}{{Ding}
  et~al.}{2018}]{ding2018}
{Ding} X.,  et~al., 2018, arXiv e-prints, \href
  {http://adsabs.harvard.edu/abs/2018arXiv180101506D} {}

\bibitem[\protect\citeauthoryear{{Dom{\'{\i}}nguez} et~al.,}{{Dom{\'{\i}}nguez}
  et~al.}{2019}]{dominguez2019}
{Dom{\'{\i}}nguez} A.,  et~al., 2019, arXiv e-prints, \href
  {http://adsabs.harvard.edu/abs/2019arXiv190312097D} {}

\bibitem[\protect\citeauthoryear{{Dutton} \& {Treu}}{{Dutton} \&
  {Treu}}{2014}]{dutton2014}
{Dutton} A.~A.,  {Treu} T.,  2014, \mn@doi [\mnras] {10.1093/mnras/stt2489},
  \href {http://adsabs.harvard.edu/abs/2014MNRAS.438.3594D} {438, 3594}

\bibitem[\protect\citeauthoryear{{Dutton} et~al.,}{{Dutton}
  et~al.}{2011}]{dutton2011}
{Dutton} A.~A.,  et~al., 2011, \mn@doi [\mnras]
  {10.1111/j.1365-2966.2011.18706.x}, \href
  {http://adsabs.harvard.edu/abs/2011MNRAS.417.1621D} {417, 1621}

\bibitem[\protect\citeauthoryear{{Eigenbrod}, {Courbin}, {Vuissoz}, {Meylan},
  {Saha}  \& {Dye}}{{Eigenbrod} et~al.}{2005}]{eigenbrod2005}
{Eigenbrod} A.,  {Courbin} F.,  {Vuissoz} C.,  {Meylan} G.,  {Saha} P.,   {Dye}
  S.,  2005, \mn@doi [\aap] {10.1051/0004-6361:20042422}, \href
  {http://adsabs.harvard.edu/abs/2005A%26A...436...25E} {436, 25}

\bibitem[\protect\citeauthoryear{{Eigenbrod}, {Courbin}, {Meylan}, {Vuissoz}
  \& {Magain}}{{Eigenbrod} et~al.}{2006}]{Eigenbrod2006b}
{Eigenbrod} A.,  {Courbin} F.,  {Meylan} G.,  {Vuissoz} C.,   {Magain} P.,
  2006, \mn@doi [\aap] {10.1051/0004-6361:20054454}, 451, 759

\bibitem[\protect\citeauthoryear{{Falco}, {Gorenstein}  \& {Shapiro}}{{Falco}
  et~al.}{1985}]{falco1985}
{Falco} E.~E.,  {Gorenstein} M.~V.,   {Shapiro} I.~I.,  1985, \mn@doi [\apjl]
  {10.1086/184422}, \href {http://adsabs.harvard.edu/abs/1985ApJ...289L...1F}
  {289, L1}

\bibitem[\protect\citeauthoryear{{Fassnacht}, {Pearson}, {Readhead}, {Browne},
  {Koopmans}, {Myers}  \& {Wilkinson}}{{Fassnacht}
  et~al.}{1999}]{fassnacht1999}
{Fassnacht} C.~D.,  {Pearson} T.~J.,  {Readhead} A.~C.~S.,  {Browne} I.~W.~A.,
  {Koopmans} L.~V.~E.,  {Myers} S.~T.,   {Wilkinson} P.~N.,  1999, \mn@doi
  [\apj] {10.1086/308118}, \href
  {http://adsabs.harvard.edu/abs/1999ApJ...527..498F} {527, 498}

\bibitem[\protect\citeauthoryear{{Fassnacht}, {Xanthopoulos}, {Koopmans}  \&
  {Rusin}}{{Fassnacht} et~al.}{2002}]{fassnacht2002}
{Fassnacht} C.~D.,  {Xanthopoulos} E.,  {Koopmans} L.~V.~E.,   {Rusin} D.,
  2002, \mn@doi [\apj] {10.1086/344368}, \href
  {http://adsabs.harvard.edu/abs/2002ApJ...581..823F} {581, 823}

\bibitem[\protect\citeauthoryear{{Fassnacht}, {Gal}, {Lubin}, {McKean},
  {Squires}  \& {Readhead}}{{Fassnacht} et~al.}{2006}]{fassnacht2006}
{Fassnacht} C.~D.,  {Gal} R.~R.,  {Lubin} L.~M.,  {McKean} J.~P.,  {Squires}
  G.~K.,   {Readhead} A.~C.~S.,  2006, \mn@doi [\apj] {10.1086/500927}, \href
  {http://adsabs.harvard.edu/abs/2006ApJ...642...30F} {642, 30}

\bibitem[\protect\citeauthoryear{{Fassnacht}, {Koopmans}  \&
  {Wong}}{{Fassnacht} et~al.}{2011}]{fassnacht2011}
{Fassnacht} C.~D.,  {Koopmans} L.~V.~E.,   {Wong} K.~C.,  2011, \mn@doi
  [\mnras] {10.1111/j.1365-2966.2010.17591.x}, \href
  {http://adsabs.harvard.edu/abs/2011MNRAS.410.2167F} {410, 2167}

\bibitem[\protect\citeauthoryear{Fazio et~al.,}{Fazio et~al.}{2004}]{fazio04}
Fazio G.~G.,  et~al., 2004, \mn@doi [\apjs] {10.1086/422843}, 154, 10

\bibitem[\protect\citeauthoryear{{Feeney}, {Peiris}, {Williamson}, {Nissanke},
  {Mortlock}, {Alsing}  \& {Scolnic}}{{Feeney} et~al.}{2019}]{feeney2019}
{Feeney} S.~M.,  {Peiris} H.~V.,  {Williamson} A.~R.,  {Nissanke} S.~M.,
  {Mortlock} D.~J.,  {Alsing} J.,   {Scolnic} D.,  2019, \mn@doi [Physical
  Review Letters] {10.1103/PhysRevLett.122.061105}, \href
  {http://adsabs.harvard.edu/abs/2019PhRvL.122f1105F} {122, 061105}

\bibitem[\protect\citeauthoryear{Flaugher et~al.,}{Flaugher
  et~al.}{2015}]{flaugher15}
Flaugher B.,  et~al., 2015, \mn@doi [\aj] {10.1088/0004-6256/150/5/150}, 150,
  150

\bibitem[\protect\citeauthoryear{{Gao} et~al.,}{{Gao} et~al.}{2016}]{gao2016}
{Gao} F.,  et~al., 2016, \mn@doi [\apj] {10.3847/0004-637X/817/2/128}, \href
  {http://adsabs.harvard.edu/abs/2016ApJ...817..128G} {817, 128}

\bibitem[\protect\citeauthoryear{{Gavazzi}, {Treu}, {Rhodes}, {Koopmans},
  {Bolton}, {Burles}, {Massey}  \& {Moustakas}}{{Gavazzi}
  et~al.}{2007}]{gavazzi2007}
{Gavazzi} R.,  {Treu} T.,  {Rhodes} J.~D.,  {Koopmans} L.~V.~E.,  {Bolton}
  A.~S.,  {Burles} S.,  {Massey} R.~J.,   {Moustakas} L.~A.,  2007, \mn@doi
  [\apj] {10.1086/519237}, \href
  {http://adsabs.harvard.edu/abs/2007ApJ...667..176G} {667, 176}

\bibitem[\protect\citeauthoryear{{Gavazzi}, {Treu}, {Koopmans}, {Bolton},
  {Moustakas}, {Burles}  \& {Marshall}}{{Gavazzi} et~al.}{2008}]{gavazzi2008}
{Gavazzi} R.,  {Treu} T.,  {Koopmans} L.~V.~E.,  {Bolton} A.~S.,  {Moustakas}
  L.~A.,  {Burles} S.,   {Marshall} P.~J.,  2008, \mn@doi [\apj]
  {10.1086/529541}, \href {http://adsabs.harvard.edu/abs/2008ApJ...677.1046G}
  {677, 1046}

\bibitem[\protect\citeauthoryear{{Golse} \& {Kneib}}{{Golse} \&
  {Kneib}}{2002}]{golse2002}
{Golse} G.,  {Kneib} J.-P.,  2002, \mn@doi [\aap] {10.1051/0004-6361:20020639},
  \href {http://adsabs.harvard.edu/abs/2002A%26A...390..821G} {390, 821}

\bibitem[\protect\citeauthoryear{{Gorenstein}, {Shapiro}  \&
  {Falco}}{{Gorenstein} et~al.}{1988}]{gorenstein1988}
{Gorenstein} M.~V.,  {Shapiro} I.~I.,   {Falco} E.~E.,  1988, \mn@doi [\apj]
  {10.1086/166226}, \href {http://adsabs.harvard.edu/abs/1988ApJ...327..693G}
  {327, 693}

\bibitem[\protect\citeauthoryear{{Greene} et~al.,}{{Greene}
  et~al.}{2013}]{greene2013}
{Greene} Z.~S.,  et~al., 2013, \mn@doi [\apj] {10.1088/0004-637X/768/1/39},
  \href {http://adsabs.harvard.edu/abs/2013ApJ...768...39G} {768, 39}

\bibitem[\protect\citeauthoryear{{Grillo} et~al.,}{{Grillo}
  et~al.}{2018}]{grillo2018}
{Grillo} C.,  et~al., 2018, \mn@doi [\apj] {10.3847/1538-4357/aac2c9}, \href
  {http://adsabs.harvard.edu/abs/2018ApJ...860...94G} {860, 94}

\bibitem[\protect\citeauthoryear{{Hernquist}}{{Hernquist}}{1990}]{hernquist1990}
{Hernquist} L.,  1990, \mn@doi [\apj] {10.1086/168845}, \href
  {http://adsabs.harvard.edu/abs/1990ApJ...356..359H} {356, 359}

\bibitem[\protect\citeauthoryear{{Heymans} et~al.,}{{Heymans}
  et~al.}{2012}]{heymans2012}
{Heymans} C.,  et~al., 2012, \mn@doi [\mnras]
  {10.1111/j.1365-2966.2012.21952.x}, \href
  {http://adsabs.harvard.edu/abs/2012MNRAS.427..146H} {427, 146}

\bibitem[\protect\citeauthoryear{{Hilbert}, {Hartlap}, {White}  \&
  {Schneider}}{{Hilbert} et~al.}{2009}]{hilbert2009}
{Hilbert} S.,  {Hartlap} J.,  {White} S.~D.~M.,   {Schneider} P.,  2009,
  \mn@doi [\aap] {10.1051/0004-6361/200811054}, \href
  {http://adsabs.harvard.edu/abs/2009A%26A...499...31H} {499, 31}

\bibitem[\protect\citeauthoryear{{Hook}, {J{\o}rgensen}, {Allington-Smith},
  {Davies}, {Metcalfe}, {Murowinski}  \& {Crampton}}{{Hook}
  et~al.}{2004}]{Hook2004}
{Hook} I.~M.,  {J{\o}rgensen} I.,  {Allington-Smith} J.~R.,  {Davies} R.~L.,
  {Metcalfe} N.,  {Murowinski} R.~G.,   {Crampton} D.,  2004, \mn@doi [\pasp]
  {10.1086/383624}, 116, 425

\bibitem[\protect\citeauthoryear{{Hu}}{{Hu}}{2005}]{hu2005}
{Hu} W.,  2005, in {Wolff} S.~C.,  {Lauer} T.~R.,  eds,  Astronomical Society
  of the Pacific Conference Series Vol. 339, Observing Dark Energy. p.~215
  (\mn@eprint {} {astro-ph/0407158})

\bibitem[\protect\citeauthoryear{{Hunter}}{{Hunter}}{2007}]{hunter2007}
{Hunter} J.~D.,  2007, \mn@doi [Computing in Science and Engineering]
  {10.1109/MCSE.2007.55}, \href
  {http://adsabs.harvard.edu/abs/2007CSE.....9...90H} {9, 90}

\bibitem[\protect\citeauthoryear{{Jaffe}}{{Jaffe}}{1983}]{jaffe1983}
{Jaffe} W.,  1983, \mn@doi [\mnras] {10.1093/mnras/202.4.995}, \href
  {http://adsabs.harvard.edu/abs/1983MNRAS.202..995J} {202, 995}

\bibitem[\protect\citeauthoryear{{Kassiola} \& {Kovner}}{{Kassiola} \&
  {Kovner}}{1993}]{kassiola1993}
{Kassiola} A.,  {Kovner} I.,  1993, \mn@doi [\apj] {10.1086/173325}, \href
  {http://adsabs.harvard.edu/abs/1993ApJ...417..450K} {417, 450}

\bibitem[\protect\citeauthoryear{{Keeton}}{{Keeton}}{2003}]{keeton2003}
{Keeton} C.~R.,  2003, \mn@doi [\apj] {10.1086/345717}, \href
  {http://adsabs.harvard.edu/abs/2003ApJ...584..664K} {584, 664}

\bibitem[\protect\citeauthoryear{{Keeton} \& {Kochanek}}{{Keeton} \&
  {Kochanek}}{1997}]{keeton1997}
{Keeton} C.~R.,  {Kochanek} C.~S.,  1997, \apj, \href
  {http://adsabs.harvard.edu/abs/1997ApJ...487...42K} {487, 42}

\bibitem[\protect\citeauthoryear{Kissler-Patig et~al.,}{Kissler-Patig
  et~al.}{2008}]{kissler08}
Kissler-Patig M.,  et~al., 2008, \mn@doi [\aap] {10.1051/0004-6361:200809910},
  491, 941

\bibitem[\protect\citeauthoryear{{Kochanek}}{{Kochanek}}{2002}]{kochanek2002}
{Kochanek} C.~S.,  2002, \mn@doi [\apj] {10.1086/342476}, \href
  {http://adsabs.harvard.edu/abs/2002ApJ...578...25K} {578, 25}

\bibitem[\protect\citeauthoryear{{Kochanek} \& {Apostolakis}}{{Kochanek} \&
  {Apostolakis}}{1988}]{kochanek1988}
{Kochanek} C.~S.,  {Apostolakis} J.,  1988, \mn@doi [\mnras]
  {10.1093/mnras/235.4.1073}, \href
  {http://adsabs.harvard.edu/abs/1988MNRAS.235.1073K} {235, 1073}

\bibitem[\protect\citeauthoryear{{Kochanek}, {Morgan}, {Falco}, {McLeod},
  {Winn}, {Dembicky}  \& {Ketzeback}}{{Kochanek} et~al.}{2006}]{kochanek2006}
{Kochanek} C.~S.,  {Morgan} N.~D.,  {Falco} E.~E.,  {McLeod} B.~A.,  {Winn}
  J.~N.,  {Dembicky} J.,   {Ketzeback} B.,  2006, \mn@doi [\apj]
  {10.1086/499766}, \href {http://adsabs.harvard.edu/abs/2006ApJ...640...47K}
  {640, 47}

\bibitem[\protect\citeauthoryear{{Koopmans}}{{Koopmans}}{2004}]{koopmans2004}
{Koopmans} L.~V.~E.,  2004, preprint, \href
  {http://adsabs.harvard.edu/abs/2004astro.ph.12596K} {} (\mn@eprint {arXiv}
  {astro-ph/0412596})

\bibitem[\protect\citeauthoryear{{Koopmans}, {Treu}, {Fassnacht}, {Blandford}
  \& {Surpi}}{{Koopmans} et~al.}{2003}]{koopmans2003}
{Koopmans} L.~V.~E.,  {Treu} T.,  {Fassnacht} C.~D.,  {Blandford} R.~D.,
  {Surpi} G.,  2003, \mn@doi [\apj] {10.1086/379226}, \href
  {http://adsabs.harvard.edu/abs/2003ApJ...599...70K} {599, 70}

\bibitem[\protect\citeauthoryear{{Kovner}}{{Kovner}}{1987}]{kovner1987}
{Kovner} I.,  1987, \mn@doi [\apj] {10.1086/165179}, \href
  {http://adsabs.harvard.edu/abs/1987ApJ...316...52K} {316, 52}

\bibitem[\protect\citeauthoryear{{Kreisch}, {Cyr-Racine}  \&
  {Dor{\'e}}}{{Kreisch} et~al.}{2019}]{kreisch2019}
{Kreisch} C.~D.,  {Cyr-Racine} F.-Y.,   {Dor{\'e}} O.,  2019, arXiv e-prints,
  \href {http://adsabs.harvard.edu/abs/2019arXiv190200534K} {}

\bibitem[\protect\citeauthoryear{{Lemon}, {Auger}  \& {McMahon}}{{Lemon}
  et~al.}{2019}]{lemon2019}
{Lemon} C.~A.,  {Auger} M.~W.,   {McMahon} R.~G.,  2019, \mn@doi [\mnras]
  {10.1093/mnras/sty3366}, \href
  {http://adsabs.harvard.edu/abs/2019MNRAS.483.4242L} {483, 4242}

\bibitem[\protect\citeauthoryear{{Lewis} \& {Bridle}}{{Lewis} \&
  {Bridle}}{2002}]{lewis2002}
{Lewis} A.,  {Bridle} S.,  2002, \mn@doi [\prd] {10.1103/PhysRevD.66.103511},
  \href {http://adsabs.harvard.edu/abs/2002PhRvD..66j3511L} {66, 103511}

\bibitem[\protect\citeauthoryear{{Liao} et~al.,}{{Liao}
  et~al.}{2015}]{liao2015}
{Liao} K.,  et~al., 2015, \mn@doi [\apj] {10.1088/0004-637X/800/1/11}, \href
  {http://adsabs.harvard.edu/abs/2015ApJ...800...11L} {800, 11}

\bibitem[\protect\citeauthoryear{{Macaulay} et~al.,}{{Macaulay}
  et~al.}{2019}]{macaulay2019}
{Macaulay} E.,  et~al., 2019, \mn@doi [\mnras] {10.1093/mnras/stz978}, \href
  {http://adsabs.harvard.edu/abs/2019MNRAS.486.2184M} {486, 2184}

\bibitem[\protect\citeauthoryear{{Magain}, {Courbin}  \& {Sohy}}{{Magain}
  et~al.}{1998}]{magain1998}
{Magain} P.,  {Courbin} F.,   {Sohy} S.,  1998, \mn@doi [\apj]
  {10.1086/305187}, \href {http://adsabs.harvard.edu/abs/1998ApJ...494..472M}
  {494, 472}

\bibitem[\protect\citeauthoryear{{Mamon} \& {{\L}okas}}{{Mamon} \&
  {{\L}okas}}{2005}]{mamonlokas2005}
{Mamon} G.~A.,  {{\L}okas} E.~L.,  2005, \mn@doi [\mnras]
  {10.1111/j.1365-2966.2005.09400.x}, \href
  {http://adsabs.harvard.edu/abs/2005MNRAS.363..705M} {363, 705}

\bibitem[\protect\citeauthoryear{{McCully}, {Keeton}, {Wong}  \&
  {Zabludoff}}{{McCully} et~al.}{2014}]{mccully2014}
{McCully} C.,  {Keeton} C.~R.,  {Wong} K.~C.,   {Zabludoff} A.~I.,  2014,
  \mn@doi [\mnras] {10.1093/mnras/stu1316}, \href
  {http://adsabs.harvard.edu/abs/2014MNRAS.443.3631M} {443, 3631}

\bibitem[\protect\citeauthoryear{{McCully}, {Keeton}, {Wong}  \&
  {Zabludoff}}{{McCully} et~al.}{2017}]{mccully2017}
{McCully} C.,  {Keeton} C.~R.,  {Wong} K.~C.,   {Zabludoff} A.~I.,  2017,
  \mn@doi [\apj] {10.3847/1538-4357/836/1/141}, \href
  {http://adsabs.harvard.edu/abs/2017ApJ...836..141M} {836, 141}

\bibitem[\protect\citeauthoryear{{Merritt}}{{Merritt}}{1985}]{merritt1985}
{Merritt} D.,  1985, \mn@doi [\aj] {10.1086/113810}, \href
  {http://adsabs.harvard.edu/abs/1985AJ.....90.1027M} {90, 1027}

\bibitem[\protect\citeauthoryear{{Moffat}}{{Moffat}}{1969}]{moffat1969}
{Moffat} A.~F.~J.,  1969, \aap, \href
  {http://adsabs.harvard.edu/abs/1969A%26A.....3..455M} {3, 455}

\bibitem[\protect\citeauthoryear{{Momcheva}, {Williams}, {Keeton}  \&
  {Zabludoff}}{{Momcheva} et~al.}{2006}]{momcheva2006}
{Momcheva} I.,  {Williams} K.,  {Keeton} C.,   {Zabludoff} A.,  2006, \mn@doi
  [\apj] {10.1086/500382}, \href
  {http://adsabs.harvard.edu/abs/2006ApJ...641..169M} {641, 169}

\bibitem[\protect\citeauthoryear{{Momcheva}, {Williams}, {Cool}, {Keeton}  \&
  {Zabludoff}}{{Momcheva} et~al.}{2015}]{momcheva2015}
{Momcheva} I.~G.,  {Williams} K.~A.,  {Cool} R.~J.,  {Keeton} C.~R.,
  {Zabludoff} A.~I.,  2015, \mn@doi [\apjs] {10.1088/0067-0049/219/2/29}, \href
  {http://adsabs.harvard.edu/abs/2015ApJS..219...29M} {219, 29}

\bibitem[\protect\citeauthoryear{{Morgan}, {Caldwell}, {Schechter}, {Dressler},
  {Egami}  \& {Rix}}{{Morgan} et~al.}{2004}]{Morgan2004}
{Morgan} N.~D.,  {Caldwell} J.~A.~R.,  {Schechter} P.~L.,  {Dressler} A.,
  {Egami} E.,   {Rix} H.-W.,  2004, \mn@doi [\aj] {10.1086/383295}, 127, 2617

\bibitem[\protect\citeauthoryear{{Morgan}, {Hyer}, {Bonvin}, {Mosquera},
  {Cornachione}, {Courbin}, {Kochanek}  \& {Falco}}{{Morgan}
  et~al.}{2018}]{morgan2018}
{Morgan} C.~W.,  {Hyer} G.~E.,  {Bonvin} V.,  {Mosquera} A.~M.,  {Cornachione}
  M.,  {Courbin} F.,  {Kochanek} C.~S.,   {Falco} E.~E.,  2018, \mn@doi [\apj]
  {10.3847/1538-4357/aaed3e}, \href
  {http://adsabs.harvard.edu/abs/2018ApJ...869..106M} {869, 106}

\bibitem[\protect\citeauthoryear{{Motta}, {Mediavilla}, {Rojas}, {Falco},
  {Jim{\'e}nez-Vicente}  \& {Mu{\~n}oz}}{{Motta} et~al.}{2017}]{motta2017}
{Motta} V.,  {Mediavilla} E.,  {Rojas} K.,  {Falco} E.~E.,
  {Jim{\'e}nez-Vicente} J.,   {Mu{\~n}oz} J.~A.,  2017, \mn@doi [\apj]
  {10.3847/1538-4357/835/2/132}, \href
  {http://adsabs.harvard.edu/abs/2017ApJ...835..132M} {835, 132}

\bibitem[\protect\citeauthoryear{{Navarro}, {Frenk}  \& {White}}{{Navarro}
  et~al.}{1996}]{navarro1996}
{Navarro} J.~F.,  {Frenk} C.~S.,   {White} S.~D.~M.,  1996, \mn@doi [\apj]
  {10.1086/177173}, \href {http://adsabs.harvard.edu/abs/1996ApJ...462..563N}
  {462, 563}

\bibitem[\protect\citeauthoryear{{Oguri}}{{Oguri}}{2007}]{oguri2007}
{Oguri} M.,  2007, \mn@doi [\apj] {10.1086/513093}, \href
  {http://adsabs.harvard.edu/abs/2007ApJ...660....1O} {660, 1}

\bibitem[\protect\citeauthoryear{{Osipkov}}{{Osipkov}}{1979}]{osipkov1979}
{Osipkov} L.~P.,  1979, Pis ma Astronomicheskii Zhurnal, \href
  {http://adsabs.harvard.edu/abs/1979PAZh....5...77O} {5, 77}

\bibitem[\protect\citeauthoryear{{Petters}, {Levine}  \&
  {Wambsganss}}{{Petters} et~al.}{2001}]{petters2001}
{Petters} A.~O.,  {Levine} H.,   {Wambsganss} J.,  2001, {Singularity theory
  and gravitational lensing}.
Birkhauser

\bibitem[\protect\citeauthoryear{Pirard et~al.,}{Pirard
  et~al.}{2004}]{pirard04}
Pirard J.-F.,  et~al., 2004, in {Moorwood} A.~F.~M.,  {Iye} M.,  eds,
  \procspie Vol. 5492, Ground-based Instrumentation for Astronomy. pp
  1763--1772, \mn@doi{10.1117/12.578293}, \url
  {http://adsabs.harvard.edu/abs/2004SPIE.5492.1763P}

\bibitem[\protect\citeauthoryear{{Planck Collaboration} et~al.,}{{Planck
  Collaboration} et~al.}{2015}]{planck2015}
{Planck Collaboration} et~al., 2015, preprint, \href
  {http://adsabs.harvard.edu/abs/2015arXiv150201589P} {} (\mn@eprint {arXiv}
  {1502.01589})

\bibitem[\protect\citeauthoryear{{Planck Collaboration} et~al.,}{{Planck
  Collaboration} et~al.}{2018}]{planck2018}
{Planck Collaboration} et~al., 2018, preprint, \href
  {http://adsabs.harvard.edu/abs/2018arXiv180706209P} {} (\mn@eprint {arXiv}
  {1807.06209})

\bibitem[\protect\citeauthoryear{{Poulin}, {Smith}, {Karwal}  \&
  {Kamionkowski}}{{Poulin} et~al.}{2018}]{poulin2018}
{Poulin} V.,  {Smith} T.~L.,  {Karwal} T.,   {Kamionkowski} M.,  2018, arXiv
  e-prints, \href {http://adsabs.harvard.edu/abs/2018arXiv181104083P} {}

\bibitem[\protect\citeauthoryear{{Refsdal}}{{Refsdal}}{1964}]{refsdal1964}
{Refsdal} S.,  1964, \mnras, \href
  {http://adsabs.harvard.edu/cgi-bin/nph-bib_query?bibcode=1964MNRAS.128..307R&db_key=AST}
  {128, 307}

\bibitem[\protect\citeauthoryear{{Richard}, R.  \& J.}{{Richard}
  et~al.}{2017}]{Richard2017}
{Richard} J.,  R. B.,   J. V.,  2017, MUSE Pipeline User Manual,
  VLT-MAN-ESO-261650, Issue 7.0.
ESO, 7.0 edn, \url
  {https://www.eso.org/sci/activities/vltsv/muse/ESO-261650_7_MUSE_User_Manual.pdf}

\bibitem[\protect\citeauthoryear{{Riess} et~al.,}{{Riess}
  et~al.}{2016}]{riess2016}
{Riess} A.~G.,  et~al., 2016, \mn@doi [\apj] {10.3847/0004-637X/826/1/56},
  \href {http://adsabs.harvard.edu/abs/2016ApJ...826...56R} {826, 56}

\bibitem[\protect\citeauthoryear{{Riess}, {Casertano}, {Yuan}, {Macri}  \&
  {Scolnic}}{{Riess} et~al.}{2019}]{riess2019}
{Riess} A.~G.,  {Casertano} S.,  {Yuan} W.,  {Macri} L.~M.,   {Scolnic} D.,
  2019, arXiv e-prints, \href
  {http://adsabs.harvard.edu/abs/2019arXiv190307603R} {}

\bibitem[\protect\citeauthoryear{{Rusu} et~al.,}{{Rusu}
  et~al.}{2017}]{rusu2017}
{Rusu} C.~E.,  et~al., 2017, \mn@doi [\mnras] {10.1093/mnras/stx285}, \href
  {http://adsabs.harvard.edu/abs/2017MNRAS.467.4220R} {467, 4220}

\bibitem[\protect\citeauthoryear{{Saha}}{{Saha}}{2000}]{saha2000}
{Saha} P.,  2000, \mn@doi [\aj] {10.1086/301581}, \href
  {http://adsabs.harvard.edu/abs/2000AJ....120.1654S} {120, 1654}

\bibitem[\protect\citeauthoryear{{Schechter} et~al.,}{{Schechter}
  et~al.}{1997}]{schechter1997}
{Schechter} P.~L.,  et~al., 1997, \mn@doi [\apjl] {10.1086/310478}, \href
  {http://adsabs.harvard.edu/abs/1997ApJ...475L..85S} {475, L85}

\bibitem[\protect\citeauthoryear{{Schneider}}{{Schneider}}{2014}]{schneider2014a}
{Schneider} P.,  2014, \mn@doi [\aap] {10.1051/0004-6361/201424450}, \href
  {http://adsabs.harvard.edu/abs/2014A%26A...568L...2S} {568, L2}

\bibitem[\protect\citeauthoryear{{Schneider} \& {Sluse}}{{Schneider} \&
  {Sluse}}{2013}]{schneider2013}
{Schneider} P.,  {Sluse} D.,  2013, \mn@doi [\aap]
  {10.1051/0004-6361/201321882}, \href
  {http://adsabs.harvard.edu/abs/2013A%26A...559A..37S} {559, A37}

\bibitem[\protect\citeauthoryear{{Schneider} \& {Sluse}}{{Schneider} \&
  {Sluse}}{2014}]{schneider2014}
{Schneider} P.,  {Sluse} D.,  2014, \mn@doi [\aap]
  {10.1051/0004-6361/201322106}, \href
  {http://adsabs.harvard.edu/abs/2014A%26A...564A.103S} {564, A103}

\bibitem[\protect\citeauthoryear{{Schneider}, {Ehlers}  \& {Falco}}{{Schneider}
  et~al.}{1992}]{schneider1992}
{Schneider} P.,  {Ehlers} J.,   {Falco} E.~E.,  1992, {Gravitational Lenses}.
Springer, \mn@doi{10.1007/978-3-662-03758-4}

\bibitem[\protect\citeauthoryear{{Seljak}}{{Seljak}}{1994}]{seljak1994}
{Seljak} U.,  1994, \mn@doi [\apj] {10.1086/174924}, \href
  {http://adsabs.harvard.edu/abs/1994ApJ...436..509S} {436, 509}

\bibitem[\protect\citeauthoryear{{Shajib}, {Treu}  \& {Agnello}}{{Shajib}
  et~al.}{2018}]{shajib2018}
{Shajib} A.~J.,  {Treu} T.,   {Agnello} A.,  2018, \mn@doi [\mnras]
  {10.1093/mnras/stx2302}, \href
  {http://adsabs.harvard.edu/abs/2018MNRAS.473..210S} {473, 210}

\bibitem[\protect\citeauthoryear{{Shakura} \& {Sunyaev}}{{Shakura} \&
  {Sunyaev}}{1973}]{shakura1973}
{Shakura} N.~I.,  {Sunyaev} R.~A.,  1973, \aap, \href
  {http://adsabs.harvard.edu/abs/1973A%26A....24..337S} {24, 337}

\bibitem[\protect\citeauthoryear{{Sluse}, {Hutsem{\'e}kers}, {Courbin},
  {Meylan}  \& {Wambsganss}}{{Sluse} et~al.}{2012}]{Sluse2012}
{Sluse} D.,  {Hutsem{\'e}kers} D.,  {Courbin} F.,  {Meylan} G.,   {Wambsganss}
  J.,  2012, \mn@doi [\aap] {10.1051/0004-6361/201219125}, \href
  {http://adsabs.harvard.edu/abs/2012A%26A...544A..62S} {544, A62}

\bibitem[\protect\citeauthoryear{{Sluse} et~al.,}{{Sluse}
  et~al.}{2019}]{sluse2019}
{Sluse} D.,  et~al., 2019, \mn@doi [\mnras] {10.1093/mnras/stz2483}, \href
  {https://ui.adsabs.harvard.edu/abs/2019MNRAS.tmp.2136S} {}

\bibitem[\protect\citeauthoryear{{Springel} et~al.,}{{Springel}
  et~al.}{2005}]{springel2005}
{Springel} V.,  et~al., 2005, \mn@doi [\nat] {10.1038/nature03597}, \href
  {http://adsabs.harvard.edu/abs/2005Natur.435..629S} {435, 629}

\bibitem[\protect\citeauthoryear{{Starkey}, {Horne}  \& {Villforth}}{{Starkey}
  et~al.}{2016}]{starkey2016}
{Starkey} D.~A.,  {Horne} K.,   {Villforth} C.,  2016, \mn@doi [\mnras]
  {10.1093/mnras/stv2744}, \href
  {http://adsabs.harvard.edu/abs/2016MNRAS.456.1960S} {456, 1960}

\bibitem[\protect\citeauthoryear{{Suyu}}{{Suyu}}{2012}]{suyu2012a}
{Suyu} S.~H.,  2012, \mn@doi [\mnras] {10.1111/j.1365-2966.2012.21661.x}, \href
  {http://adsabs.harvard.edu/abs/2012MNRAS.426..868S} {426, 868}

\bibitem[\protect\citeauthoryear{{Suyu} \& {Halkola}}{{Suyu} \&
  {Halkola}}{2010}]{suyu2010a}
{Suyu} S.~H.,  {Halkola} A.,  2010, \mn@doi [\aap]
  {10.1051/0004-6361/201015481}, \href
  {http://adsabs.harvard.edu/abs/2010A%26A...524A..94S} {524, A94}

\bibitem[\protect\citeauthoryear{{Suyu}, {Marshall}, {Hobson}  \&
  {Blandford}}{{Suyu} et~al.}{2006}]{suyu2006}
{Suyu} S.~H.,  {Marshall} P.~J.,  {Hobson} M.~P.,   {Blandford} R.~D.,  2006,
  \mn@doi [\mnras] {10.1111/j.1365-2966.2006.10733.x}, \href
  {http://adsabs.harvard.edu/abs/2006MNRAS.371..983S} {371, 983}

\bibitem[\protect\citeauthoryear{{Suyu}, {Marshall}, {Auger}, {Hilbert},
  {Blandford}, {Koopmans}, {Fassnacht}  \& {Treu}}{{Suyu}
  et~al.}{2010}]{suyu2010b}
{Suyu} S.~H.,  {Marshall} P.~J.,  {Auger} M.~W.,  {Hilbert} S.,  {Blandford}
  R.~D.,  {Koopmans} L.~V.~E.,  {Fassnacht} C.~D.,   {Treu} T.,  2010, \mn@doi
  [\apj] {10.1088/0004-637X/711/1/201}, \href
  {http://adsabs.harvard.edu/abs/2010ApJ...711..201S} {711, 201}

\bibitem[\protect\citeauthoryear{{Suyu} et~al.,}{{Suyu}
  et~al.}{2012a}]{suyu2012c}
{Suyu} S.~H.,  et~al., 2012a, preprint, \href
  {http://adsabs.harvard.edu/abs/2012arXiv1202.4459S} {} (\mn@eprint {arXiv}
  {1202.4459})

\bibitem[\protect\citeauthoryear{{Suyu} et~al.,}{{Suyu}
  et~al.}{2012b}]{suyu2012b}
{Suyu} S.~H.,  et~al., 2012b, \mn@doi [\apj] {10.1088/0004-637X/750/1/10},
  \href {http://adsabs.harvard.edu/abs/2012ApJ...750...10S} {750, 10}

\bibitem[\protect\citeauthoryear{{Suyu} et~al.,}{{Suyu}
  et~al.}{2013}]{suyu2013}
{Suyu} S.~H.,  et~al., 2013, \mn@doi [\apj] {10.1088/0004-637X/766/2/70}, \href
  {http://adsabs.harvard.edu/abs/2013ApJ...766...70S} {766, 70}

\bibitem[\protect\citeauthoryear{{Suyu} et~al.,}{{Suyu}
  et~al.}{2014}]{suyu2014}
{Suyu} S.~H.,  et~al., 2014, \mn@doi [\apjl] {10.1088/2041-8205/788/2/L35},
  \href {http://adsabs.harvard.edu/abs/2014ApJ...788L..35S} {788, L35}

\bibitem[\protect\citeauthoryear{{Suyu} et~al.,}{{Suyu}
  et~al.}{2017}]{suyu2017}
{Suyu} S.~H.,  et~al., 2017, \mn@doi [\mnras] {10.1093/mnras/stx483}, \href
  {http://adsabs.harvard.edu/abs/2017MNRAS.468.2590S} {468, 2590}

\bibitem[\protect\citeauthoryear{{Tewes}, {Courbin}  \& {Meylan}}{{Tewes}
  et~al.}{2013}]{tewes2013a}
{Tewes} M.,  {Courbin} F.,   {Meylan} G.,  2013, \mn@doi [\aap]
  {10.1051/0004-6361/201220123}, \href
  {http://adsabs.harvard.edu/abs/2013A%26A...553A.120T} {553, A120}

\bibitem[\protect\citeauthoryear{{Tie} \& {Kochanek}}{{Tie} \&
  {Kochanek}}{2018}]{tie2018}
{Tie} S.~S.,  {Kochanek} C.~S.,  2018, \mn@doi [\mnras]
  {10.1093/mnras/stx2348}, \href
  {http://adsabs.harvard.edu/abs/2018MNRAS.473...80T} {473, 80}

\bibitem[\protect\citeauthoryear{{Tihhonova} et~al.,}{{Tihhonova}
  et~al.}{2018}]{tihhonova2018}
{Tihhonova} O.,  et~al., 2018, \mn@doi [\mnras] {10.1093/mnras/sty1040}, \href
  {http://adsabs.harvard.edu/abs/2018MNRAS.477.5657T} {477, 5657}

\bibitem[\protect\citeauthoryear{{Treu} \& {Koopmans}}{{Treu} \&
  {Koopmans}}{2002}]{treu2002}
{Treu} T.,  {Koopmans} L.~V.~E.,  2002, \mn@doi [\apj] {10.1086/341216}, \href
  {http://adsabs.harvard.edu/abs/2002ApJ...575...87T} {575, 87}

\bibitem[\protect\citeauthoryear{{Treu} \& {Marshall}}{{Treu} \&
  {Marshall}}{2016}]{treu2016}
{Treu} T.,  {Marshall} P.~J.,  2016, \mn@doi [\aapr]
  {10.1007/s00159-016-0096-8}, \href
  {http://adsabs.harvard.edu/abs/2016A%26ARv..24...11T} {24, 11}

\bibitem[\protect\citeauthoryear{{Unruh}, {Schneider}  \& {Sluse}}{{Unruh}
  et~al.}{2017}]{unruh2017}
{Unruh} S.,  {Schneider} P.,   {Sluse} D.,  2017, \mn@doi [\aap]
  {10.1051/0004-6361/201629048}, \href
  {http://adsabs.harvard.edu/abs/2017A%26A...601A..77U} {601, A77}

\bibitem[\protect\citeauthoryear{{Vanderriest}, {Schneider}, {Herpe},
  {Chevreton}, {Moles}  \& {Wlerick}}{{Vanderriest}
  et~al.}{1989}]{vanderriest1989}
{Vanderriest} C.,  {Schneider} J.,  {Herpe} G.,  {Chevreton} M.,  {Moles} M.,
  {Wlerick} G.,  1989, \aap, \href
  {http://adsabs.harvard.edu/abs/1989A%26A...215....1V} {215, 1}

\bibitem[\protect\citeauthoryear{{Vattis}, {Koushiappas}  \& {Loeb}}{{Vattis}
  et~al.}{2019}]{vattis2019}
{Vattis} K.,  {Koushiappas} S.~M.,   {Loeb} A.,  2019, arXiv e-prints, \href
  {http://adsabs.harvard.edu/abs/2019arXiv190306220V} {}

\bibitem[\protect\citeauthoryear{{Vuissoz} et~al.,}{{Vuissoz}
  et~al.}{2008}]{vuissoz2008}
{Vuissoz} C.,  et~al., 2008, \mn@doi [\aap] {10.1051/0004-6361:200809866},
  \href {http://adsabs.harvard.edu/abs/2008A%26A...488..481V} {488, 481}

\bibitem[\protect\citeauthoryear{{Weinberg}, {Mortonson}, {Eisenstein},
  {Hirata}, {Riess}  \& {Rozo}}{{Weinberg} et~al.}{2013}]{weinberg2013}
{Weinberg} D.~H.,  {Mortonson} M.~J.,  {Eisenstein} D.~J.,  {Hirata} C.,
  {Riess} A.~G.,   {Rozo} E.,  2013, \mn@doi [\physrep]
  {10.1016/j.physrep.2013.05.001}, \href
  {http://adsabs.harvard.edu/abs/2013PhR...530...87W} {530, 87}

\bibitem[\protect\citeauthoryear{{Williams}, {Momcheva}, {Keeton}, {Zabludoff}
  \& {Leh{\'a}r}}{{Williams} et~al.}{2006}]{williams2006}
{Williams} K.~A.,  {Momcheva} I.,  {Keeton} C.~R.,  {Zabludoff} A.~I.,
  {Leh{\'a}r} J.,  2006, \mn@doi [\apj] {10.1086/504788}, \href
  {http://adsabs.harvard.edu/abs/2006ApJ...646...85W} {646, 85}

\bibitem[\protect\citeauthoryear{{Wilson}, {Zabludoff}, {Ammons}, {Momcheva},
  {Williams}  \& {Keeton}}{{Wilson} et~al.}{2016}]{wilson2016}
{Wilson} M.~L.,  {Zabludoff} A.~I.,  {Ammons} S.~M.,  {Momcheva} I.~G.,
  {Williams} K.~A.,   {Keeton} C.~R.,  2016, \mn@doi [\apj]
  {10.3847/1538-4357/833/2/194}, \href
  {http://adsabs.harvard.edu/abs/2016ApJ...833..194W} {833, 194}

\bibitem[\protect\citeauthoryear{{Wong}, {Keeton}, {Williams}, {Momcheva}  \&
  {Zabludoff}}{{Wong} et~al.}{2011}]{wong2011}
{Wong} K.~C.,  {Keeton} C.~R.,  {Williams} K.~A.,  {Momcheva} I.~G.,
  {Zabludoff} A.~I.,  2011, \mn@doi [\apj] {10.1088/0004-637X/726/2/84}, \href
  {http://adsabs.harvard.edu/abs/2011ApJ...726...84W} {726, 84}

\bibitem[\protect\citeauthoryear{{Wong} et~al.,}{{Wong}
  et~al.}{2017}]{wong2017}
{Wong} K.~C.,  et~al., 2017, \mn@doi [\mnras] {10.1093/mnras/stw3077}, \href
  {http://adsabs.harvard.edu/abs/2017MNRAS.465.4895W} {465, 4895}

\bibitem[\protect\citeauthoryear{{Wong} et~al.,}{{Wong}
  et~al.}{2019}]{wong2019}
{Wong} K.~C.,  et~al., 2019, arXiv e-prints, \href
  {https://ui.adsabs.harvard.edu/abs/2019arXiv190704869W} {}

\bibitem[\protect\citeauthoryear{{Wucknitz}}{{Wucknitz}}{2002}]{wucknitz2002}
{Wucknitz} O.,  2002, \mn@doi [\mnras] {10.1046/j.1365-8711.2002.05426.x},
  \href {http://adsabs.harvard.edu/abs/2002MNRAS.332..951W} {332, 951}

\bibitem[\protect\citeauthoryear{{Y{\i}ld{\i}r{\i}m}, {Suyu}  \&
  {Halkola}}{{Y{\i}ld{\i}r{\i}m} et~al.}{2019}]{yildirim2019}
{Y{\i}ld{\i}r{\i}m} A.,  {Suyu} S.~H.,   {Halkola} A.,  2019, arXiv e-prints
  (1904.07237), \href {http://adsabs.harvard.edu/abs/2019arXiv190407237Y} {}

\bibitem[\protect\citeauthoryear{{de Vaucouleurs}}{{de
  Vaucouleurs}}{1948}]{vaucouleurs1948}
{de Vaucouleurs} G.,  1948, Annales d'Astrophysique, \href
  {http://adsabs.harvard.edu/abs/1948AnAp...11..247D} {11, 247}

\makeatother
\end{thebibliography}
\bibliographystyle{../common/mnras}


\appendix

\section{Residuals without power-law weighting} \label{app:noplwht}
\begin{figure*}
\includegraphics[width=\textwidth]{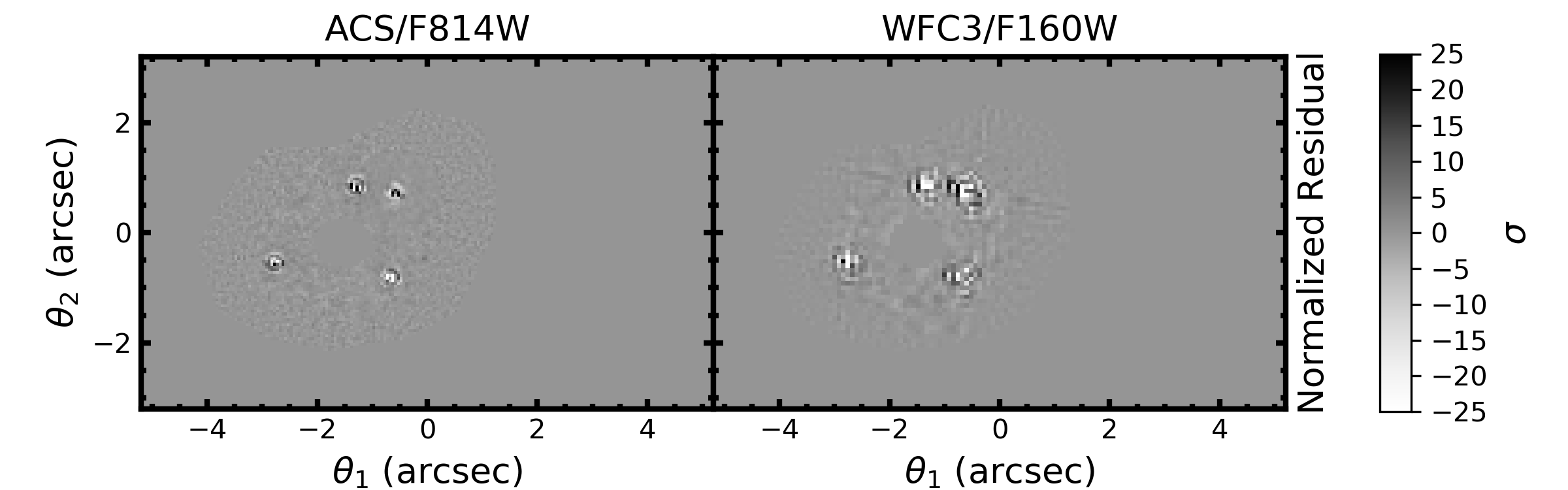}
\caption{
Normalized residual for the fiducial SPEMD model without power-law weighting.
\label{fig:spemd_nres_noplwht}}
\end{figure*}

\begin{figure*}
\includegraphics[width=\textwidth]{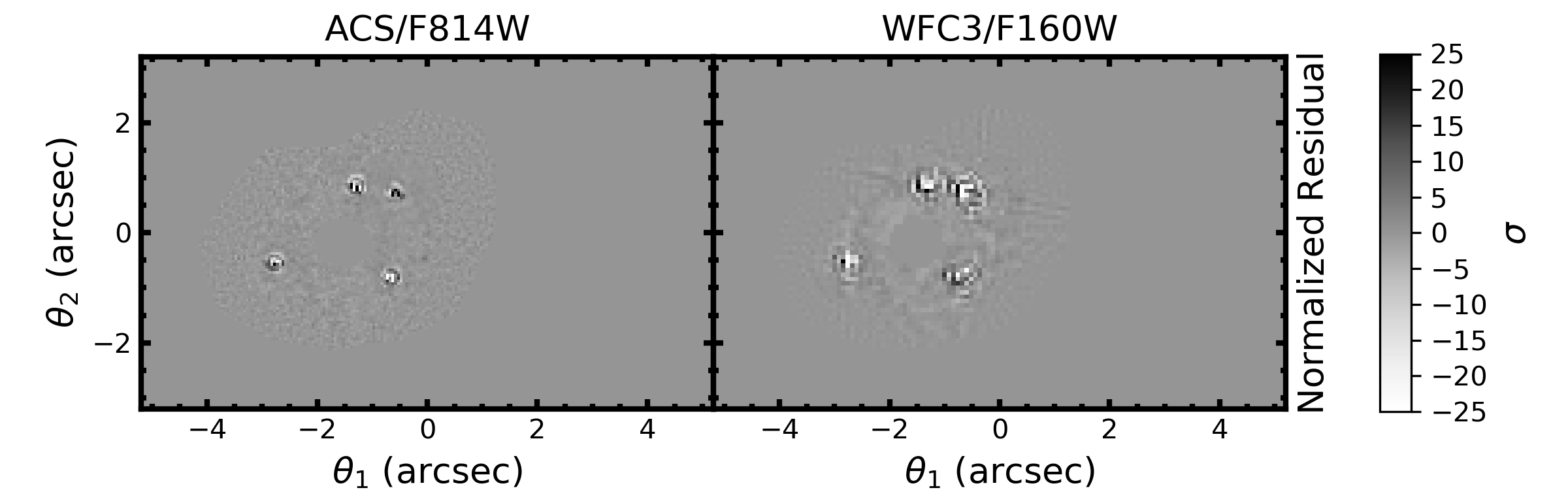}
\caption{
Same as Figure~\ref{fig:spemd_nres_noplwht} but for the fiducial composite model.
\label{fig:comp_nres_noplwht}}
\end{figure*}

For completeness, we show the normalized residuals for the fiducial SPEMD model (Figure~\ref{fig:spemd_nres_noplwht}) and fiducial composite model (Figure~\ref{fig:comp_nres_noplwht}) using the weight images without the power-law weighting in the region near the AGN images.  We see that there are strong residuals due to the AGN images, which motivates our downweighting of these regions.

\section{Impact of Different Cosmologies} \label{app:cosmo}
\begin{figure}
\includegraphics[width=0.5\textwidth]{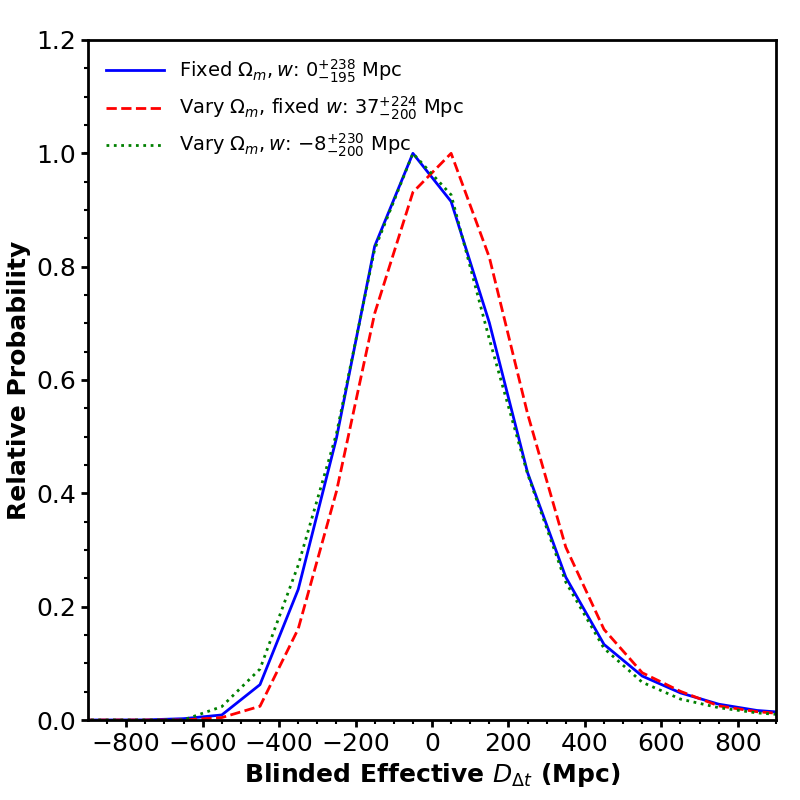}
\caption{
PDF of $\tdist$ for the various cosmologies.  We compare the fiducial SPEMD model to one in which $\Om$ is allowed to vary (with $\Om + \OL = 1$), and one in which $w$ is also allowed to vary.  The distributions are blinded by subtracting the median of the fiducial SPEMD model PDF.  The different cosmology tests are indicated by the legend, and the median and 68\% quantiles of the $\tdist$ distributions are given.  The median of the blinded effective time-delay distance PDF is insensitive to the extra cosmological parameters to within 1\%.
\label{fig:dt_cosmo}}
\end{figure}

In multi-lens-plane modeling, we need to sample the cosmological
parameters in order to carry out the ray tracing.
For computational reasons, we directly vary $H_{0}$ but keep other
cosmological parameters fixed ($\Om = 0.3$, $\OL =
0.7$, $w = -1$).  $\tdist$ has a weak dependence
on these other parameters in principle, but we found that varying these parameters affected the posterior $\tdist$ distribution by $< 1\%$ for \hequad~(H0LiCOW IV).
We perform a similar check for \wfilens~in which we run the fiducial SPEMD model while allowing either $\Om$ to vary, or allowing both $\Om$ and $w$ to vary.  The resulting effective $\tdist$ distributions are shown in Figure~\ref{fig:dt_cosmo}.  The peaks of the distribution are consistent to within $1\%$ of the absolute value, which shows that the results are insensitive to these extra cosmological parameters at the level of accuracy that we are currently working at, similar to \hequad.

In the case of multiple lens planes \citep[e.g.,][]{blandford1986,kovner1987,kochanek1988,schneider1992,petters2001,collett2014,mccully2014,schneider2014a}, there is not a unique time-delay distance for the system, but instead multiple time-delay distances between planes $i$ and $j$,
\be \label{eq:tdist_mp}
\tdist^{ij} \equiv (1+z_i) \frac{D_i D_j}{D_{ij}},
\ee
with $z_i$ being the redshift of plane $i$.  The multi-plane time delay is given by
\be \label{eq:td_mp}
t = \sum_{i=1}^{s-1} \frac{\tdist^{i,i+1}}{c} \left[ \frac{(\bm{\theta}_{i+1} - \bm{\theta}_{i})^{2}}{2} - \beta_{i,i+1}\psi_{i}( \bm{\theta}_{i}) \right],
\ee
where $s$ is the index of the source plane (starting with $i=1$ as the lowest-redshift lens plane and counting up towards the source plane).  $\psi_{i}$ is the lens potential related to the scaled deflection angle via $\nabla \psi_{i}=\bm{\alpha}_i$, and $\beta$ is a ratio of angular diameter distances among deflector planes and the source,
\be \label{eq:beta}
\beta_{ij} = \frac{D_{ij} D_s}{D_j D_{is}}.
\ee

From \eref{eq:td_mp}, we
see that the time delay depends on the multiple time-delay distances and
$\beta$ terms.  In general, it is difficult to constrain all of these
quantities independently, so we adopt specific cosmological models
to compute the distances for the ray tracing,
then compare the time-delay distance measurements from these different background
cosmologies.  For the case of \wfilens, where G2 is at a different redshift from the main lens plane and is not strongly
lensing the background source but merely perturbing it, the effect on
the time delays is weak.  The
lack of sensitivity to $\Om$ and $w$ seen in \fref{fig:dt_cosmo} suggests that \wfilens~is not
sensitive to the extra $\beta$ terms at an interesting level to probe it
directly in the same way as a double source plane lens
\citep[e.g.,][]{gavazzi2008,collett2014}.  Since the time delays are mostly set by the main lens plane, we
can measure the ``effective" $\tdist$ (which is $\tdist^{is}$ with $i$ as the main lens plane and $s$ as the source plane) that is
independent of assumptions on the background cosmology.  This robust distance
determination then permits us to constrain any reasonable cosmological model via the distance-redshift
relation.

\section{Model Parameters}\label{app:params}
We show the marginalized parameter constraints for each of the SPEMD models in Table~\ref{tab:spemd_params} and for each of the composite models in Table~\ref{tab:comp_params}.

\renewcommand\tabcolsep{2pt}
\renewcommand*\arraystretch{1.5}
\begin{table*}
\caption{SPEMD Model Parameters \label{tab:spemd_params}}
\begin{minipage}{\linewidth}
\begin{tabular}{l|ccccccc}
\hline
Parameter &
\multicolumn{7}{c}{Marginalized Constraints}
\\
\hline
 &
Fiducial &
AGNmask+1 &
AGNwht=0 &
Arc+1,60src &
Group &
Group + $z$=0.49 group &
Chameleon
\\
\hline
$\theta_{\mathrm{E}}~(\arcsec)$\footnote{Spherical-equivalent Einstein radius} &
$0.944_{-0.004}^{+0.004}$ & $0.943_{-0.004}^{+0.004}$ &$0.950_{-0.005}^{+0.004}$ &$0.933_{-0.004}^{+0.004}$ &$0.912_{-0.003}^{+0.003}$ &$0.927_{-0.006}^{+0.005}$ &$0.946_{-0.003}^{+0.003}$
\\
$q$ &
$0.80_{-0.01}^{+0.01}$ &$0.80_{-0.01}^{+0.01}$ &$0.81_{-0.01}^{+0.01}$ &$0.79_{-0.01}^{+0.01}$ &$0.79_{-0.01}^{+0.01}$ &$0.78_{-0.01}^{+0.01}$ &$0.78_{-0.01}^{+0.01}$
\\
$\theta_{q}$ ($^{\circ}$) &
$32.7_{-0.9}^{+0.8}$ &$32.8_{-0.9}^{+0.8}$ &$34.7_{-1.0}^{+1.1}$ &$31.9_{-0.7}^{+0.8}$ &$32.9_{-0.7}^{+0.6}$ &$33.5_{-0.7}^{+0.8}$ &$31.5_{-0.7}^{+0.7}$
\\
$\gamma^{\prime}$ &
$1.96_{-0.02}^{+0.02}$ &$1.98_{-0.02}^{+0.02}$ &$2.01_{-0.02}^{+0.01}$ &$1.90_{-0.01}^{+0.02}$ &$1.94_{-0.01}^{+0.01}$ &$1.95_{-0.01}^{+0.01}$ &$2.02_{-0.01}^{+0.01}$
\\
$\gext$ &
$0.117_{-0.004}^{+0.004}$ &$0.120_{-0.004}^{+0.003}$ &$0.125_{-0.004}^{+0.004}$ &$0.109_{-0.003}^{+0.003}$ &$0.110_{-0.004}^{+0.003}$ &$0.110_{-0.003}^{+0.003}$ &$0.126_{-0.003}^{+0.003}$
\\
$\theta_{\gamma}$ ($^{\circ}$) &
$-89.6_{-1.3}^{+0.9}$ &$-89.1_{-1.3}^{+0.8}$ &$-87.7_{-0.6}^{+0.5}$ &$88.0_{-0.6}^{+0.7}$ &$82.4_{-1.0}^{+0.9}$ &$82.6_{-1.9}^{+1.8}$ &$-88.4_{-0.5}^{+0.5}$
\\
X $\theta_{\mathrm{E}}~(\arcsec)$ &
$0.001_{-0.001}^{+0.001}$ &$0.001_{-0.001}^{+0.001}$ &$0.001_{-0.001}^{+0.001}$ &$0.001_{-0.001}^{+0.001}$ &$0.001_{-0.001}^{+0.001}$ &$0.001_{-0.001}^{+0.001}$ &$0.001_{-0.000}^{+0.001}$
\\
G2 $\theta_{\mathrm{E}}~(\arcsec)$ &
$0.926_{-0.027}^{+0.029}$ &$0.939_{-0.025}^{+0.023}$ &$0.930_{-0.036}^{+0.036}$ &$0.945_{-0.027}^{+0.021}$ &$0.868_{-0.008}^{+0.008}$ &$0.947_{-0.014}^{+0.017}$ &$0.929_{-0.014}^{+0.015}$
\\
G2 $q$ &
$0.66_{-0.02}^{+0.02}$ &$0.67_{-0.02}^{+0.02}$ &$0.67_{-0.03}^{+0.02}$ &$0.68_{-0.01}^{+0.02}$ &$0.70_{-0.02}^{+0.01}$ &$0.65_{-0.02}^{+0.01}$ &$0.69_{-0.02}^{+0.01}$
\\
G2 $\theta_{q}$ ($^{\circ}$) &
$41.5_{-4.7}^{+4.0}$ &$39.8_{-3.4}^{+4.1}$ &$38.2_{-5.3}^{+6.0}$ &$35.2_{-3.4}^{+2.9}$ &$40.4_{-1.9}^{+2.6}$ &$35.4_{-2.1}^{+2.4}$ &$46.2_{-4.0}^{+3.7}$
\\
\hline
\end{tabular}
\\
{\footnotesize Reported values are medians, with errors corresponding to the 16th and 84th percentiles.}
\\
{\footnotesize Angles are measured east of north.}
\end{minipage}
\end{table*}
\renewcommand*\arraystretch{1.0}
\renewcommand\tabcolsep{6pt}

\renewcommand*\arraystretch{1.5}
\begin{table*}
\caption{Composite Model Parameters \label{tab:comp_params}}
\begin{minipage}{\linewidth}
\begin{tabular}{l|cccccc}
\hline
Parameter &
\multicolumn{6}{c}{Marginalized Constraints}
\\
\hline
 &
Fiducial &
AGNmask+1 &
AGNwht=0 &
Arc+1,60src &
Group &
Group + $z$=0.49 group
\\
\hline
Stellar M/L ($\mathrm{M_{\odot}/L_{\odot}}$)\footnote{M/L within $\theta_{\mathrm{E}}$ for rest-frame $V$ band.  The point source component of the lens light is assumed to be from low-level AGN emission as opposed to stellar light and is not included in the calculation.  The given uncertainties are a combination of statistical effects and a systematic uncertainty equal to the difference between the calculated M/L with and without the point source contribution.  The stellar mass is calculated assuming $H_{0} = 70~\mathrm{km~s^{-1}~Mpc^{-1}}$, $\Omega_{\mathrm{m}} = 0.3$, $\Omega_{\Lambda} = 0.7$, but changes in the cosmology affect the M/L by a negligible amount.} &
$2.2_{-0.2}^{+0.2}$ &$2.1_{-0.2}^{+0.2}$ &$2.1_{-0.2}^{+0.2}$ &$2.1_{-0.2}^{+0.2}$ &$2.1_{-0.2}^{+0.2}$ &$2.1_{-0.2}^{+0.2}$
\\
Chameleon1 $q$ &
$0.759_{-0.001}^{+0.003}$ &$0.761_{-0.001}^{+0.001}$ &$0.761_{-0.001}^{+0.001}$ &$0.758_{-0.001}^{+0.001}$ &$0.763_{-0.001}^{+0.001}$ &$0.765_{-0.000}^{+0.000}$
\\
Chameleon1 $\theta_{q}$ ($^{\circ}$) &
$23.1_{-0.2}^{+0.2}$ &$23.1_{-0.2}^{+0.2}$ &$23.3_{-0.1}^{+0.2}$ &$23.3_{-0.2}^{+0.1}$ &$23.3_{-0.1}^{+0.1}$ &$23.4_{-0.1}^{+0.1}$
\\
Chameleon2 $q$ &
$0.771_{-0.001}^{+0.001}$ &$0.772_{-0.001}^{+0.001}$ &$0.774_{-0.001}^{+0.001}$ &$0.770_{-0.001}^{+0.001}$ &$0.770_{-0.001}^{+0.001}$ &$0.767_{-0.001}^{+0.001}$
\\
Chameleon2 $\theta_{q}$ ($^{\circ}$) &
$26.1_{-0.3}^{+0.3}$ &$26.4_{-0.3}^{+0.3}$ &$25.5_{-0.3}^{+0.3}$ &$25.8_{-0.3}^{+0.3}$ &$26.4_{-0.2}^{+0.2}$ &$26.3_{-0.2}^{+0.2}$
\\
NFW $\kappa_{0,\mathrm{h}}$ &
$0.143_{-0.007}^{+0.003}$ &$0.148_{-0.003}^{+0.003}$ &$0.147_{-0.004}^{+0.003}$ &$0.159_{-0.004}^{+0.005}$ &$0.136_{-0.003}^{+0.002}$ &$0.149_{-0.001}^{+0.001}$
\\
NFW $r_{\mathrm{s}}~(\arcsec)$ &
$10.46_{-0.14}^{+0.07}$ &$10.31_{-0.08}^{+0.12}$ &$10.53_{-0.08}^{+0.08}$ &$10.32_{-0.08}^{+0.07}$ &$11.35_{-0.10}^{+0.05}$ &$11.10_{-0.05}^{+0.05}$
\\
NFW $q$ &
$0.88_{-0.01}^{+0.02}$ &$0.90_{-0.00}^{+0.00}$ &$0.92_{-0.01}^{+0.01}$ &$0.88_{-0.01}^{+0.01}$ &$0.88_{-0.01}^{+0.01}$ &$0.89_{-0.02}^{+0.01}$
\\
NFW $\theta_{q}$ ($^{\circ}$) &
$71.6_{-1.0}^{+1.1}$ &$73.5_{-0.4}^{+0.4}$ &$73.6_{-0.7}^{+0.5}$ &$71.3_{-0.8}^{+0.7}$ &$71.7_{-0.4}^{+0.3}$ &$73.4_{-0.4}^{+0.4}$
\\
$\gext$ &
$0.138_{-0.002}^{+0.001}$ &$0.137_{-0.001}^{+0.001}$ &$0.135_{-0.001}^{+0.001}$ &$0.137_{-0.001}^{+0.002}$ &$0.134_{-0.002}^{+0.002}$ &$0.133_{-0.001}^{+0.002}$
\\
$\theta_{\gamma}$ ($^{\circ}$) &
$-89.1_{-0.3}^{+0.3}$ &$-88.6_{-0.2}^{+0.1}$ &$-88.1_{-0.1}^{+0.1}$ &$89.9_{-0.2}^{+0.3}$ &$89.5_{-0.4}^{+0.3}$ &$89.2_{-0.4}^{+0.5}$
\\
X $\theta_{\mathrm{E}}~(\arcsec)$ &
$0.018_{-0.002}^{+0.002}$ &$0.015_{-0.002}^{+0.002}$ &$0.010_{-0.002}^{+0.002}$ &$0.021_{-0.002}^{+0.002}$ &$0.018_{-0.002}^{+0.003}$ &$0.018_{-0.002}^{+0.002}$
\\
G2 $\theta_{\mathrm{E}}~(\arcsec)$ &
$1.034_{-0.005}^{+0.004}$ &$1.036_{-0.003}^{+0.004}$ &$1.046_{-0.005}^{+0.005}$ &$1.034_{-0.004}^{+0.004}$ &$1.026_{-0.002}^{+0.002}$ &$1.005_{-0.003}^{+0.002}$
\\
G2 $q$ &
$0.94_{-0.01}^{+0.01}$ &$0.91_{-0.01}^{+0.01}$ &$0.90_{-0.01}^{+0.01}$ &$0.95_{-0.01}^{+0.01}$ &$0.93_{-0.01}^{+0.01}$ &$0.93_{-0.00}^{+0.00}$
\\
G2 $\theta_{q}$ ($^{\circ}$) &
$39.7_{-2.1}^{+1.3}$ &$40.5_{-0.7}^{+0.6}$ &$40.3_{-0.8}^{+0.8}$ &$40.5_{-0.8}^{+1.0}$ &$29.4_{-0.8}^{+1.0}$ &$40.5_{-1.0}^{+0.8}$
\\
\hline
\end{tabular}
\\
{\footnotesize Reported values are medians, with errors corresponding to the 16th and 84th percentiles.}
\\
{\footnotesize Angles are measured east of north.}
\end{minipage}
\end{table*}
\renewcommand*\arraystretch{1.0}

\section{Source resolution changes}\label{app:srcres}
In Table~\ref{tab:dt_srctest}, we show the BIC and BIC weight values for the source resolution tests described in Section~\ref{subsec:bic}.  All models are the fiducial models run with different source resolutions.  The BIC weights have been renormalized.

\renewcommand*\arraystretch{1.5}
\begin{table*}
\caption{Effective time-delay distance and BIC weighting for different source resolutions \label{tab:dt_srctest}}
\begin{minipage}{\linewidth}
\begin{tabular}{l|lll}
\hline
Model &
$\tdist$ (Mpc) &
$\Delta$BIC &
Relative BIC weight
\\
\hline
SPEMD fiducial, $47\times47$ source &
$4711_{-184}^{+232}$ &
92 &
0.316
\\
SPEMD fiducial, $48\times48$ source &
$4706_{-180}^{+227}$ &
76 &
0.470
\\
SPEMD fiducial, $49\times49$ source &
$4685_{-210}^{+241}$ &
139 &
0.049
\\
SPEMD fiducial, $50\times50$ source &
$4640_{-195}^{+238}$ &
86 &
0.365
\\
SPEMD fiducial, $51\times51$ source &
$4631_{-184}^{+229}$ &
62 &
0.605
\\
SPEMD fiducial, $52\times52$ source &
$4712_{-193}^{+244}$ &
94 &
0.295
\\
SPEMD fiducial, $53\times53$ source &
$4691_{-202}^{+243}$ &
59 &
0.641
\\
SPEMD fiducial, $54\times54$ source &
$4645_{-172}^{+230}$ &
32 &
0.864
\\
SPEMD fiducial, $56\times56$ source &
$4686_{-186}^{+231}$ &
0 &
1.000
\\
SPEMD fiducial, $58\times58$ source &
$4724_{-188}^{+238}$ &
14 &
0.956
\\
SPEMD fiducial, $60\times60$ source &
$4844_{-183}^{+243}$ &
12 &
0.964
\\
\hline
Composite fiducial, $47\times47$ source &
$4933_{-253}^{+401}$ &
0 &
1.000
\\
Composite fiducial, $48\times48$ source &
$4753_{-251}^{+418}$ &
10 &
0.507
\\
Composite fiducial, $49\times49$ source &
$5011_{-259}^{+420}$ &
114 &
0.000
\\
Composite fiducial, $50\times50$ source &
$4732_{-247}^{+417}$ &
26 &
0.175
\\
Composite fiducial, $51\times51$ source &
$4810_{-254}^{+417}$ &
118 &
0.000
\\
Composite fiducial, $52\times52$ source &
$4956_{-261}^{+422}$ &
60 &
0.013
\\
Composite fiducial, $53\times53$ source &
$4893_{-258}^{+431}$ &
69 &
0.006
\\
Composite fiducial, $54\times54$ source &
$4819_{-258}^{+422}$ &
114 &
0.000
\\
Composite fiducial, $56\times56$ source &
$4886_{-251}^{+423}$ &
112 &
0.000
\\
Composite fiducial, $58\times58$ source &
$4885_{-250}^{+424}$ &
129 &
0.000
\\
Composite fiducial, $60\times60$ source &
$4843_{-248}^{+415}$ &
189 &
0.000
\\
\hline
\end{tabular}
\\
{\footnotesize Reported values are medians, with errors corresponding to the 16th and 84th percentiles.}
\end{minipage}
\end{table*}
\renewcommand*\arraystretch{1.0}

\section{Further tests on the selection and implementation of the weighted number count constraints} \label{app:kappa}

\subsection{The use of $\gamma_\mathrm{ext}$ constraints} \label{app:closestruct}

In \sref{sec:lensmod} we have shown that a large value of $\gamma_\mathrm{ext}\sim0.10-0.14$ is required to model the lens system, even after the nearby perturbers are being accounted for. In addition, the different inner mass profiles of the lens, explored in \sref{sec:lensmod}, all require a large external shear. This means that the large-scale environment and/or LOS structures must be responsible for this effect. 

The specific large-scale structures responsible for this effect have not been identified, since the mass models incorporating the two large galaxy groups identified in H0LiCOW X still require a large $\gamma_\mathrm{ext}$. This was also pointed out in previous studies, based on more limited ancillary data and less detailed modeling, both superseeded in our work. For example, \citet{wong2011} compute an expected $\gamma_\mathrm{ext}=0.08\pm0.03$ based on a spectroscopic galaxy catalogue, where galaxies are treated as SIS, and galaxy groups as NFW halos. They note that the orientation of this shear caused by the environment does not match the one obtained from the mass models of the lens. \citet{vuissoz2008}, based on mass models constrained only by the positions of the quasar images (and in some models by the measured time delays as well), find $\gamma_\mathrm{ext}$ as large as 0.3, but as small as 0.06 if they incorporate the galaxy group which includes the lens. However, our more complete spectroscopic catalogue shows that the group centroid is more distant from the lens ($\sim30\arcsec$ compared to $\sim10\arcsec$), leading to a larger amount of external shear necessary to model the system, after accounting for the group.

Based on the above, while we concede that we cannot find a mass model which explains most of the shear, we are confident that the external shear values measured for the various models in \sref{sec:lensmod} are robust, and that we are justified to attribute the shear to the lens environment and/or LOS structures which are not captured in our mass models, but are captured in our statistical approach to computing $\kappa_\mathrm{ext}$. This justifies our use of the $\gamma_\mathrm{ext}$ constraints to infer $\kappa_\mathrm{ext}$ in Figure~\ref{fig:kappa}. As we will show in Appendix~\ref{app:kappachoices}, this constraint has a dominant effect on our inference, which is expected from results of ray-tracing through the MS \citep[e.g., see Figure~8 in][]{collett2016}, where $|\kappa|$ and $|\gamma|$ are found to correlate. We note that the lens \rxjlens , which has also been modeled as part of H0LiCOW \citep{suyu2013,chen2019}, also has a fairly large measured shear of $\sim0.08$, which was used to constrain its $\kappa_\mathrm{ext}$. Independently, its $\kappa_\mathrm{ext}$ was measured using a different methodology by \citet{mccully2017}, which has found $\kappa_\mathrm{ext}$ to be offset to smaller, but nonetheless consistent values. 

After unblinding our analysis, we checked what the impact of using $\kappa_\mathrm{ext}$ inferred without the shear constraint would have been on our analysis. We obtain $H_{0}=76_{-3.0}^{+2.9}~\mathrm{km~s^{-1}~Mpc^{-1}}$, a value 6\% larger than the one in \sref{sec:results}, and with significantly increased statistical precision due to the tighter $\kappa_\mathrm{ext}$ distributions (see Figure~\ref{fig:kappabar}).

In order to avoid biases, we must ensure that we construct $P(\kappa_\mathrm{ext}|\gamma,...)$ such that it is consistent with the discussion above. In particular, we must ensure that when we select LOS with large shear from the MS, the shear is not due to galaxies very close to the LOS, or to galaxy groups/clusters so massive that we would incorporate them in our mass models, in the real data. We address each of these in the following. 

\subsubsection{Galaxies close to the LOS} \label{app:nearbystruct}
In order to ensure that the weighted number counts are not dominated by the galaxies very close to the LOS, in \citet[][]{greene2013}; H0LiCOW III, IX we have used a $5\arcsec$-radius inner mask, and set an upper limit to the weights incorporating $1/r$ of $1/10\arcsec$ for each galaxy. While the usage of a mask when computing weighted number counts is agnostic to the actual existence of galaxies inside of it, and therefore their contribution to $\kappa_\mathrm{ext}$ at that particular spatial location, here and in \citet{chen2019} we only select LOS from the MS which have no galaxies inside the $5\arcsec$-radius. Such galaxies, if above the magnitude threshold, would be modeled explicitly in the real data. This radius corresponds to the inner $\gtrsim4$ pixels of the $(\kappa,\gamma)$ map from \citet{hilbert2009}. The addition of this constraint has the effect of lowering $\kappa_\mathrm{ext}$ by approximately $0.1\kappa_\mathrm{ext}$.

\subsubsection{Massive large-scale structures} \label{app:largestruct}

For the FOV around \wfilens, we used our spectroscopic catalogue to identify large galaxy groups, which we incorporated in our mass models, in cases where they exceeded the flexion shift threshold. We must take this fact into account when we infer $\kappa_\mathrm{ext}$ by selecting LOS from the MS. The easiest way to account for this in the MS is to use a complete catalogue of galaxy groups, and to test at each location of the $\kappa$ map whether their flexion shifts exceed the threshold. If so, those LOS are removed. In this way, we only use the $\kappa_\mathrm{ext}$ distribution free of the contribution of massive structures, which is what is needed. 

In practice, we use a catalogue of galaxies from the MS, which identifies their parent halos and the masses of those halos. We impose the observational constraints from \wfilens, that the groups (in this case the parent halos) we model have at least 5 galaxies within our magnitude threshold. For these haloes, we convert their masses into velocity dispersions (assuming the SIS mass profile, for simplicity), and compute the radius of the circle around the halo centroid inside which the flexion shift for each of these structures exceeds the threshold. We then remove the LOS inside those circles from the $\kappa_\mathrm{ext}$ computation. We have compared the $\kappa_\mathrm{ext}$ distributions with and without removing these groups, and have found them to be indistinguishable, given the rarity of these groups. We therefore safely ignore the effect of massive but rare large scale structures on our analysis.

\subsection{The choice of conjoined constraints} \label{app:kappachoices}

\begin{figure*}
\includegraphics[width=\textwidth]{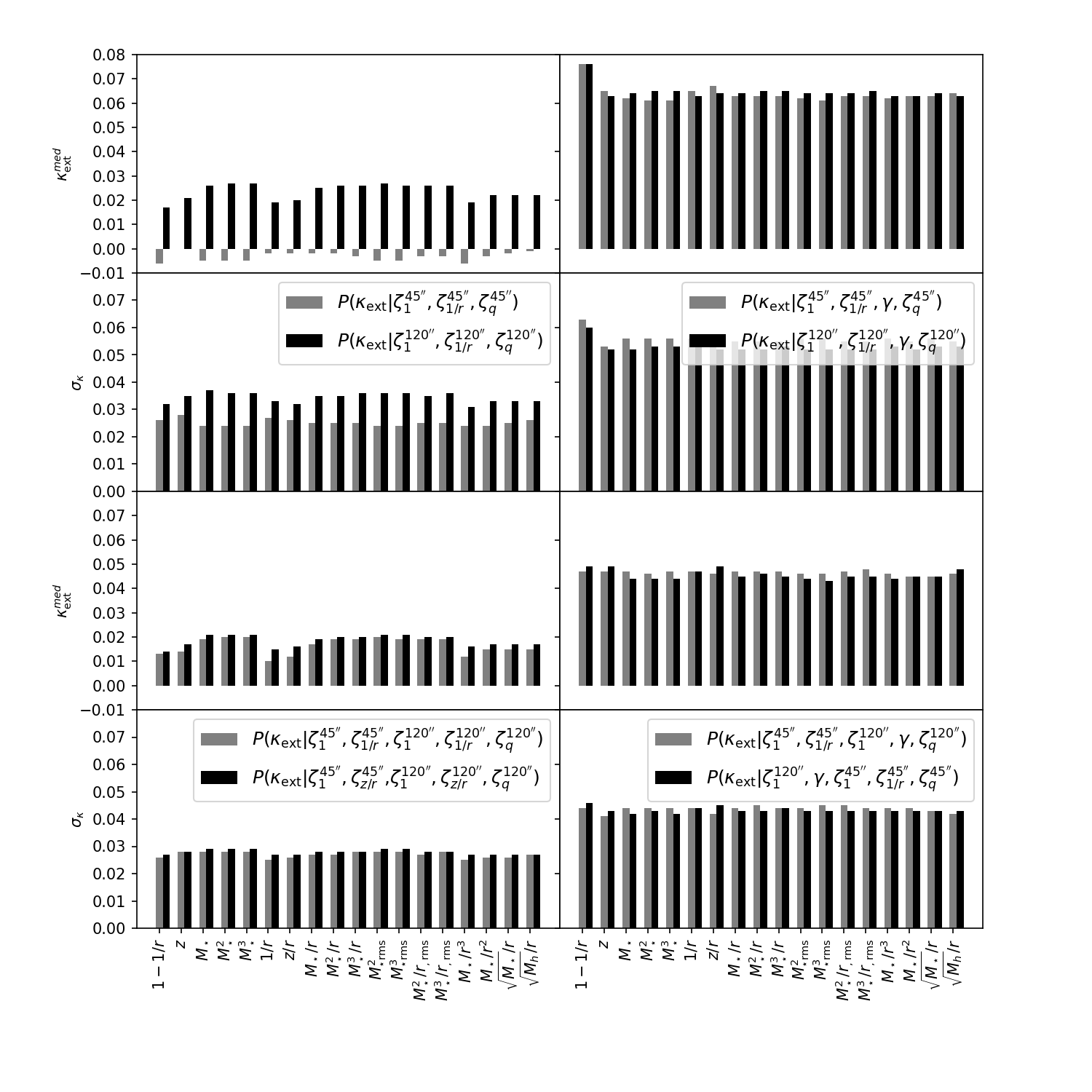}
\caption{Statistics of $P(\kappa_\mathrm{ext})$ for various combinations of weighted counts with or without shear, in the case of the fiducial lensing model. ``$1-1/r$'' means that the constraint from '$\zeta_{1/r}$' is not used. 
  \label{fig:kappabar}}
\end{figure*}

\citet{greene2013} and H0LiCOW III have explored different combinations of weighted number count ratio constraints, most of which incorporate unweighted number counts $\zeta_1$, as well as $\zeta_{1/r}$. These correspond to the most robust constraints which can be determined from imaging data, usually with the tightest uncertainties. In Figure~\ref{fig:kappabar} we explore the median and standard deviation of the $\kappa_\mathrm{ext}$ distributions for various combinations of constraints, typically incorporating the two constraints above. We explore combinations of constraints measured inside the same aperture, as well as measured inside both the $45\arcsec$- and $120\arcsec$-radii apertures. The latter is because the weighted number count ratios we measure in Table~\ref{tab:overdens} appear larger inside the $120\arcsec$-radius aperture, suggesting that \wfilens~is positioned on the outskirts of a galaxy overdensity, and this observation might include statistical information useful for tightening the $\kappa_\mathrm{ext}$ distribution. 

\begin{figure*}
\includegraphics[width=\textwidth]{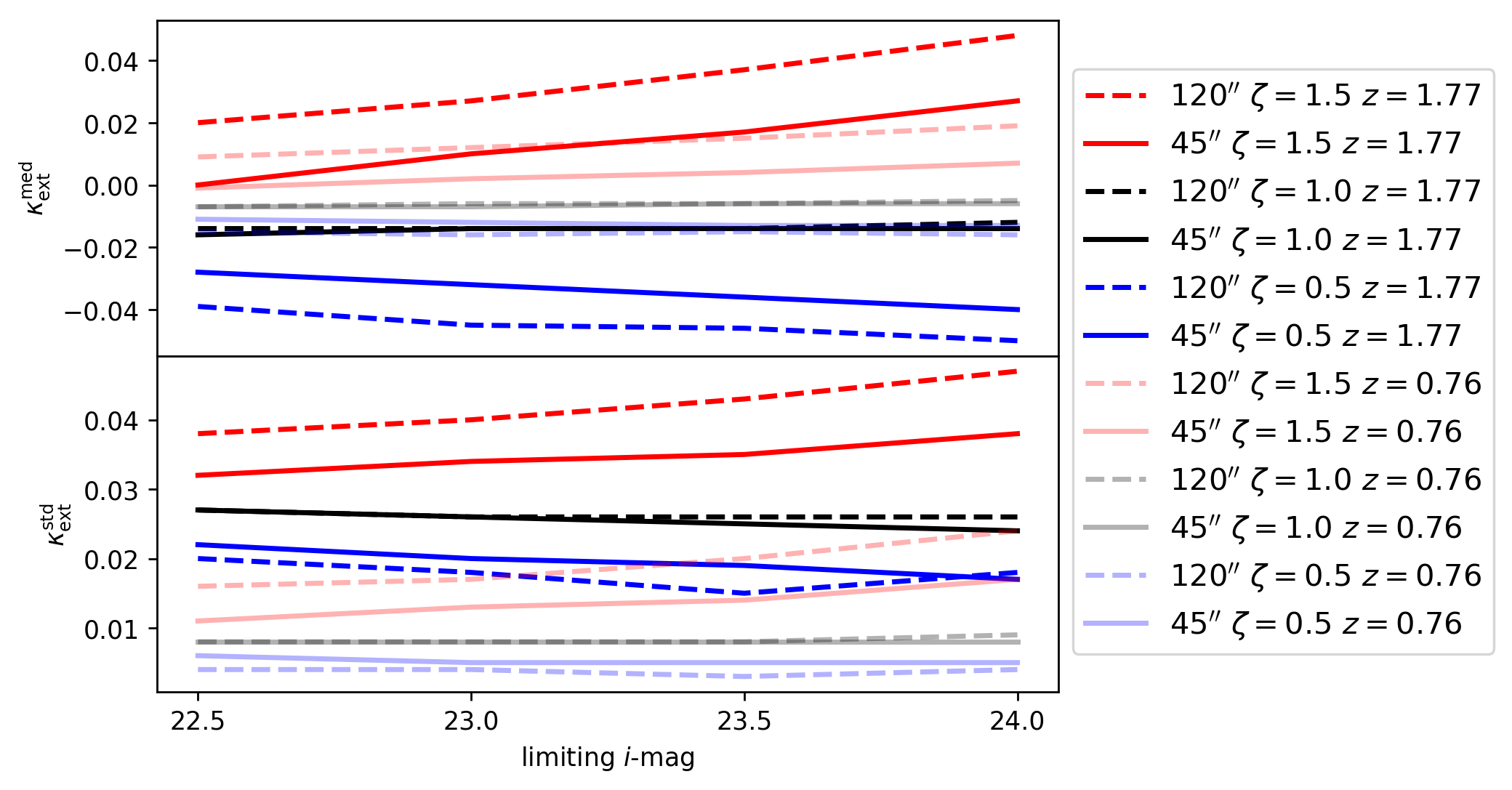}
\caption{{\it Upper plot:} Behavior of the median of $\kappa_\mathrm{ext}$ as a function of the limiting magnitude $i=22.5, 23.0, 23.5$ and $24.0$ mag, for two choices of the aperture radii and three choices of the relative (unweighted) number counts. For completeness, we show the behavior for two source redshifts in the MS, $z=1.77$ (solid symbols) and $z=0.76$ (transparent symbols). {\it Lower plot:} Behavior of the semi-difference between the 16th and 84th percentiles for the same distributions. 
  \label{fig:chrischeck}}
\end{figure*}

Comparing the distributions constrained by $(\zeta^{45\arcsec}_{1},\zeta^{45\arcsec}_{1/r},\zeta^{45\arcsec}_{q})$ and $(\zeta^{120\arcsec}_{1},\zeta^{120\arcsec}_{1/r},\zeta^{120\arcsec}_{q})$ we see that the former produce lower $\kappa^\mathrm{med}_\mathrm{ext}$. This is partly because of the somewhat lower weighted number counts inside the smaller aperture, but mostly due to the fact that the same overdensity over a larger aperture implies a larger structure, and therefore larger convergence. This is demonstrated in Figure~\ref{fig:chrischeck}. In the case of constraints from both apertures, such as $(\zeta^{45\arcsec}_{1},\zeta^{45\arcsec}_{1/r},\zeta^{120\arcsec}_{1},\zeta^{120\arcsec}_{1/r},\zeta^{120\arcsec}_{q})$, $\kappa^\mathrm{med}_\mathrm{ext}$ is brought closer to the average value obtained from the two individual apertures. Amongst the various $\zeta_q$, $\kappa^\mathrm{med}_\mathrm{ext}$ varies by $\sim 0.01$, or at the $1\%$ level.

Once the large $\gamma_\mathrm{ext}$ is used as a constraint, $\kappa^\mathrm{med}_\mathrm{ext}$ reaches much larger values, as expected. All distributions including the $\zeta_{1/r}$ constraint are in agreement to within the $1\%$ level, even if we compare constraints from different aperture radii. Interestingly, once the constraints from different apertures are combined, the distributions for different $\zeta_q$ remain in agreement, but $\kappa^\mathrm{med}_\mathrm{ext}$ decreases by almost $\sim2\%$, or $\sim0.3\sigma$. We cannot fully explain this result, but we note that we have explored the $\kappa_\mathrm{ext}-\gamma_\mathrm{ext}$ plane constrained by either, as well as both apertures. It appears that this behavior is a result of both the very large $\gamma_\mathrm{ext}$ value and of the large weighted count constraints, and disappears for smaller values of either.

We do not use more than 5 conjoined constraints, because the number of MS LOS surviving the cut becomes too small, and the distributions are noisy. Our conclusions regarding the convergence distributions from different constraints are similar to those from H0LiCOW III, wherever a direct comparison is possible.

\subsection{Tests for bias and scatter} \label{app:bias}

\begin{figure*}
\includegraphics[width=\textwidth]{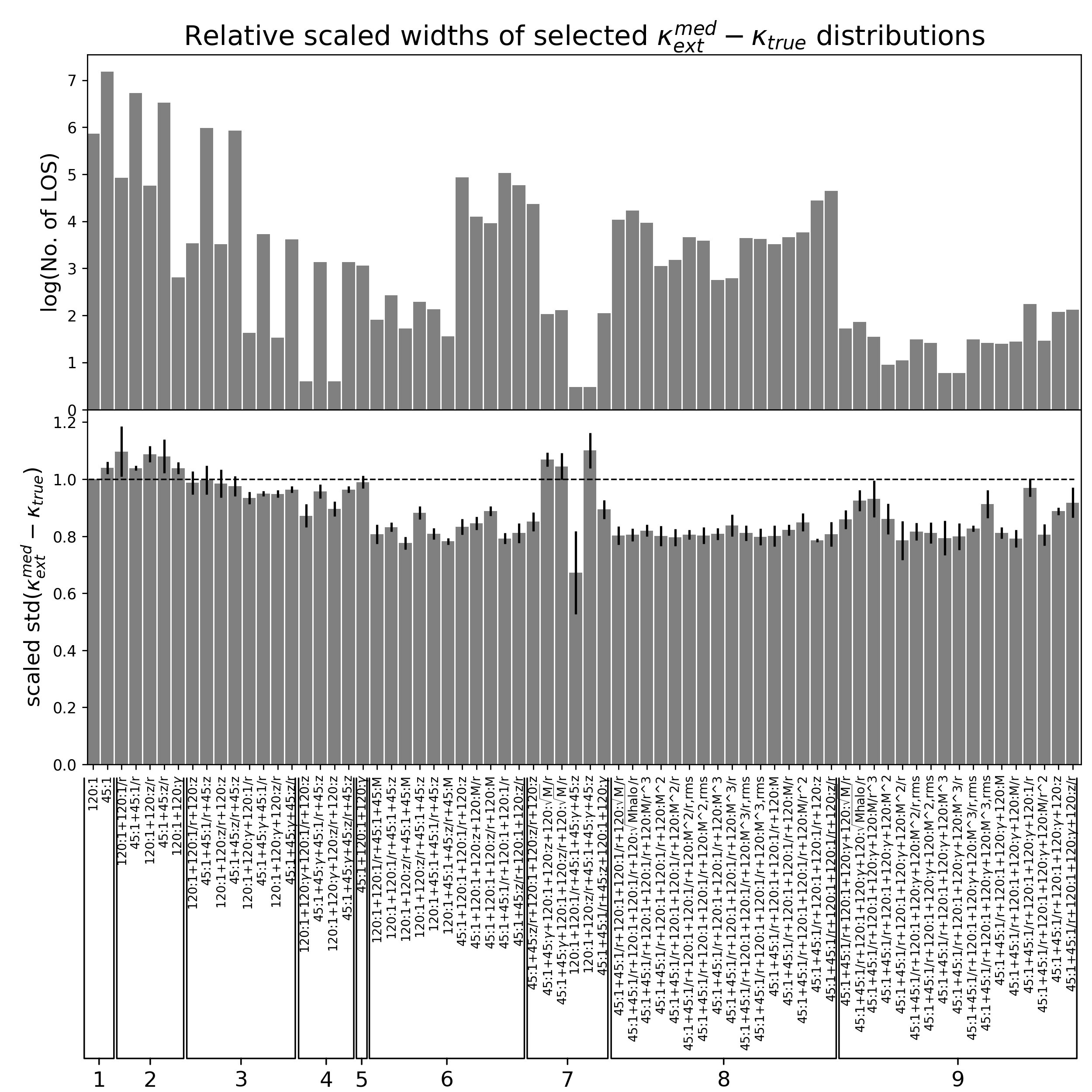}
\caption{\textit{Lower plot:} Standard deviations of 70 representative $\kappa^{med}_\mathrm{ext}-\kappa_\mathrm{true}$ distributions, relative to the standard deviation of $P(\kappa^{med}_\mathrm{ext}-\kappa_\mathrm{true}|\zeta_1^{120''})$, and scaled with respect to $\kappa^\mathrm{med}_\mathrm{ext}$. See text for details. Error bars represent the scatter resulting from running the simulations 3 times. Eight groups of distributions are identified: 1) single apertures, single constraint, 2) single apertures, two joint constraints, 3) single apertures, three joint constraints, 4) single apertures, four joint constraints, 5) two apertures, three joint constraints, 6) two apertures, four joint constraints, 7) two apertures, five joint constraints, 8) distributions of the type $P(\kappa^\mathrm{med}_\mathrm{ext}-\kappa_\mathrm{true}|\zeta_1^{45''},\zeta_{1/r}^{45''},\zeta_{1}^{120''},\zeta_{1/r}^{120''},\zeta_{q}^{120''})$, where $q$ stands for various constraints; 9) distributions of the type $P(\kappa^\mathrm{med}_\mathrm{ext}-\kappa_\mathrm{true}|\zeta_1^{45''},\zeta_{1/r}^{45''},\zeta_{1}^{120''},\zeta_{q}^{120''},\gamma)$. \textit{Upper plot:} median number of LOS which survived the given constraints and were used to compute each distribution. The more LOS, the more reliable the results of this simulations are.
  \label{fig:scaledstd}}
\end{figure*}

In H0LiCOW III, we ran two types of simulations using the MS data, to check whether or not our combination of constraints biases $P(\kappa_\mathrm{ext})$. We ran these for the case of \wfilens~as well, and found that $P(\kappa^\mathrm{med}_\mathrm{ext}-\kappa^\mathrm{true}_\mathrm{ext}|\zeta_q,...)$ is centered on zero for all $\zeta_q,...$ combinations in Figures~\ref{fig:kappabar} and \ref{fig:scaledstd}. All our $P(\kappa_\mathrm{ext})$ distributions are, therefore, unbiased.

We also wish to determine which combinations of constraints produces the tightest $P(\kappa_\mathrm{ext})$. However, from \citet{greene2013} as well as Figure~\ref{fig:kappabar} we see that the medians and standard deviations of  $P(\kappa_\mathrm{ext})$ are always correlated, such that smaller $\kappa^\mathrm{med}_\mathrm{ext}$ implies smaller $\sigma_\kappa$. We therefore follow a different approach, where we use the scatter in $P(\kappa^\mathrm{med}_\mathrm{ext}-\kappa^\mathrm{true}_\mathrm{ext}|\zeta_q,...)$. We normalize the scatter in  $P(\kappa^\mathrm{med}_\mathrm{ext}-\kappa^\mathrm{true}_\mathrm{ext}|\zeta^{45\arcsec}_1)$\footnote{Using instead $P(\kappa^\mathrm{med}_\mathrm{ext}-\kappa^\mathrm{true}_\mathrm{ext}|\zeta^{120\arcsec}_1)$ produces consistent results.} to the unit value, and we fit the standard deviation of $P(\kappa_\mathrm{ext}|\zeta^{45\arcsec}_1)$ as a function of its median with a linear function, while we vary $\zeta^{45\arcsec}_1$. Finally, we divide the standard deviation of $P(\kappa^\mathrm{med}_\mathrm{ext}-\kappa^\mathrm{true}_\mathrm{ext}|\zeta_q,...)$ by the value of this linear function at the corresponding $\kappa^\mathrm{med}_\mathrm{ext}$, and we show the results, for selected combinations of constraints, in Figure~\ref{fig:scaledstd}. We can see that combining the results from the two different aperture radii typically results in scaled standard deviations reduced by up to $\sim20\%$, whether the $\gamma_\mathrm{ext}$ constraint is used or not. While the figure shows a spike in value for the scaled standard deviation corresponding to the combination of weights we choose to consider as fiducial in this work, $(\zeta^{45\arcsec}_{1},\zeta^{45\arcsec}_{1/r},\zeta^{120\arcsec}_{1},\zeta^{120\arcsec}_{1/r},\gamma)$, most of the surrounding similar distributions (with only one of the $\zeta_q$ constraints replaced) show small values, and we know from Figure~\ref{fig:kappabar} that the resulting $P(\kappa_\mathrm{ext})$ for these constraints are consistent with each other.

\bsp
\label{lastpage}
\end{document}